\documentclass[prb,twocolumn]{revtex4}
\usepackage[dvips]{graphicx}
\usepackage{latexsym,amsmath,amssymb,bm,euscript}
\usepackage[dvips]{color}

\def\etal{~\textit{et~al.}} 
\def\ra{\rangle} 
\def\la{\langle} 
\def\up{\uparrow}
\def\dn{\downarrow}

\begin{document}

\title{D-wave correlated Critical Bose Liquids in two dimensions}

\author{Olexei I. Motrunich}
\affiliation{Department of Physics, California Institute of Technology,
Pasadena, CA 91125}
\author{Matthew P. A. Fisher}
\affiliation{Kavli Institute for Theoretical Physics, 
University of California, Santa Barbara, CA 93106}

\date{March 9, 2007}

\begin{abstract}

We develop a description of a new quantum liquid phase of interacting 
bosons confined in two dimensions which possesses relative D-wave 
two-body correlations.  We refer to this stable quantum phase as a 
D-wave Bose Liquid (DBL).  The DBL has no broken symmetries, 
supports gapless boson excitations which reside on ``Bose surfaces" 
in momentum space, and exhibits power law correlation functions 
characterized by a manifold of continuously variable exponents.
While the DBL can be constructed for bosons moving in the 2d continuum,
the state only respects the point group symmetries of the square lattice.
On the square lattice the DBL respects all symmetries and does not 
require a particular lattice filling.  But lattice effects do allow for 
the possibility of a second distinct phase, a quasi-local variant which 
we refer to as a D-wave Local Bose Liquid (DLBL).
Remarkably, the DLBL has short-range boson correlations and hence no 
Bose surfaces, despite sharing gapless excitations and other critical 
signatures with the DBL.  Moreover, both phases are metals with a 
resistance that vanishes as a power of the temperature.
We establish these results by constructing a class of many-particle 
wavefunctions for the DBL, which are time reversal invariant analogs of 
Laughlin's quantum Hall wavefunction for bosons in a half-filled 
Landau level.
A gauge theory formulation leads to a simple mean field theory and a 
suitable $N-$flavor generalization enables incorporation of gauge field
fluctuations to deduce the properties of the DBL/DLBL in a controlled 
and systematic fashion.
Various equal time correlation functions thereby obtained are in 
qualitative accord with the properties inferred from the variational 
wavefunctions.  
We also identify a promising microscopic Hamiltonian which might manifest
the DBL or DLBL, and perform a variational energetics study comparing to 
other competing phases including the superfluid.
We suggest how the D-wave Bose Liquid wavefunction can be suitably 
generalized to describe an itinerant non-Fermi liquid phase of electrons 
on the square lattice with a no double occupancy constraint, 
a D-wave metal phase.

\end{abstract}

\maketitle

\section{Introduction}
\label{sec:intro}

The principal roadblock impeding progress in disentangling the physics 
of the cuprate superconductors is arguably our inability to access 
quantum ground states of two-dimensional itinerant electrons which are 
qualitatively distinct from a Landau Fermi liquid.
Overcoming this obstruction is of paramount importance, indispensable 
in explaining the strange metal behavior observed near optimal doping 
and a likely requisite to account for the emergent pseudo-gap at lower 
energies.
A putative underlying paramagnetic Mott insulator provides the 
scaffolding for a popular class of theories, which view the pseudo-gap 
as a lightly doped spin liquid.
\cite{Anderson, Baskaran, KotliarLiu, IoffeLarkin, LeeNagaosaWen}  
Significant progress has been made in developing the groundwork and 
there now exists a well established theoretical framework to describe a 
myriad of distinct spin liquids.  For the cuprates the most promising
spin liquids are described in terms of fermionic spinons minimally 
coupled to a compact $U(1)$ gauge field and moving in various 
background fluxes.
Upon doping formidable challenges arise.  Bosonic holons carrying the 
electron charge become mobile carriers and lead to electrical conduction.
But at low temperatures Bose condensation appears inevitable and this 
leads to Fermi liquid behavior, either a metal with conventional Landau 
quasiparticles or a  BCS superconductor if the spinons are paired.
Accessing a pseudo-gap or a strange metal which conduct electricity 
despite the absence of long-lived Landau quasiparticles requires doped 
holons that form an uncondensed quantum Bose fluid rather than a 
condensed superfluid.  But is this possible, even in principle?
If possible, what properties would such a putative ``Bose metal'' 
exhibit?\cite{Feigelman, Dalidovich, Galitski, Alicea}
And what theoretical framework is appropriate?
The slave-particle gauge theory approach has been stymied by this 
stumbling block for over 15 years.  
In this paper we provide (some) answers to these questions by 
constructing explicit examples of such unusual phases of bosons
that may offer some routes out of the conundrum.

Our goal, then, is to access and explore uncondensed quantum phases of
2d bosons which are conducting fluids but not superfluids.
Specifically we have in mind hard core bosons moving on a 2d square 
lattice, but seek to construct states which do not require particular 
commensurate densities.  While our construction can be implemented for 
bosons moving in the 2d continuum Euclidean plane, the states will only 
possess the reduced point group symmetry of the square lattice.  
Despite our interest in time reversal invariant quantum ground states, 
our technical approach will be strongly informed by theories of the 
fractional quantum Hall effect (FQHE).  In some regards, the quantum 
phases that we construct are time reversal invariant analogs of the 
Laughlin state for bosons in a half-filled Landau level.  
But the physical properties of the phases will be dramatically different 
from the incompressible FQHE states, and will have gapless excitations 
and metallic transport for example.

To motivate and illustrate our approach, it will be helpful to briefly 
revisit the bosonic FQHE.  Consider the Laughlin wavefunction for bosons 
in a half-filled Landau level,
\begin{equation}
\Psi_{\nu = 1/2} (z_1, z_2, \dots, z_N) = \prod_{i<j} (z_i - z_j)^2 ~.
\label{Psi_nu_1/2}
\end{equation}
Much of the physics of the Laughlin state has its origin in the structure
of zeroes, which reveals that any two particles upon close approach in 
real space are in a relative two-body $d+id$ state, $(z_i-z_j)^2$.  
Our first objective is to construct a time reversal invariant (real) 
wavefunction in which particle pairs are similarly in a relative 
$d_{xy}$ state.  
A clue is offered by noting that the Laughlin wavefunction is the square 
of a Vandermonde determinant, $\Psi_{\nu = 1/2} = (\det {\bf V})^2$.  
In the Vandermonde determinant, all pairs of particles are in a 
relative $p+ip$ state, $(z_i-z_j)$.  Upon squaring, the two $p+ip$ states
combine to form a single $d+id$ state, which is essentially just 
addition of angular momentum.

This suggests constructing a time reversal invariant boson wavefunction 
in zero magnetic field by simply squaring a determinant constructed from 
momentum states within a Fermi sea,
\begin{equation}
\Psi({\bf r}_1, {\bf r}_2, \dots, {\bf r}_N) 
= ( \det e^{i {\bf k}_i \cdot {\bf r}_j} )^2 ~, \quad
\text{(S-type)}.
\label{Psi_det-squared}
\end{equation}
Let's examine the nodal structure.  The generic behavior of each 
fermion determinant when any two particles are taken close together 
is dictated by Fermi statistics and reality of the wavefunction
and has the functional form $({\bf r}_i -{\bf r}_j) \cdot \hat{\bm l}$.
The unit vector $\hat{\bm l}$ will depend in a complicated way on the 
location of all the other particles 
(see Ref.~\onlinecite{Ceperley91} for a discussion and illustrations of 
the free fermion nodes).
When $\hat{\bm l} = \hat{\bf y}$, this is a $p_x$ form vanishing along a 
nodal line (in the relative coordinate) parallel to the $x$-axis.
Unfortunately, in contrast with the Laughlin case, 
squaring the determinant leads here to an ``extended" $s$-wave form 
with a quadratic nodal line rather than the desired $d$-wave.

\begin{figure}
\centerline{\includegraphics[width=\columnwidth]{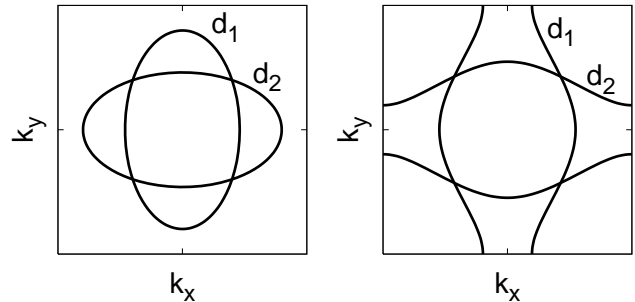}}
\vskip -2mm
\caption{DBL wavefunction Eq.~(\ref{Psi_DBL}) is obtained by multiplying 
two Slater determinants corresponding to distinct but symmetry related
Fermi surfaces, which can be viewed as Fermi surfaces of $d_1$ and 
$d_2$ slave fermions.
The left panel shows an example with closed elliptical Fermi surfaces, 
while the right panel is a case with open Fermi surfaces in the first 
Brillouin zone of the square lattice.
}
\label{fig:introFS}
\end{figure}

\begin{figure}
\centerline{\includegraphics[width=3.0in]{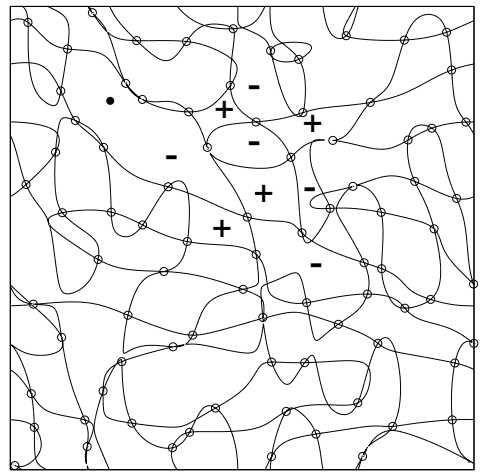}}
\vskip -2mm
\caption{``Nodal picture'' of the continuum DBL wavefunction as probed 
by moving a single particle (whose initial location is marked by the 
filled circle) while keeping the rest of the particles fixed
(open circles).
The signs of the wavefunction are indicated in several places to 
bring out the $+ - + -$ pattern upon encircling a target particle.
}
\label{fig:nodes}
\end{figure}

But consider instead multiplying together two fermion determinants, 
each constructed by filling up a Fermi sea of momentum states,
but with different Fermi surfaces.  
A wavefunction with $d_{xy}$ two-particle correlations
can be constructed by choosing two elliptical Fermi surfaces, one with 
its long axis along the $x$-axis and the other rotated by 90 degrees,
as illustrated in Fig.~\ref{fig:introFS},
\begin{equation}
\Psi({\bf r}_1, {\bf r}_2, \dots, {\bf r}_N) = 
(\det)_x \times (\det)_y ~, \quad \text{(D$_{xy}$-type)} ~,
\label{Psi_DBL}
\end{equation}
where the short-hands $(\det)_x$ and $(\det)_y$ represent the 
corresponding Slater determinants.
In the limit of extreme eccentricity of the ellipses, the wavefunction 
will have the desired $d_{xy}$ form when two particles are brought 
close together, $(x_i-x_j)(y_i-y_j)$.  Away from this limit, the two 
nodal lines will not align precisely with the $x$ and $y$ axes,
but upon taking one particle around the other the wavefunction will 
exhibit the same sign structure as a $d_{xy}$ orbital, $+ - + -$,
changing sign twice. 
A picture of such sign structure as seen by a test particle is shown in 
Fig.~\ref{fig:nodes}.

Our principal thesis is that this wavefunction captures qualitative 
features of a new quantum liquid phase of bosons, which we call
``D-wave-correlated Bose Liquid'' (DBL).
But a variational wavefunction does not provide a complete 
characterization of a quantum phase, and cannot be used to address its 
stability.  As for the FQHE, a field theoretic approach, such as 
Chern-Simons gauge theory, is both desirable and ultimately necessary.
Since we require time reversal invariance, Chern-Simons is inappropriate,
but a tractable field theoretic framework for the DBL will nevertheless
turn out to be a gauge theory.

Indeed, we follow closely the gauge theory approaches to spin liquids in 
quantum antiferromagnets,\cite{LeeNagaosaWen, WenPSG} which after all are 
lattice bosonic systems.  But there are some notable and important differences 
for the DBL which we discuss below.  The variational wavefunction in 
Eq.~(\ref{Psi_DBL}), being a product of two fermion determinants, naturally 
suggests expressing the boson creation operator as a product of two fermion 
operators,
\begin{equation}
b^\dagger ({\bf r}) = d_1^\dagger ({\bf r}) d_2^\dagger ({\bf r}) ~,
\end{equation}
with the 2d position ${\bf r}$ either continuous or denoting the 
discrete sites of a square lattice.
In a general case, this decomposition introduces a local $SU(2)$ gauge 
redundancy well-known in the slave-fermion treatments of the 
spin-1/2 antiferromagnet 
(see Ref.~\onlinecite{WenPSG} for a recent comprehensive discussion).
The S-type wavefunction corresponds to a so-called $SU(2)$ liquid,
while the DBL wavefunctions are $U(1)$ liquids and require a $U(1)$ 
gauge field which is minimally coupled to the two fermions with 
opposite gauge charges (the product $d_1^\dagger d_2^\dagger$
is then a gauge neutral composite, the physical boson).  

Within slave particle theory a mean field state with a fixed background 
gauge ``magnetic" flux is chosen.
This choice is constrained by symmetry:  All the physical symmetries 
of the boson model, even if not respected by the mean field ansatz, 
must be present in the full gauge theory once gauge transformations are 
allowed.
The collection of these symmetry transformations in the gauge theory
is called the Projective Symmetry Group (PSG).\cite{WenPSG}  
For lattice boson theories many different mean field states with
different PSG's are possible, while little is offered to guide
which mean field state is the correct one.

In the continuum, the symmetries of the Euclidean plane together with 
time reversal invariance are more restrictive and we can only offer 
two PSG mean field ansatze in this case.
One is a mean field state with zero gauge flux and spherical Fermi seas
for both fermions; this is an $SU(2)$ ansatz and corresponds to the 
S-type wavefunction considered earlier.
A different $U(1)$ ansatz is obtained by considering $d_1$ and $d_2$ 
fermions each moving in a uniform field but of opposite signs for the 
two; in this case, the boson wavefunction is schematically
$\Psi = (\det) \times (\det)^*$,
e.g., $\Psi = |\det {\bf V}|^2 = |\Psi_{\nu=1/2}|$ to give a concrete 
example;
this wavefunction is again non-negative and has off-diagonal
quasi-long-range order.\cite{GirvinMacDonald, Kane}

For bosons moving in continuous space but with only the square lattice 
point group symmetry, we can offer only four PSG mean fields; 
all of them have zero flux, but differ in the way the 90 degree rotation 
and mirror symmetries are realized.
One example has both $d_1$ and $d_2$ Fermi surfaces individually
respecting the symmetries but otherwise independent of each other.  
Two more examples are the already introduced D$_{xy}$ state,
Eq.~(\ref{Psi_DBL}) and Fig.~\ref{fig:introFS}, with elliptical 
Fermi surfaces elongated along the $x$ and $y$ axes, 
and a similarly constructed D$_{x^2-y^2}$ state with ellipses along 
$\hat{\bf x} \pm \hat{\bf y}$;
in both cases, the $d_1$ and $d_2$ Fermi surfaces are related by
the 90 degree rotation, but the two states differ in the way the mirror 
symmetries are realized.
In the fourth example, each Fermi surface is invariant under the rotation
but not under the mirrors, which instead are realized by interchanging 
the two fermions.

Each of the four presented ansatze can be loosely called a D-wave Bose 
liquid as far as the nodal pictures like that in Fig.~\ref{fig:nodes}
are concerned; the gauge theory analysis would also be rather similar.
From now on, we will focus on the D$_{xy}$ boson liquid.
The associated mean field wavefunction once Gutzwiller projected into
the physical Hilbert space, $d_1^\dagger d_1 = d^\dagger_2 d_2$, 
is precisely the variational wavefunction in Eq.~(\ref{Psi_DBL}).
A full incorporation of gauge fluctuations about the mean field 
performs, in principle, this projection exactly.  But in practice the gauge theory approach allows incorporation of 
slowly varying gauge fluxes and is different in this respect from
the wavefunction.  
Provided the gauge theory is not in a confined state and no physical 
symmetries are broken, an exact description of the putative DBL phase 
is thereby obtained.  
Herein, we introduce a large $N$ generalization of the gauge 
theory which allows a controlled and systematic treatment of the gauge 
fluctuations order by order in powers of $1/N$.
We can then perturbatively extract physical properties of the DBL phase.
Whether the DBL phase survives down to $N=1$ is not something that 
we can reliably address.

In describing the D$_{xy}$ liquid phase for lattice bosons there is 
considerable freedom when choosing the precise form of the slave fermion 
hopping in the mean field Hamiltonian.  The simplest choice is near 
neighbor hopping with different amplitudes in the $x$ and $y$ directions.
But even with this restriction, there are two different possible 
Fermi surface topologies, closed or open, as depicted in 
Fig.~\ref{fig:introFS}.  
The Fermi surface topology has rather dramatic consequences 
for the nature of the associated D-wave Bose fluid.
When the Fermi surfaces are open and have no parallel Fermi surface 
tangents, the resulting phase has a quasi-local character -- 
the boson Green's function is found to fall off exponentially in space.
This is a distinct quantum phase which we refer to as a 
D-Wave Local Bose liquid (DLBL).
The DLBL is intrinsically very stable against gauge fluctuations and 
we are fairly confident that when present it will be a stable quantum 
phase.  The effect of gauge fluctuations on the DBL phase, however, 
are more subtle.   Within our systematic large $N$ approach we do 
find a regime where the DBL is stable, but it is less clear that 
this will survive down to $N=1$.
The Gutzwiller wavefunction does appear to describe a DBL phase with 
properties that are consistent with those inferred from the gauge theory.
This provides some support for the stability of the DBL phase.

Together the gauge theory and the variational wavefunction provide a 
consistent and rather complete picture of both the DBL and the DLBL.
Here we briefly highlight some of the important characteristics.
In the DBL, the single particle boson Green's function, 
$G_b({\bf r}, \tau) \equiv \la b^\dagger ({\bf r},\tau) b({\bf 0},0) \ra$,
decays as an oscillatory power law at equal times,
\begin{eqnarray}
G_b({\bf r},0) &\sim & 
\frac{\cos[({\bf k}_{F_1} - {\bf k}_{F_2}) \cdot {\bf r}] }
     {|{\bf r}|^{4-\eta}} \nonumber \\
&+& 
\frac{\cos[({\bf k}_{F_1} + {\bf k}_{F_2}) \cdot {\bf r} - 3\pi/2]} 
     {|{\bf r}|^4} ~,
\label{Gb_final}
\end{eqnarray}
where the two wave vectors ${\bf k}_{F_{1,2}}$ depend on the observation 
direction $\hat{\bf r}$, as does the anomalous exponent $\eta > 0$.
As the direction of $\hat{\bf r}$ is rotated in real space, the 
wave vectors ${\bf k}_{F_{1,2}}$ and ${\bf k}_{F_1} \pm {\bf k}_{F_2}$ 
trace out closed momentum space curves, and for the DBL in the continuum 
these curves will have the topology of a circle.

If one measures the boson Green's function and finds it to be of the
above form, it is natural to refer to the 
${\bf k}_{F_1} \pm {\bf k}_{F_2}$ as ``Bose surfaces.''
Within the gauge theory description the origin of these singular surfaces
can be traced to the Fermi surfaces of the constituent fermions.
Specifically, ${\bf k}_{F_{1,2}}$ are locations on the two Fermi 
surfaces where the surface normals are along the observation 
direction $\hat{\bf r}$, and the momentum space areas enclosed by 
each of these surfaces will equal $(2\pi)^2 \rho_b$.
Note that the ${\bf k}_{F_{1,2}}$ surfaces can be uniquely
reconstructed from the measured Bose surfaces 
${\bf k}_{F_1} \pm {\bf k}_{F_2}$.
In this sense, the Bose surfaces in the DBL contain information
analogous to Luttinger's Fermi surface volume theorem in the 
Fermi liquid.  The Bose surface can also be extracted directly 
from the DBL wavefunction by using variational Monte Carlo to compute 
the equal time boson Green's function, and can also be indirectly 
inferred by using the shape of the two Fermi surfaces input into the 
$(\det)_x$ and $(\det)_y$ factors.  It is both reassuring and quite 
remarkable that these two coincide.


The spatially local Green's function, $G_b({\bf 0}, \tau)$, in the DBL 
falls off in time as a power law which is at least as slow as 
$1/\tau^2$ corresponding to the local boson tunneling density of states 
$A_b(E) \sim E$;
these are mean field results obtained by combining two fermions each 
with a finite density of states, but we suspect the time decay may
actually be slower upon including gauge fluctuations.

The behavior of the boson Green's function in the DLBL is quite 
different.
The time decay at ${\bf r}=0$ is a power law, 
$G_b({\bf 0}, \tau) \sim \tau^{-2}$,
which is an exact result insensitive to lattice scale details.
But the equal time Green's function falls off exponentially in space
in the DLBL, $G_b({\bf r}, 0) \sim \exp(-r/\xi)$.
Despite this, a two boson box correlator,
\begin{equation}
{\cal B}_b (x) \equiv
\la b^\dagger(0,0) b^\dagger(x,x) b(x,0) b(0,x) \ra ~,
\label{Bb_def}
\end{equation}
falls off as a (non-oscillatory) power law at large distances in the 
DLBL, ${\cal B}(x) \sim -x^{-8}$, with both the sign (negative) and the 
exponent being universal and insensitive to lattice scale physics.
Paradoxically, this seems to imply that a pair of bosons injected into 
the DLBL on opposite corners of a square box can move more readily 
than a single injected boson.
But in a strongly interacting quantum state such a single-particle 
interpretation can be misleading -- the dynamics of an injected boson 
is not that of a weakly interacting quasiparticle.

In both the DBL and the DLBL the density-density correlator
$D_b({\bf r}) \equiv
 \la \hat{\rho}_b({\bf r}) \hat{\rho}_b({\bf 0}) \ra$ 
behaves at large distances as,
\begin{equation}
D_b({\bf r}) \sim - \sum_{\alpha = 1,2} 
\frac{\cos[2{\bf k}_{F_\alpha} \cdot {\bf r} - 3\pi/2]}
      {|{\bf r}|^{4-\gamma_\alpha}}
+ \frac{1}{|{\bf r}|^4} ~,
\label{Db_final}
\end{equation}
where again both the wave vectors ${\bf k}_{F_{1,2}}$ and the anomalous 
exponent $\gamma_\alpha$ depend on the observation direction 
$\hat{\bf r}$.   Measurement of such correlation can be used to extract 
${\bf k}_{F_{1,2}}$ in the DLBL, which are not accessible from the boson 
Green's function here.
Upon rotating the unit vector $\hat{\bf r}$ in real space, 
${\bf k}_{F_{1,2}}$ will again trace out the Fermi surfaces of the 
underlying fermions which for the DLBL will be open.
Both the DBL and the DLBL are conductors with a resistance varying with 
temperature as $R \sim T^{4/3}$.  This is particularly striking for the
DLBL where the boson Green's function is short-ranged.  
Evidently, it is not possible to understand the metallic transport in 
terms of the motion of weakly interacting bosons.  In a sense it is the 
fermionic constituents which transport the charge, but even this is not 
quite right since the $d_1,d_2$ fermions do not exist as well defined
quasiparticle excitations, being strongly scattered by the gauge 
fluctuations.

The lattice gauge theory formulation can be used to motivate a lattice
boson Hamiltonian which might plausibly have the DBL/DLBL as a ground 
state.  The simplest such model consists of a near neighbor boson 
hopping term supplemented by a four-site ``ring" exchange term:
\begin{eqnarray}
\label{Hring}
H &=& H_J + H_4 ~,\\
H_J &=& -J \sum_{{\bf r};\, \hat{\mu} = \hat{\bf x}, \hat{\bf y}}
(b^\dagger_{\bf r} b_{{\bf r} + \hat{\mu}} + h.c.) ~,\\
H_4 &=& K_4 \sum_{\bf r} 
(b^\dagger_{\bf r} b_{{\bf r} + \hat{\bf x}}
b^\dagger_{{\bf r} + \hat{\bf x} + \hat{\bf y}} b_{{\bf r} + \hat{\bf y}}
+ h.c.) ~,
\end{eqnarray}
with $J,K_4 \ge 0$.  This model is fully specified by two dimensionless 
numbers, the ratio $J/K$ and the boson filling $0 \le \rho_b \le 1$.
This ring Hamiltonian with $J>0$ and $K_4<0$ was introduced and studied 
by Paramekanti\etal,\cite{Paramekanti} and later studied extensively
with quantum Monte Carlo by Sandvik\etal,\cite{Sandvik}
Melko\etal,\cite{Melko} and Rousseau\etal\cite{Rousseau}
There is no sign problem in this regime, and it was possible to
access large system sizes and low temperatures.
But for $J>0$ and $K_4>0$, there is a sign problem since the Hamiltonian 
does not satisfy the Marshall sign conditions, and one expects the 
ground state wavefunction to take both positive and negative values.
For this ring model we have evaluated the variational energetics
for the DBL/DLBL wavefunctions and compared these to the energy of a 
superfluid wavefunction of the usual Jastrow form.
Within this necessarily limited energetics study we do find a region 
of parameter space with small $J/K_4$ and near half filling where the 
DLBL wavefunction has the lowest energy.  In view of its quasi-local 
character,  we suspect that DMRG studies could be fruitful in helping 
establish whether or not the DLBL is present in the phase diagram of 
this (or related) ring Hamiltonian.

The paper is organized as follows.
In Sec.~\ref{sec:motiv_Wavefnc} we discuss in more detail the 
wavefunction motivation for considering the DBL states.
In Sec.~\ref{sec:MFT} we introduce the lattice gauge theory
description and study properties of the DBL state in the 
mean field theory that ignores gauge fluctuations.
This provides an initial guide to the singular (Bose) surfaces,
and is followed in Sec.~\ref{sec:Wavefnc_props} with numerical
characterization of the properties of the actual DBL wavefunctions.
In Sec.~\ref{sec:Gauge_fluct}, we consider the full gauge theory
description, focusing on the effects of the gauge fluctuations on
the singularities across the Bose surfaces, and in particular
obtain the long-distance properties such as
Eq.~(\ref{Gb_final}) and Eq.~(\ref{Db_final}) to order $1/N$ in our systematic
large $N$ approach.
In Sec.~\ref{sec:eff_FT} we address the issue of the stability
of the DBL by considering a putative fixed point theory.
We conclude in Sec.~\ref{sec:concl} with a discussion of physical
properties and possible future directions.

\section{Boson Wavefunctions and Nodes}
\label{sec:motiv_Wavefnc}

In this section, we expand on the initial motivation for considering
the DBL states as a way to perform (a kind of) flux attachment 
transformation in a time reversal invariant manner.
We also discuss the nodal structure of the bosonic wavefunctions,
providing some justification for the qualifier ``D-wave'' in the 
suggested wavefunction names in this work.  
To this end, we consider a ``relative single particle wavefunction"
-- more precisely, a cross-section of the many-body wavefunction --
defined as follows.
Fixing the positions of all the particles except one, we define a 
function which depends explicitly on the coordinates of the one 
``test" particle, and implicitly on the coordinates of all the other 
particles,
\begin{equation}
\Phi_{{\bf r}_2, \dots, {\bf r}_N}({\bf r}) \equiv
\Psi({\bf r}, {\bf r}_2, \dots, {\bf r}_N) ~.
\label{Phi_b}
\end{equation}
For notational ease we will henceforth drop the implicit dependence 
on the $2(N-1)$ spatial coordinates and just use the notation 
$\Phi_b({\bf r})$.

\subsection{Laughlin $\nu=1/2$ revisited}
\label{subsec:Laughlin}
Consider the Laughlin wavefunction for bosons in a half-filled Landau 
level, Eq.~(\ref{Psi_nu_1/2}).  In this case, the relative single
particle wavefunction $\Phi(z)$ is complex and has double strength 
zeroes at the positions of all the other $N-1$ particles.
Upon close approach to a specific particle, say $z_i$, the function
$\Phi(z) \sim (z-z_i)^2$ is of a $d+id$ form.  It is in this sense
that the $\nu=1/2$ Laughlin state can be viewed as a D-wave fluid:
All pairs of particles, upon close approach in real space, are in a 
two-body $d+id$ state.
Our goal is to construct a time reversal invariant analog of the 
$\nu=1/2$ Laughlin state in which particle pairs are similarly in
a relative $d_{xy}$ or $d_{x^2-y^2}$ state.

In thinking about Laughlin states, it has been particularly instructive 
to view them in terms of composite particles created by 
``flux attachment."
For example, if one re-expresses the $\nu=1/2$ Laughlin state as,
\begin{equation}
\Psi_{\nu=1/2} = \prod_{i<j} (z_i - z_j) \, \Psi_{cf} ~,
\end{equation}
the ``composite fermion" wavefunction is simply a (Vandermonde) 
determinant, $\det {\bf V}$, of filled Landau level orbitals,
\begin{equation}
\Psi_{cf} = \det {\bf V} = \prod_{i<j} (z_i - z_j) ~,
\end{equation}
with $V_{ij} = (z_i)^j$.
In a second quantized framework, flux attachment can be achieved by 
introducing a Chern-Simons gauge field\cite{Zhang_CS}.
Chern-Simons gauge theory has been a useful tool to describe the full 
Haldane-Halperin hierarchy of fractional quantum Hall states,
encoding the fractional charge and statistics of the quasiparticles as
well as the structure of the edge excitations\cite{MPAF_Lee, Wen_edge}.  But in discussing bosons
in a time reversal invariant setting (zero magnetic field), 
flux-attachment techniques are problematic.  For instance, the usual 
flux smearing mean field approximation is likely to inadvertently
break time reversal invariance.  

In seeking to avoid Chern-Simons theory, it is worth noting that
the $\nu=1/2$ state is a perfect square -- a square of the Vandermonde 
determinant,
\begin{equation}
\Psi_{\nu =1/2} = (\det {\bf V})^2 ~.
\label{detV2}
\end{equation}
This suggests that the $\nu=1/2$ Laughlin state can be fruitfully 
described within a slave particle framework.\cite{Wen_proj4FQHE}
Consider decomposing the boson creation operator as a product of two 
fermions,
$b^\dagger ({\bf r}) = d_1^\dagger ({\bf r}) d_2^\dagger ({\bf r})$.
As discussed in the Introduction, there is a local gauge redundancy,
and a correct treatment requires the presence of a gauge field 
minimally coupled to both fermions.
In the slave particle mean field approach to the $\nu=1/2$ boson problem,
one simply drops the gauge field, obtaining a problem of two fermion 
flavors each moving in a magnetic field.
If the electrical charge of the boson is divided equally, each fermion 
flavor is effectively in a full Landau level.
The mean field wavefunction is,
\begin{equation}
\Psi^{MF}_{\nu = 1/2} = \prod_{i<j} (z^{(1)}_i - z^{(1)}_j) \times 
\prod_{i<j} (z^{(2)}_i - z^{(2)}_j) ~.
\end{equation}
To obtain a wavefunction for the bosons it is necessary to project
into the physical Hilbert space, and this is achieved here by simply 
setting $z^{(1)}_i = z^{(2)}_i = z_i$.
One thereby recovers the Laughlin state as a square of the 
Vandermonde determinant, Eq.~(\ref{detV2}).

\subsection{Time Reversal Invariant Wavefunctions}
\label{subsec:det-squared}

Within a first quantized framework, the relative simplicity of a time 
reversal invariant bosonic superfluid as compared to the Laughlin state 
is the nodelessness of the ground state wavefunction.  
As Feynman argued many years ago, for non-relativistic bosons moving 
in the continuum with an interaction only depending on the particle 
positions, the kinetic energy of any bosonic wavefunction which has 
sign changes could be reduced by making all the signs positive while 
keeping the magnitude of the wavefunction fixed to leave the potential 
energy unchanged.
The ground state wavefunction should thus be everywhere non-negative.  
For the non-interacting Bose gas the ground state wavefunction is just
$\Psi =1$, but in the presence of interactions a popular variational
wavefunction is of the Jastrow form,\cite{Kane}
\begin{eqnarray}
\Psi_{\rm Jastrow}({\bf r}_1, {\bf r}_2, \dots, {\bf r}_N) 
\;\propto\; e^{-\sum_{i<j} u({\bf r}_i - {\bf r}_j)} ~,
\label{Psi_Jastrow}
\end{eqnarray}
with variational freedom in the two-particle pseudopotential 
$u({\bf r}_i - {\bf r}_j)$, which is usually taken to approach zero as 
$1/|{\bf r}|^p$ at large separations.

As in the case with a magnetic field present, we again consider
the relative single particle wavefunction.
For a Bose condensed  superfluid in a time reversal invariant system 
$\Phi_b({\bf r})$ can be taken as real and is  
everywhere non-negative.  If there are repulsive interactions between
the bosons in the superfluid, the amplitude of $\Phi_b({\bf r})$ will be 
reduced when the test particle is taken nearby another particle, 
which is implemented by the Jastrow pseudopotential in 
Eq.~(\ref{Psi_Jastrow}).
But the sign of $\Phi_b({\bf r})$ will remain positive, so in some sense 
all of the particle pairs are in a relative S-type state.
In the special case of a hard core interaction,
$\Phi_b({\bf r}) \to 0$ for ${\bf r} \to {\bf r}_i$, 
so one can then view this as an ``extended" S-type wavefunction.

Motivated by the preceding discussion of the $\nu=1/2$ Laughlin state,
in Eq.~(\ref{Psi_det-squared}) we introduced a simple time reversal 
invariant wavefunction for hardcore bosons which is the square of a 
determinant for free fermions filling a Fermi sea.
By construction this wavefunction is non-negative.  
Moreover, the wavefunction will have zeroes which coincide with the 
nodes of the filled Fermi sea,
$\Psi_{FS}({\bf r}_1, {\bf r}_2, \dots, {\bf r}_N) 
= \det e^{i {\bf k}_i \cdot {\bf r}_j}$.  
With time reversal invariance, a relative single fermion wavefunction, 
$\Phi_f({\bf r}) \equiv \Psi_{FS}({\bf r}, {\bf r}_2, \dots, {\bf r}_N)$, 
can be taken as real.
As a result, the nodal structure of $\Phi_f({\bf r})$ will be 
qualitatively different than for the filled Landau level state,
vanishing along nodal lines rather than at isolated points.  
These nodal lines will pass through the positions of all of the other 
fermions.\cite{Ceperley91}  
Upon taking the ``test" particle across a nodal line, the function 
$\Phi_f({\bf r})$ changes sign, vanishing linearly upon approaching the 
nodal line.
If one takes the position of the test particle close to another particle,
$\Phi_f({\bf r})$ will have a $p$-wave character -- in particular a 
$p_x$ form if we define the $x$-axis as being parallel to the 
nodal line.

Despite the $p$-wave character of the free fermion determinant, the 
determinant-squared wavefunction will not have a $d$-wave character. 
Rather, the relative single boson wavefunction will vanish 
quadratically upon crossing the nodal lines of the fermion determinant.
When the test particle is taken near another particle ${\bf r}_i$, 
the function vanishes quadratically,
$\Phi_b({\bf r}) \sim (x - x_i)^2$.  
The ``pair" wavefunction is thus of an ``extended S-type" form, 
vanishing along the residual nodal line.

Let us now consider the DBL many-body wavefunction Eq.~(\ref{Psi_DBL}), 
which is the product of two different Slater determinants for fermions 
that fill elliptical Fermi surfaces as shown in Fig.~\ref{fig:introFS}.
The nodal structure is revealed by exploring $\Phi_b({\bf r})$.
This function will have two sets of nodal lines, 
one from each of the determinants.  Due to the elliptical nature of the 
respective Fermi surfaces, the two sets of lines, both of which pass 
through all of the particles, will generally not coincide with one 
another.  Indeed, the fermion determinant coming from an elliptical 
Fermi surface will have nodal lines running preferentially perpendicular 
to the long axis of the ellipse.  Focusing on the behavior of 
$\Phi_b({\bf r})$ near a target particle ${\bf r}_i$, one anticipates a 
behavior of the $d_{xy}$ form,
$\Phi_b({\bf r}) \sim (x - x_i)(y - y_i)$.
Here we have assumed that the two nodal lines actually coincide with the
$x$ and $y$ axes.  In general, for a typical configuration of fixed 
particle coordinates and a given target particle, this will not 
precisely be the case.  More generically, the two nodal lines will 
intersect a particle at two angles which are not aligned with the axes.
But the sign of the relative wavefunction when the test particle 
encircles the target particle will still behave as $+ - + -$, 
the same sign structure as a D$_{xy}$ or D$_{x^2-y^2}$ orbital.  
This is illustrated in Fig.~\ref{fig:nodes}.

Since the DBL many-body wavefunction is not nodeless, it cannot be the 
ground state of a continuum Hamiltonian of bosons.  
If we put the coordinates on the sites of a 2d square lattice, 
the ground state wavefunction is only assured to be non-negative if the 
sign and form of the hopping matrix in the lattice tight binding 
Hamiltonian is such that it satisfies the Marshall sign conditions -- 
the requirement that a choice of gauge is possible to make all of the 
off-diagonal matrix elements negative.  
In Sec.~\ref{sec:Energetics} we consider a particular Marshall sign 
violating lattice boson Hamiltonian which might exhibit a ground state 
of the proposed D-wave form.

\subsection{Precedents of $(\det)_1 \times (\det)_2$ wavefunctions for 
spin liquids}
Before focusing solely on the DBL states, 
we want to mention that our construction of time reversal invariant 
bosonic wavefunctions as a product of two distinct determinants has 
nice precedents in the studies of spin liquids on the triangular 
lattice.  One can view the triangular Heisenberg antiferromagnet as a 
system of hard-core bosons at half-filling in the background field of 
flux $\pi$ through each triangle.  
Kalmeyer and Laughlin\cite{KL} proposed to view this in the continuum, 
obtaining a boson system at $\nu=1/2$, and their chiral spin liquid
wavefunction is precisely the lattice analog of the $\nu=1/2$ state 
Eq.~(\ref{Psi_nu_1/2}).  Alternatively, using a slave fermion 
approach,\cite{WenWilczekZee, LaughlinZou}
$b^\dagger = d_1^\dagger d_2^\dagger$,
we divide the boson charge equally between the two fermions,
so each sees on average a flux of $\pi/2$ per triangle;
the mean field where the $d_1$ and $d_2$ see the same static flux of 
$\pi/2$ per triangle gives a filled Landau level for each fermion
and reproduces precisely the Laughlin-Kalmeyer chiral spin liquid.
  
However, we can be more creative about the fluxes seen by the slave 
fermions while maintaining the average of $\pi/2$ flux per triangle.  
One example is to take different flux patterns for the two species as 
follows:  For the $d_1$ fermions, put $0$ flux through all up-pointing 
triangles and $\pi$ flux through all down-pointing triangles, 
while for the $d_2$ fermions interchange the locations of the $0$ and 
$\pi$ fluxes.
This state is in fact identical to the so-called 
$U1B \tau^1 \tau_-^0 \tau_+^1$ spin liquid found in 
Ref.~\onlinecite{ZhouWen}.
It is a time reversal invariant gapless algebraic spin liquid (ASL)
with Dirac nodes in the spinon spectrum;
it has very good energetics for the nearest neighbor triangular 
antiferromagnet, starting from the isotropic lattice and all the way to 
the limit of weakly coupled chains
(e.g., Ref.~\onlinecite{Yunoki06} found a different gauge-equivalent 
formulation of this state without realizing its ASL character).

Another example is obtained by taking yet different flux patterns:
Select one lattice direction -- chain direction in the anisotropic
lattice case.  For the $d_1$ fermions, put $0$ flux through triangles 
siding even chains and $\pi$ flux through triangles siding odd chains, 
while for the $d_2$ fermions interchange the locations of the $0$ 
and $\pi$ fluxes.  This is again a time reversal invariant ASL and is 
listed as $U1C \tau_+^0 \tau_-^0 \tau^1$ in 
Ref.~\onlinecite{ZhouWen}.
Unlike the $U1B$ state, there is no isotropic $U1C$ liquid, but
the energetics performance of $U1C$ and $U1B$ spin liquids is almost 
indistinguishable for weakly coupled chains and matches that of competing
magnetically ordered states; in this context, the time reversal 
invariant $U1B$ and $U1C$ liquids are significantly better than the 
chiral Laughlin-Kalmeyer variant.

Our construction of the DBL states differs in that we do not require
any special filling for the bosons and there is no special Heisenberg
spin symmetry.  Neither of these special conditions of the spin model 
setting are needed for the construction and subsequent gauge theory 
analysis to go through.  
Given the growing belief\cite{Rantner, WenPSG, Hermele} 
that critical spin liquids do exist, we do not see any reasons why the 
situation should be any different for our boson liquids at arbitrary 
incommensurate densities.  In our construction, such liquids will 
generically have some underlying partially filled bands and therefore 
Fermi surfaces of slave fermions.  
Of course, whether a particular state is realized in a given model 
requires a detailed case by case study, and in Sec.~\ref{sec:Energetics} 
we suggest some frustrated boson models that may stabilize the DBL phase.
Our primary goal in the next Secs.~\ref{sec:MFT}-\ref{sec:eff_FT}
will be to characterize the DBL states without worrying where to find 
them.
\\

\section{Mean Field Theory for the D-wave Bose Liquid}
\label{sec:MFT}

\subsection{Gauge Theory formulation}
\label{subsec:Gauge_formulation}

We next consider the challenge of constructing a field theory which 
can access such a $D_{xy}$-Bose liquid (DBL) state.
Due to our inability to implement flux attachment in a tractable 
time reversal invariant manner, we follow instead the slave particle 
approach.  As above, we decompose the hard core lattice boson as a 
product of two fermions, $b^\dagger = d_1^\dagger d_2^\dagger$.
Consider then a lattice $U(1)$ gauge theory on the square lattice 
in terms of these slave particles:
\begin{equation}
H_{U(1)} = H_t + H_a ~,
\label{HU1}
\end{equation}
with the fermion hopping Hamiltonian of the form,
\begin{widetext}
\begin{eqnarray}
\label{Ht}
H_t &=& -\sum_{\bf r}
\left[ 
t_\parallel e^{i a_x({\bf r})} d_1^\dagger({\bf r}) d_1({\bf r} + \hat{\bf x}) + 
t_\perp e^{i a_y({\bf r})} d_1^\dagger({\bf r}) d_1({\bf r} + \hat{\bf y})+ h.c. \right]\\
&& -\sum_{\bf r}
\left[
t_\perp e^{-i a_x({\bf r})} d_2^\dagger({\bf r}) d_2({\bf r} + \hat{\bf x}) + 
t_\parallel e^{-i a_y({\bf r})} d_2^\dagger({\bf r}) d_2({\bf r} + \hat{\bf y})+ h.c. \right] ~.
\end{eqnarray}
\end{widetext}
The $d_1$ fermion hopping amplitudes in the $\hat{\bf x}$ and 
$\hat{\bf y}$ directions are $t_\parallel$ and $t_\perp$ respectively,
while the two amplitudes are interchanged for the $d_2$ fermions.
In the following, we take $t_\parallel \ge t_\perp$ for concreteness.
Note also that the two fermion species carry opposite gauge charges.
The gauge field Hamiltonian is simply
\begin{equation}
H_a = h \sum_{\bf r} \sum_{\mu = x, y} e^2_\mu({\bf r}) 
- K \sum_{\bf r} \cos[ (\nabla \times a)_{\bf r}  ] ~,
\label{Hgauge}
\end{equation}
where the lattice ``magnetic" field is
\begin{equation}
(\nabla \times a)_{\bf r} = a_x({\bf r}) + a_y({\bf r} + \hat{\bf x}) 
- a_x({\bf r}+ \hat{\bf y}) - a_y({\bf r}) ~.
\end{equation}
The integer-valued ``electric" field $e_\mu({\bf r})$ is canonically 
conjugate to the compact gauge field $a_\mu({\bf r})$ on the same link.  
The above Hamiltonian is supplemented by a gauge constraint on the 
physical states, which must satisfy Gauss' law,
\begin{equation}
(\nabla \cdot e)_{\bf r} = 
d^\dagger_1({\bf r}) d_1({\bf r}) - d^\dagger_2({\bf r}) d_2({\bf r}) ~,
\end{equation} with
\begin{equation}
(\nabla \cdot e)_{\bf r} = e_x({\bf r}) - e_x({\bf r}-\hat{\bf x})
 + e_y({\bf r}) - e_y({\bf r}-\hat{\bf y}) ~.
\end{equation} 

In the limit $h \gg K, t_\parallel, t_\perp$, the electric field 
vanishes, and Gauss' law reduces to $d^\dagger_1 d_1 = d^\dagger_2 d_2$ 
which projects back into the physical boson Hilbert space.  
In this strong coupling limit, it is possible to perturbatively 
eliminate the gauge field to obtain a Hamiltonian for hard core bosons 
hopping on the square lattice with additional ring exchange terms.
We pursue this in Section~\ref{sec:Energetics} where we compare the 
energetics of the DBL wavefunction with other states.  
Here we instead focus on the weak coupling limit, $K \gg h$, 
which suppresses the magnetic flux.
The usual slave particle mean field treatment corresponds to simply 
setting $(\nabla \times a)_{\bf r}$ equal to a constant.  
The simplest mean field state, and the one which should correspond to 
the wavefunctions in the previous section, is with zero flux through all 
plaquettes.  The corresponding mean field Hamiltonian describes 
non-interacting slave fermions.  
Each species has anisotropic near-neighbor hopping amplitudes,
but the two are related under the 90 degree rotation,
thus producing a boson liquid that respects the symmetries of the square 
lattice.

\subsection{Mean field results for the DBL}
\label{subsec:MF_props}

In order to focus on the effects of the underlying Fermi surfaces
without the complications of lattice physics such as Brillouin zone 
folding, we first consider fermions in the 2d continuum with anisotropic 
effective masses.
The resulting elliptical Fermi surfaces are,
\begin{eqnarray}
\label{wparam}
d_1: \quad (w k_x)^2 + (k_y/w)^2 &=& k_F^2 ~,\\
d_2: \quad (k_x/w)^2 + (w k_y)^2 &=& k_F^2 ~,
\end{eqnarray}
where the parameter $w \ge 1$ characterizes the degree of eccentricity
(the conventional eccentricity of the ellipses is given by 
$\epsilon = \sqrt{1 - 1/w^2}$).
In the D-wave Bose liquid, $w$ signifies the mismatch between the
two Fermi surfaces, and we will refer to this measure as
``D-eccentricity".
We study equal time correlation functions since these can be compared 
directly with the properties of the wavefunctions, which is done in 
Sec.~\ref{sec:Wavefnc_props};
we also consider temporal dependencies within the mean field as a
measure of spectral properties.
It is useful to have in mind that much of the following analysis of 
long-distance properties needs only the knowledge of relevant Fermi
surface patches and not of the full surfaces.

Consider first the one-particle off-diagonal density matrix 
(or Green's function) for the boson,
$G_b({\bf r}, \tau) = 
 \la b^\dagger({\bf r}, \tau) b({\bf 0}, \tau) \ra$, with
$G_b({\bf 0}, 0) = \bar\rho$ -- the average boson density.
It will also be of interest to consider the momentum occupation 
probability,
\begin{eqnarray}
\la n_{\bf k} \ra = \la b^\dagger_{\bf k} b_{\bf k} \ra =
\int d{\bf r}\; G_b({\bf r}) e^{-i {\bf k} \cdot {\bf r}} ~. 
\end{eqnarray}
Within the mean field theory, the natural approximation for the order 
parameter correlation is
\begin{equation}
\label{Gb_MFdef}
G_b^{MF}({\bf r}, \tau) = 
G^{MF}_{d_1}({\bf r}, \tau) G^{MF}_{d_2}({\bf r}, \tau) / \bar\rho ~,
\end{equation}
where $G^{MF}_{d_\alpha}$ are the mean field (bare) fermion Green's 
functions.  This approximation satisfies 
$G_b({\bf 0}) = G_{d_\alpha}({\bf 0}) = \bar\rho$.
Here and below, the imaginary time $\tau$ is understood to be zero
if not explicitly present.

The fermion Green's functions are readily calculated.
Thus, at long distances $r \gg k_F^{-1}$, the main contribution to 
$G^{MF}_{d_\alpha}({\bf r})$ comes from the Fermi surface patches where 
the group velocity is parallel or antiparallel to the observation 
direction $\hat{\bf r} = {\bf r}/|{\bf r}|$.  
With inversion symmetry, we can denote the corresponding patch 
locations as $\pm {\bf k}_{F_\alpha}$ and the Fermi surface 
curvature as $c_\alpha$, and obtain
\begin{equation}
G^{MF}_{d_\alpha}({\bf r}) \approx \frac{1}{2^{1/2} \pi^{3/2}}
\frac{\cos({\bf k}_{F_\alpha} \cdot {\bf r} - 3\pi/4)} 
     {c_\alpha^{1/2} |{\bf r}|^{3/2}} ~.
\label{Gd_MF}
\end{equation}
It is important to remember that ${\bf k}_{F_1}, {\bf k}_{F_2}$, 
and $c_1, c_2$ depend implicitly on the direction $\hat{\bf r}$ and are 
different in general when the $d_1$ and $d_2$ Fermi surfaces do not 
coincide.

\begin{figure}
\centerline{\includegraphics[width=\columnwidth]{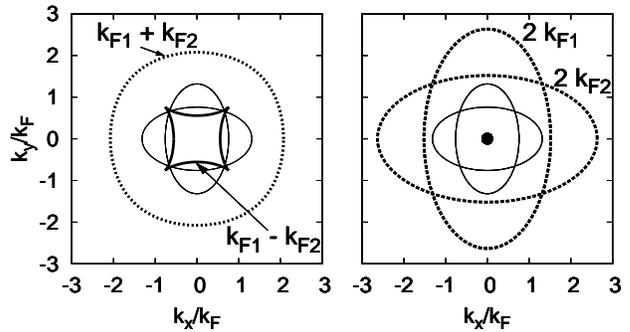}}
\vskip -2mm
\caption{Singular lines in the momentum space for the boson
Green's function (left) and the density correlation (right).
In each panel, the thin lines show the elliptical $d_1$ and $d_2$
Fermi surfaces for the parameter $w^4 = 3$, cf.~Eq.~(\ref{wparam}).
In the left panel, the ${\bf k}_{F_1} \pm {\bf k}_{F_2}$ loci are 
constructed by considering a pair of $d_1$ and $d_2$ fermions with 
parallel ($+$ sign) or antiparallel ($-$ sign) group velocities.
The axes are scaled by an ``average'' $k_F$ that would obtain if the 
Fermi surfaces were circular for the same particle density.
}
\label{fig:loci2kF}
\end{figure}

For the equal-time boson Green's function we thus get
\begin{eqnarray}
G_b^{MF}({\bf r}) &\sim& 
\frac{\cos[({\bf k}_{F_1} - {\bf k}_{F_2}) \cdot {\bf r}]}
     {c_1^{1/2} c_2^{1/2} |{\bf r}|^3} \nonumber \\
&+& 
\frac{\cos[({\bf k}_{F_1} + {\bf k}_{F_2}) \cdot {\bf r} - 3\pi/2]}
     {c_1^{1/2} c_2^{1/2} |{\bf r}|^3} ~,
\label{Gb_MF}
\end{eqnarray}
which decays algebraically while oscillating with 
$\hat{\bf r}$-direction-dependent wave vectors 
${\bf k}_{F_1} + {\bf k}_{F_2}$ and ${\bf k}_{F_1} - {\bf k}_{F_2}$.
Such wave vectors, which are constructed by considering patches on the 
two Fermi surfaces that are parallel to each other, will span some 
new loci in the momentum space as illustrated in Fig.~\ref{fig:loci2kF}.
The above large-distance behavior corresponds to singularities 
$n_{\bf k} \sim |\delta k|^{3/2}$ across these lines.
In the zero eccentricity limit (i.e., when $w=1$), the two Fermi 
surfaces coincide, giving
\begin{equation}
G^{MF}_b (r)  \sim \frac{1 + \cos[2 k_F r - 3\pi/2]}{r^3}~,
\quad\quad \text{(S-type)} ~.
\end{equation}
Compared to the general case with $w>1$,
the ${\bf k}_{F_1} - {\bf k}_{F_2}$ locus shrinks here to zero momentum, 
while the singularity in $n_{\bf k}$ is hardened to $|k|$.

As we show in Sec.~\ref{sec:Gauge_fluct}, the power law decay of the 
mean field boson Green's function survives in the presence of gauge 
fluctuations (but with modified exponents).
The algebraic decay is a consequence (and a measurable indication) 
of the ``criticality'' of the DBL and its gapless excitations.
Of course, by the very construction we expect many gapless excitations
because of the underlying Fermi surfaces.  
Some measure of the low-energy spectrum is contained in the time 
dependence of the local boson Green's function,
\begin{equation}
G_b^{MF}({\bf 0}, \tau) = \frac{\nu_0^2}{\tau^2} ~,
\label{Gtau}
\end{equation}
where $\nu_0$ is the density of states at the Fermi energy for each
species.  The corresponding local boson spectral function is
\begin{equation}
A_b^{MF}({\bf r}={\bf 0}, E) = 
\int_{\bf k} A_b^{MF}({\bf k}, E) = \nu_0^2 E ~.
\label{Aloc}
\end{equation}
In the discussion of the DBL phase on the lattice in 
Sec.~\ref{subsec:openFS}, we will encounter a situation where the 
boson Green's function decays exponentially in space because of the 
topology of the Fermi surfaces, while the above spectral signature 
of the low-energy excitations depends only on the density of states
which is a property of the entire Fermi surface and is insensitive 
to the topology otherwise.

The most natural instability of the DBL is towards a superfluid state.  
As we discuss in Section~\ref{sec:eff_FT}, in the limit of 
vanishing D-eccentricity ($w=1)$, the resulting S-type Bose Liquid phase 
obtained in mean field theory is, in the presence of gauge fluctuations,
most probably unstable to superfluidity.  
Moreover, our Sec.~\ref{sec:Wavefnc_props} analysis of the 
determinant squared wavefunction appropriate to the S-Bose Liquid
is consistent with off-diagonal long-range order.  
These results strongly suggest that the mean field S-type Bose liquid 
phase probably can not exist as a stable quantum phase.  
However, with non-vanishing D-eccentricity, both the gauge theory 
analysis in Sec.~\ref{sec:eff_FT} and the properties of 
the corresponding $(\det)_x (\det)_y$ wavefunction which we explore
in Sec.~\ref{sec:Wavefnc_props} suggest that the DBL is a stable 
critical quantum phase.

It is also instructive to examine several other correlation functions 
within the present mean field treatment.  
Specifically, consider the boson density-density correlation function,
\begin{eqnarray}
D_b({\bf r}) &=&
\la :\! \delta\hat\rho(0) \delta\hat\rho({\bf r}) \!: \ra  \\
&=& N(N-1) \la \delta^d({\bf r}_1) \delta^d({\bf r}_2 - {\bf r}) \ra
- \bar\rho^2 ~,
\label{densitycorr:def}
\end{eqnarray}
where $N$ is the total number of particles.
As defined, $D_b({\bf r})$ approaches zero for large separations, while 
negative values at small distances signify a correlation hole.
The density structure factor can be calculated as
\begin{equation}
D_b({\bf k}) = \int d{\bf r}\; D_b({\bf r}) e^{-i {\bf k} \cdot {\bf r}} 
= \frac{\la |\delta\hat\rho_{\bf k}|^2 \ra}{V} - \bar\rho ~,
\label{Db_k}
\end{equation}
where 
$\delta\hat\rho_{\bf k} =
\sum_j \exp[-i {\bf k} \cdot {\bf r}_j] - N\delta_{\bf k,0}$ and 
$V$ is the system volume.

Microscopically, since $b^\dagger = d_1^\dagger d_2^\dagger$, for each 
boson added to the system both a $d_1$ and a $d_2$ fermions are added.
As such, the boson and fermion densities are equal: 
$\hat\rho_b = \hat\rho_{d_1} = \hat\rho_{d_2}$.  
However, in the mean field treatment, the two fermion flavors propagate 
independently with different Fermi surfaces, so the corresponding 
fermion density-density correlation functions, which we denote
as $D_{d_\alpha}^{MF}$, do not coincide.  This ambiguity in the
density correlation is an intrinsic deficiency of the 
mean field theory.  As a crude measure, we approximate the boson density 
correlation function as an average of that of the $d_1$ and $d_2$ 
fermions, 
\begin{equation}
D_b^{MF}({\bf r}) \approx \frac{1}{2} 
[ D^{MF}_{d_1}({\bf r}) + D^{MF}_{d_2}({\bf r}) ] ~.
\label{Boson_MF_density_corr}
\end{equation}
The individual density correlation functions in the mean field theory 
are simply,
\begin{equation}
D_{d_\alpha}^{MF}({\bf r}) = -|G^{MF}_{d_\alpha}({\bf r})|^2 \sim 
- \frac{1 + \cos[2 {\bf k}_{F_\alpha} \cdot {\bf r} - 3\pi/2]}
       {c_\alpha |{\bf r}|^3} ~,
\label{DD_MF}
\end{equation}
where again ${\bf k}_{F_\alpha}$ is the place on the $d_\alpha$ Fermi 
surface where the normal is parallel to the observation direction 
$\hat{\bf r}$.
We recognize the $2 k_F$ oscillation with power law envelope, which in 
the momentum space translates to a singularity in the structure factor 
$\sim |2k_F - k|^{3/2}$ across the $2k_F$ line 
(such $2k_F$ surfaces are illustrated in Fig.~\ref{fig:loci2kF}), 
while there is also a $|k|$ singularity at zero momentum.   
As obtained from Eq.~(\ref{Boson_MF_density_corr}), the mean field 
boson density correlator $D_b^{MF}$ has singularities at both 
$2k_{F_1}$ and $2 k_{F_2}$, and ``knows" about the presence of both 
Fermi surfaces.  Remarkably, this mean field approximation is 
consistent with our analysis of the $(\det)_x (\det)_y$ boson 
wavefunction in Sec.~\ref{sec:Wavefnc_props}, which also
reveals singular behavior at both $2 k_{F_1}$ and $2 k_{F_2}$.

Let us now consider the two-boson correlator,
\begin{equation}
G_{2b}({\bf r}_1, {\bf r}_2; {\bf r}_1^\prime, {\bf r}_2^\prime) \equiv 
\la b^\dagger({\bf r}_1) b^\dagger({\bf r}_2) 
    b({\bf r}_1^\prime) b({\bf r}_2^\prime ) \ra ~,
\end{equation}
which injects a pair of bosons at ${\bf r}_1, {\bf r}_2$ and removes a 
pair at ${\bf r}_1^\prime, {\bf r}_2^\prime$.
Consider first the limit where the separation between the two 
injected bosons, ${\bf r} = {\bf r}_1 - {\bf r}_2$, and the two 
removed bosons, ${\bf r}^\prime = {\bf r}_1^\prime - {\bf r}_2^\prime$, 
are both small compared to the mean inter-particle spacing, 
$|{\bf r}|, |{\bf r}^\prime| \ll k_F^{-1}$.
Moreover, take the separation
${\bf R} = \frac{{\bf r}_1 + {\bf r}_2}{2} - 
           \frac{{\bf r}_1^\prime + {\bf r}_2^\prime}{2}$ 
between the injected and removed pair to be much larger than the 
inter-particle spacing, $|{\bf R}| \gg k_F^{-1}$.  
In this limit the mean field 2-boson correlator can be expressed as,
\begin{equation}
G_{2b}^{MF} \sim
\frac{k_F^8}{ [1 + \frac{W}{2} \sin^2 (2 \phi) ]^{5/2} }
\frac{\Psi_{xy}({\bf r}; \hat{\bf R}) 
      \Psi_{xy}^*({\bf r}^\prime; \hat{\bf R})} 
     {(k_F R)^6} ~,
\end{equation}
with ``pair wavefunction",
\begin{equation}
\Psi_{xy}({\bf r}; \hat{\bf R}) = ({\bf r} \cdot \hat{\bf R})^2 
+ W  \sin(2\phi) x y ~.
\end{equation}
Here for concreteness we specialized to the elliptical Fermi surfaces,
Eq.~(\ref{wparam}), and $W \ge 0$ characterizes the D-eccentricity,
\begin{equation}
W = \frac{ w^4 + (1/w)^4 }{2} - 1 ~;
\end{equation}
$\hat{\bf R} = {\bf R}/R = \cos(\phi) \hat{\bf x} 
+ \sin(\phi) \hat{\bf y}$, is the unit vector pointing from one pair to 
the other.  
With vanishing D-eccentricity, $W=0$, the pair wavefunction is of an
``extended-$s$" form, for example $\Psi_{xy} \sim x^2$ for $\phi = 0$.
This corresponds to a quadratic nodal line in the pair wavefunction.
With $W=0$ the directions of the nodal lines for each pair are aligned 
perpendicular to the vector connecting one pair to the other.
On the other hand, in the limit of very large D-eccentricity, 
$W \gg 1$, the pair wavefunction takes the $d_{xy}$ form, 
$\Psi_{xy} \sim xy$, corresponding to two nodal lines along the 
$x$ and $y$ axes.  Upon rotating the internal coordinate, the sign
of the pair wavefunction takes the usual $d_{xy}$ form, precisely as for
the Cooper pair wavefunction in a $d_{xy}$ superconductor.  
Thus, the DBL has quasi--long-range order in the $d_{xy}$ boson 
pair channel, although the power law decay exponent within mean field 
theory is very large.

It is also interesting to examine the time decay of a ``local" two-boson 
Green's function,
\begin{equation}
G_{2b}({\bf r}, {\bf r}^\prime; \tau-\tau^\prime) \equiv
\la b^\dagger({\bf r}, \tau) b^\dagger({\bf 0}, \tau)
    b({\bf r}^\prime, \tau^\prime) b({\bf 0}, \tau^\prime) \ra ~,
\end{equation}
with $|{\bf r}|,|{\bf r}^\prime| \ll k_F^{-1}$.  
A pair of bosons is injected at positions ${\bf r}$ and ${\bf 0}$
and is removed at later times at positions ${\bf r}^\prime$ and 
${\bf 0}$.
Within mean field theory,
\begin{equation}
G^{MF}_{2b}({\bf r}, {\bf r}^\prime; \tau) \sim \frac{1}{\tau^4} 
[ ({\bf r} \cdot {\bf r}^\prime)^2 + 2 W (xy) (x^\prime y^\prime) ] ~.
\end{equation}
For large D-eccentricity, $W \gg 1$, this factorizes into a product
of two ``pair wavefunctions" of the $d_{xy}$ form.
This correlator can be used to extract a local pair tunneling 
density of states,
\begin{equation}
\rho_{xy}(E) = \int
\la \hat{\cal D}_{xy}^\dagger(\tau) \hat{\cal D}_{xy}(0) \ra 
e^{-E |\tau|} d\tau ~,
\end{equation}
where $\hat{\cal D}_{xy}^\dagger$ injects a $d_{xy}$ pair centered 
at the origin,
\begin{equation}
\hat{\cal D}_{xy}^\dagger \equiv \int_{\bf r} \,
xy \,\exp(-|{\bf r}|/\xi) b^\dagger({\bf r}) b^\dagger({\bf 0}) ~,
\end{equation}
with pair ``size" $\xi$.
Within mean field theory one obtains a power law tunneling density of 
states $\rho^{MF}_{xy}(E) \sim W E^3$,
with an amplitude that grows with the D-eccentricity parameter $W$.  
The tunneling density of states for an $s$-wave pair also vanishes as 
$E^3$ but the amplitude is independent of $W$.  
This pair tunneling density of states is perhaps the best diagnostic for 
measuring the degree of local $d$-wave two-boson correlations in the DBL.

Finally, we consider a box correlator for bosons which is defined 
in the Introduction, Eq.~(\ref{Bb_def}).
Within mean field this factorizes as,
${\cal B}^{MF}_b(x) = {\cal B}_{d_1}(x) {\cal B}_{d_2}(x)$, and from 
Wick's theorem and 90-degree rotational invariance,
\begin{equation}
{\cal B}_{d_1}(x) = G_{d_1}^2(0,x) - G_{d_1}^2(x,0) 
= -{\cal B}_{d_2}(x) ~.
\end{equation}
Thus,
\begin{equation}
\label{Bb_MF}
{\cal B}^{MF}_b(x) = -[ G_{d_1}^2(0,x) - G_{d_1}^2(x,0) ]^2 \le 0 ~.
\end{equation}
Notice that in the case with zero eccentricity the mean field box 
correlator vanishes, but since the $(\det)^2$ wavefunction is 
non-negative, one expects ${\cal B}_b$ to be positive upon inclusion of 
gauge fluctuations.
With non-zero D-eccentricity, the mean field box correlator is 
negative for all spatial separations, which reflects the underlying 
$d_{xy}$ nodal structure, and decays as $x^{-6}$.
It is possible that the inclusion of gauge fluctuations will again
modify the mean field result in the case with closed Fermi surfaces
because of the pairing tendencies of the $d_1$ and $d_2$ fermions.
In the case with open Fermi surfaces to be discussed next,
we conjecture that the exact box correlator will be negative 
at large spatial separations while decaying to zero as the box size is 
taken to infinity.
This conjecture appears to be consistent with the box correlator 
extracted from the $(\det)_x (\det)_y$ wavefunction with open Fermi
surfaces in Sec.~\ref{subsec:lattice_wavefnc}.

\subsection{Case with open Fermi surfaces}
\label{subsec:openFS}

The preceding analysis holds for arbitrary Fermi surfaces,
in particular, for the lattice bands obtained from Eq.~(\ref{Ht}).
We only need to remember that the fermion Green's function 
Eq.~(\ref{Gd_MF}) is determined by the Fermi surface patches with the 
group velocities that are parallel or antiparallel to $\hat{\bf r}$. 
Once $G_{d_1}$ and $G_{d_2}$ are known, the other correlation functions
follow.

An interesting situation occurs on the lattice when the ratio
$t_\parallel / t_\perp$ is such that the $d_1$ and $d_2$ Fermi surfaces 
are open, which is illustrated in the right panel of 
Fig.~\ref{fig:introFS}.
In this case, for an observation direction close to, say, the $y$-axis, 
there are no $d_1$ Fermi surface patches with normals in this direction, 
so the $d_1$ Green's function has an exponential decay in this direction 
instead of the power law Eq.~(\ref{Gd_MF}).  
Since the real-space boson Green's function is the product of the two 
fermion Green's functions, Eq.~(\ref{Gb_MFdef}), we conclude that it 
decays exponentially in the directions near the $x$- and $y$-axes.  
There may still still be directions near $\hat{\bf x} \pm \hat{\bf y}$ 
in which both $G_{d_1}$ and $G_{d_2}$ show power law behavior, 
and so will the boson Green's function.

Eventually, for large enough $t_\parallel / t_\perp$, the two Fermi
surfaces will have no parallel patches, and the boson Green's
function will decay exponentially in all directions.
We will call this phase DLBL for D-wave correlated Local Boson Liquid.
Note, however, that the system is still gapless and critical,
as can be measured, e.g., from the local spectral function 
Eq.~(\ref{Aloc}).
Also, the boson density correlations still have the power law envelope, 
Eq.~(\ref{DD_MF}), if one fermion field can propagate in the 
observation direction.
Furthermore, the boson box correlator, Eq.~(\ref{Bb_MF}), exhibits 
power law behavior even though the single and pair boson Green's 
functions are exponentially decaying.
From the boson field perspective, the system is local and bosons
have hard time to propagate; 
nevertheless, the system has power law correlations in other properties 
and, in particular, charges can propagate.

Based on an analysis of gauge fluctuations in Sec.~\ref{sec:Gauge_fluct},
we conjecture that the $d_1$ and $d_2$ systems effectively decouple at 
low energies in the DLBL, and at large distances the exact box correlator
is negative and decays to zero as a non-oscillatory power law with 
an integer exponent which is independent of non-universal lattice scale 
physics, ${\cal B}_b(x) \sim - x^{-8}$.  This is the same as in the 
mean field theory except with a larger power.

It is also worth mentioning the limiting case $t_\perp=0$ that gives 
completely flat Fermi surfaces, which we will call ``extremal DLBL.''
The $d_1$ fermions can move only along the $x$-axis,
\begin{equation}
G_{d_1}(x,y) = \delta_{y,0} \frac{\sin(k_F x)}{\pi x}~, \quad 
\text{(extremal DLBL)} ~,
\end{equation}
while the $d_2$ fermions can move only along the $y$-axis.
As a result, the boson field cannot propagate at all, not even one 
lattice spacing.  However, this special system still has power law 
density-density correlations, e.g.,
\begin{equation}
D^{MF}_{d_1}(x,y) \sim -\delta_{y,0} \frac{\sin^2(k_F x)}{x^2}, \quad 
\text{(extremal DLBL)} ~,
\label{Dd_MF_XTRM}
\end{equation}
as well as power law box correlation,
\begin{equation}
{\cal B}^{MF}_b(x) \sim - \frac{\sin^4(k_F x)}{x^4}, \quad
\text{(extremal DLBL)} ~.
\label{Bb_MF_XTRM}
\end{equation}
From the numerical study of the extremal DLBL wavefunction, 
Sec.~\ref{subsubsec:xtrmDLBL}, we conjecture that these mean field 
power laws also hold upon the Gutzwiller projection.
In the gauge theory context, we would say that the fermions
remain unaffected by the gauge field fluctuations.  
This extremal DLBL wavefunction is of interest because of its 
similarity to the so-called Excitonic Bose Liquid ground state of a pure 
ring exchange model -- we discuss this in Sec.~\ref{sec:Energetics}.

\section{Properties of the wavefunctions}
\label{sec:Wavefnc_props}

In this section we study the $\Psi = (\det)_x (\det)_y$ wavefunctions 
directly and measure numerically their properties such as the boson 
Green's function and the density correlation defined in 
Sec.~\ref{subsec:MF_props}.
A detailed comparison is made with the mean field, which provides an 
initial guide.  This is followed by interpretations of the observed
deviations from the mean field using simple ``Amperean interaction'' 
rules of thumb;
the actual calculations behind these rules within the gauge 
fluctuations theory are given in Sec.~\ref{sec:Gauge_fluct}.

We first consider wavefunctions in the continuum so as to avoid 
lattice effects and focus on the consequences of the underlying Fermi 
surfaces.
It appears that the S-type wavefunction, $\Psi = (\det)^2$,
has off-diagonal long-range order, while a generic DBL wavefunction
with non-zero D-eccentricity has only power law boson correlations.
We then consider wavefunctions on the lattice where we can access 
the DLBL phase with open Fermi surfaces described in 
Sec.~\ref{subsec:openFS}; this features boson Green's function that 
decays exponentially in real space.
Finally, we discuss in some detail the extremal wavefunction 
that is obtained when the Fermi surfaces are completely flat.

The calculations with the wavefunctions are performed numerically 
using standard determinantal VMC (variational Monte Carlo) 
techniques.\cite{vmc}  The system used in all calculations is a 
square box with periodic boundary conditions.
In each case, the particle number is chosen so as to fill complete 
momentum shells under the Fermi surfaces.
To facilitate the comparison, the presented mean field is calculated 
for the same finite systems.

\subsection{Wavefunctions in the continuum}

Before proceeding with the numerics, the DBL structure factors 
can be found analytically in some ranges in the momentum space:
Thus, by expanding the boson wavefunction in terms of the orbitals that
form each determinant, it is easy to see that
$n_{\bf k}$ vanishes outside the ${\bf k}_{F1} + {\bf k}_{F2}$ 
surface of Fig.~\ref{fig:loci2kF}, while $D({\bf k})$ vanishes
outside the $2{\bf k}_{F1} + 2{\bf k}_{F2}$ surface.

\subsubsection{S-type state}
\begin{figure}
\centerline{\includegraphics[width=\columnwidth]{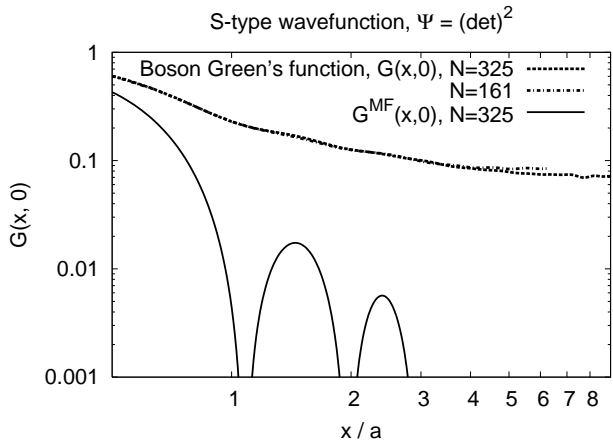}}
\vskip -2mm
\caption{Boson correlation in the $(\det)^2$ wavefunction for two 
systems with $N=161$ and $325$ particles.
The distance is measured in units of $a = \rho^{-1/2}$, 
and the plots are cut when the separation exceeds half of the box size.
Since $G_b(r)$ maintains a sizable limiting value that depends only 
weakly on $N$, we suggest that there is off-diagonal long-range order
in this wavefunction.  
The mean field result is also plotted, and the vertical scale is chosen 
to accommodate its fast decay, $G_b^{MF} \sim 1/r^3$.
}
\label{fig:S-type}
\end{figure}

We begin by considering the limit with zero D-eccentricity, 
in which case the boson wavefunction is positive everywhere 
except for the nodes.
Figure~\ref{fig:S-type} shows boson Green's function measured for 
two systems with $N=161$ and $325$ particles.
It appears that at large separations $G_b(r)$ approaches a finite 
positive value which is roughly the same for both system sizes.
To be more quantitative, the ${\bf k}=0$ mode contains
$n_{{\bf k}=0}/N = 0.10$ and $0.083$ fractions of bosons 
in the two systems.  The present data extracted from the continuum 
wavefunction cannot rule out the possibility that the Green's function 
vanishes in the large distance limit.
But at the very least, the result of the projection is clearly dramatic 
in this case with matched Fermi surfaces, since the fall-off of the 
boson correlation is very slow (if any).
We also remark that we unambiguously find off-diagonal long-range order
when this wavefunction is studied on a lattice at fixed boson density

The extremely strong enhancement of the boson correlation over the
mean field prediction seen in Fig.~\ref{fig:S-type}
can be understood qualitatively as a result of the pairing of the 
$d_1$ and $d_2$ fermions mediated by the gauge field.
Indeed, as we discuss in Sec.~\ref{sec:Gauge_fluct},
the constituents of such a (zero momentum) ``Cooper pair'' with 
oppositely directed group velocities and opposite gauge charges 
produce parallel gauge currents and therefore experience Amperean 
attraction mediated by the gauge field.  
This ``Amperean attraction'' rule of thumb for the enhancements in
correlations appears to be taken to the extreme in the $(\det)^2$ 
wavefunction, where we can crudely picture the fermions paired 
back into $b$ and condensed, giving rise to long-range order in $G_b$.
We also point out that this wavefunction has rather unusual 
density-density correlations which are singular around the 
$2 k_F$ circle.
Since the presence of the off-diagonal order makes this state 
less interesting to us, we do not consider such details any further.
The observation of the order suggests that the S-type state is unstable; 
the ultimate phase in this case is likely a conventional superfluid, 
and a good superfluid wavefunction needs to be constructed differently, 
e.g., using Jastrow approach that builds in proper density correlations.

\subsubsection{DBL state}

We now consider in detail a representative case with non-zero
D-eccentricity -- specifically, with $w^4 = 3$ in Eq.~(\ref{wparam}).
The corresponding $d_1$ and $d_2$ Fermi surfaces are shown to scale in 
Fig.~\ref{fig:loci2kF} together with the singular lines for the 
order parameter and density correlations identified in the mean field,
Sec.~\ref{sec:MFT}.

\begin{figure}
\centerline{\includegraphics[width=\columnwidth]{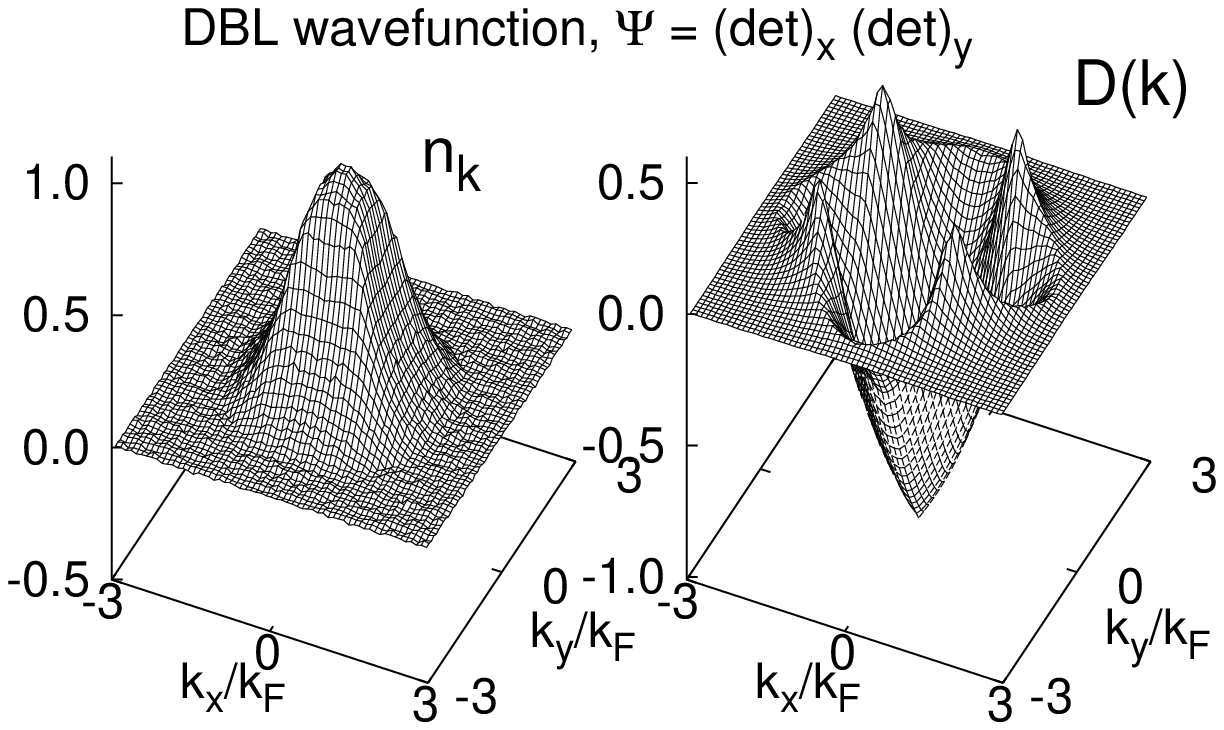}}
\vskip -10mm
\centerline{\includegraphics[width=\columnwidth]{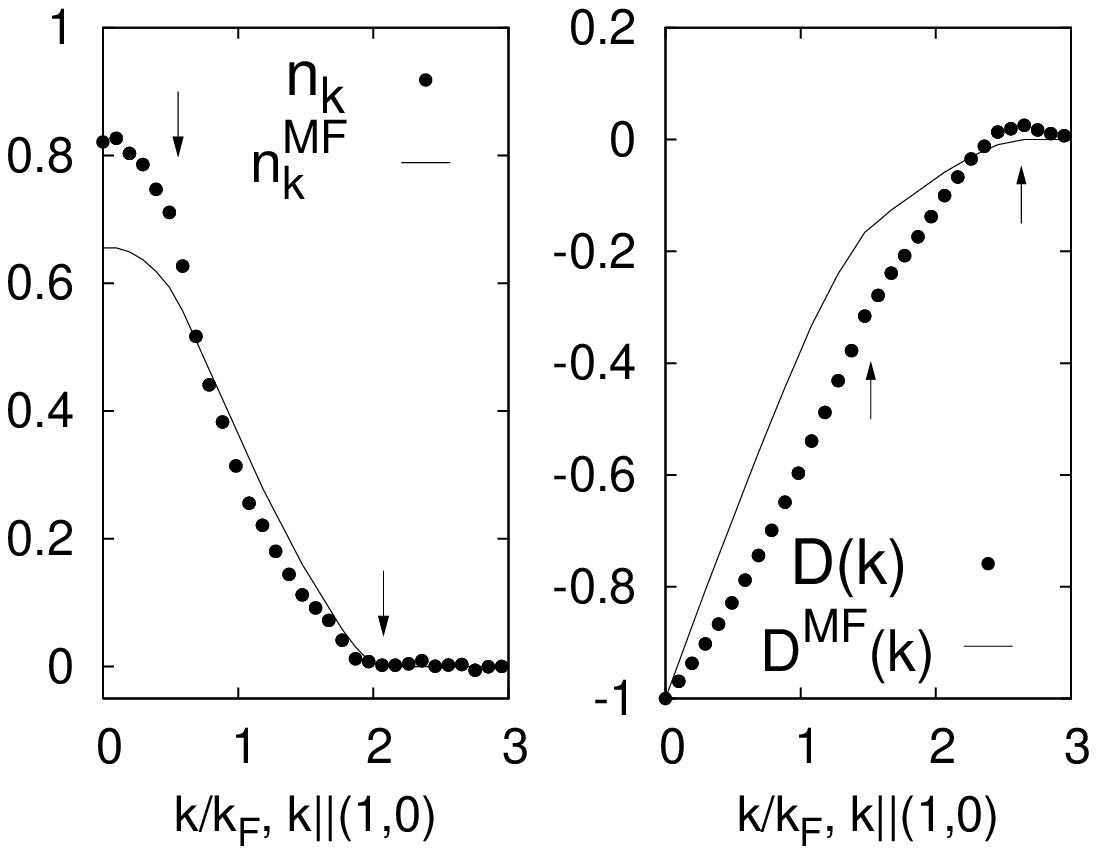}}
\vskip -2mm
\caption{
Top:  Mode occupation $n_{\bf k}$ and density structure factor
$D({\bf k})$ measured in the continuum DBL wavefunction with $w^4=3$
(note that our $D({\bf k})$ defined in Eq.~\ref{Db_k} vanishes 
for large $k$).
The boson number is $N=325$; the underlying Fermi surfaces and all 
singular lines are shown to scale in Fig.~\ref{fig:loci2kF}.
Bottom:  Cut through the data in the (1,0) direction in the k-space 
and comparison with the mean field results.
In the left panel, the arrows point the appropriate $k_{F1} \pm k_{F2}$ 
momenta (see Fig.~\ref{fig:loci2kF}), while in the right panel the 
arrows point the $2k_{F1}$ and $2k_{F2}$ momenta.
}
\label{fig:allq}
\end{figure}

Figure~\ref{fig:allq} gives an overall view of $n_{\bf k}$ and 
$D({\bf k})$ in the two-dimensional momentum space, and also shows a 
one-dimensional cut in the $(1,0)$ k-space direction together
with the mean field predictions.

Consider first the mode occupation function $n_{\bf k}$.
Upon projection, this develops a squarish top with somewhat sharper 
edges as compared with the mean field.  The latter is not shown in the 
full k-space but looks more smooth, while a (1,0) cut can be seen in the 
bottom panel of Fig.~\ref{fig:allq}.
Coming from large momenta, significant deviations from the mean field 
set in near the ${\bf k}_{F1} - {\bf k}_{F2}$ line of 
Fig.~\ref{fig:loci2kF}
(the corresponding location in the 1D cut is indicated with an arrow).
This surface is singular in the mean field, but the singularities are 
weak and are almost not visible, while they become more pronounced upon 
the projection.  On the other hand, we observe no such enhancement near
the ${\bf k}_{F1} + {\bf k}_{F2}$ line.

\begin{figure}
\centerline{\includegraphics[width=\columnwidth]{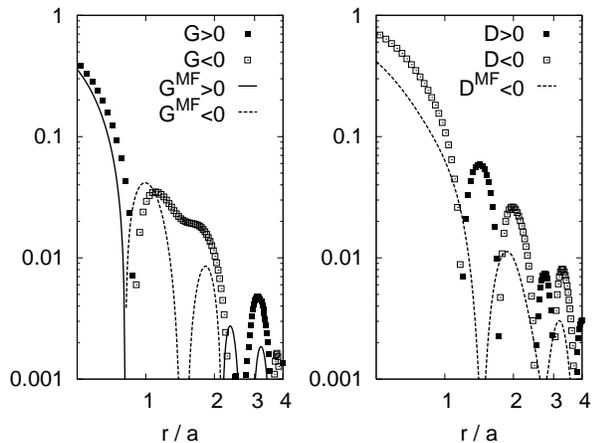}}
\vskip -2mm
\caption{Boson Green's function $G_b(r)$ and density correlation 
$D_b(r)$ measured along the $r=(x,0)$ direction for the DBL wavefunction 
with $N=325$ bosons and D-eccentricity characterized by $w^4=3$
(this is the same system as in Fig.~\ref{fig:allq}).   
The distance is measured in units of $a = \rho^{-1/2}$,
and the plots are cut roughly where the signal in $G_b(r)$ approaches
the noise level.  Note the logarithmic scales used, so it is the 
absolute values that are plotted, while the positive/negative signs
of the data are indicated with filled/open symbols respectively.
In the case of the mean field data, the signs are indicated using
solid/broken lines; $D^{MF}$ is always negative, see Eq.~(\ref{DD_MF}).
}
\label{fig:real-space-10}
\end{figure}

\begin{figure}
\centerline{\includegraphics[width=\columnwidth]{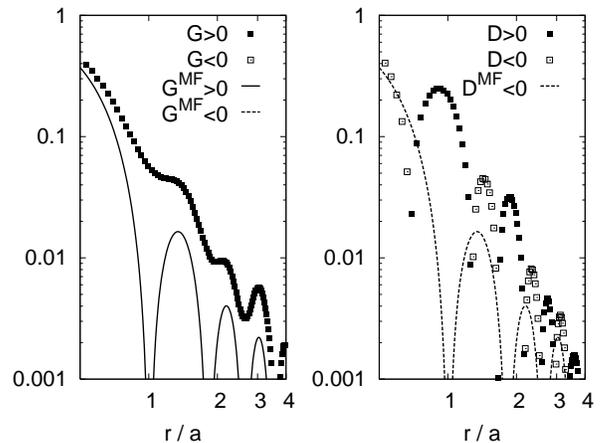}}
\vskip -2mm
\caption{The same as in Fig.~\ref{fig:real-space-10} but in the
$x = y$ diagonal direction in real space.
}
\label{fig:real-space-11}
\end{figure}

This difference between the ${\bf k}_{F1} - {\bf k}_{F2}$ and
${\bf k}_{F1} + {\bf k}_{F2}$ is also visible when examining the boson 
Green's function in real space as shown in the left panels of 
figures~\ref{fig:real-space-10}~and~\ref{fig:real-space-11}.
The mean field $G_b^{MF}({\bf r})$, Eq.~(\ref{Gb_MF}), has both
${\bf k}_{F1} - {\bf k}_{F2}$ and ${\bf k}_{F1} + {\bf k}_{F2}$ 
components equally present, while after the Gutzwiller projection one
finds that the ${\bf k}_{F1} - {\bf k}_{F2}$ component dominates.
Indeed, for the boson Green's function measured along the $x$-axis,
Fig.~\ref{fig:real-space-10}, the relevant Fermi surface patches have 
normals in the $(x, 0)$ direction and are easily located 
(the relevant momentum space cut of $n_{\bf k}$ is also shown in the 
left panel of Fig.~\ref{fig:allq}).
By comparing with the mean field Eq.~(\ref{Gb_MF}), one finds that after 
the projection the amplitude of the oscillation with the smaller 
wave vector $|k_{F1} - k_{F2}|$ wins over that with the larger wave 
vector $k_{F1} + k_{F2}$. 
Slightly more care is needed when interpreting $G_b({\bf r})$ in the 
$x=y$ diagonal direction, Fig.~\ref{fig:real-space-11}.
In this case, the appropriate locations on the 
${\bf k}_{F1} - {\bf k}_{F2}$ surface are at the ``cusp'' points with 
$k_x = -k_y$ in Fig.~\ref{fig:loci2kF}, since this is where the normal
to the singular surface is parallel to the observation direction 
$\hat{\bf r}$.  This component does not oscillate when moving along the 
$x=y$ diagonal in real space, and when it is enhanced, the oscillations 
due to the other ${\bf k}_{F1} + {\bf k}_{F2}$ component become less 
visible, which is what what we see in Fig.~\ref{fig:real-space-11}.

Unfortunately, from this data we cannot attest whether the enhanced
singularities are characterized by new exponents or whether we see
only an amplitude effect.
The locations where the enhancements occur are consistent with the 
Amperean rules of thumb, Sec.~\ref{sec:Gauge_fluct}.
Indeed, the $d_1$ and $d_2$ constituents of the boson operator 
$b^\dagger = d_1^\dagger d_2^\dagger$ move in the opposite directions 
when contributing to the correlation at ${\bf k}_{F1} - {\bf k}_{F2}$ 
and therefore experience Amperean attraction and enhancement 
(remembering that $d_1$ and $d_2$ carry opposite gauge charges), 
while they move in the same direction when contributing at 
${\bf k}_{F1} + {\bf k}_{F2}$ and therefore experience Amperean 
suppression.
These pictures are made more precise in Sec.~\ref{sec:Gauge_fluct},
where we find that within the gauge fluctuations theory, 
the enhanced correlations are characterized by new exponents that 
depend on the observation direction because of the varying degree 
of the Fermi surface curvature matching in the $d_1$-$d_2$ pairing 
channel.  The strongest such enhancement is expected along the 
diagonals $x = \pm y$, which is roughly what we find in the projected 
wavefunctions
(we repeat again that we cannot make any statement about the exponents 
from the data other than that we see increased numerical correlations 
compared with the mean field).

Consider now the density correlation $D_b(r)$ and the corresponding 
structure factor $D_b({\bf k})$. 
In Fig.~\ref{fig:allq}, we can see the singular $2{\bf k}_{F1}$ 
and $2{\bf k}_{F2}$ lines, cf.~Fig.~\ref{fig:loci2kF};
the enhancements are peaked where the two curves cross.
The singular $2k_F$ points are also visible in the (1,0) k-space cut,
bottom right panel of Fig.~\ref{fig:allq}.
On the other hand, the $|k|$ singularity near zero momentum is 
not enhanced.
Examination of the density correlation in real space, right panels of
figures~\ref{fig:real-space-10}~and~\ref{fig:real-space-11},
shows an overall increase and dominance of the oscillatory 
components over the zero-momentum component as compared with the 
mean field Eq.~(\ref{DD_MF}).
Again, we cannot tell whether there is a change in the exponents or 
just an amplitude effect.  The $2k_F$ enhancements of the density 
correlation agree qualitatively with the gauge theory expectations.  
Indeed, consider the density operator $d_1^\dagger d_1$.
The particle and hole constituents (which are oppositely gauge charged) 
have antiparallel group velocities when contributing to the $2 k_F$ 
component and therefore experience Amperean attraction, while they 
move in the same direction when contributing at zero momentum and 
therefore repel each other.  
Again, these rules of thumb are made more precise in 
Sec.~\ref{sec:Gauge_fluct}, where such enhancements in the particle-hole 
channel are characterized by new direction-dependent exponents
(the density correlations are found to decay more slowly when less 
curved patches are involved).
Note that the projection imposes 
$\hat{\rho}_b = \hat{\rho}_{d_1} = \hat{\rho}_{d_2}$, and
we expect $D_b$ to acquire both the $2k_{F1}$ and $2k_{F2}$ signatures 
(in the gauge theory, $\rho_{d_1}$ and $\rho_{d_2}$ imprint on each other
via non-singular high-energy connections that are not manifest in our 
low-energy effective description of Sec.~\ref{sec:Gauge_fluct}).

To summarize, the DBL wavefunction clearly knows about the underlying 
Fermi surfaces; it also contains germs of the gauge fluctuations theory, 
since the enhancements/suppressions of the various ``$2 k_F$'' lines 
appear to agree with the Amperean rules.
However, there are no reasons to believe that the wavefunction 
and the gauge theory will have the same long-distance properties.
In particular, our measurements can not tell whether the long-distance
power laws are changed upon the projection compared with the mean field.
Still, it is gratifying to see that the Amperean rules work for the
DBL wavefunction.

\subsection{Wavefunctions on the lattice}
\label{subsec:lattice_wavefnc}

We now describe our results for various states on the lattice.
As we have already mentioned, S-type wavefunction, $\Psi = (\det)^2$,
with finite boson density per site, have off-diagonal long-range 
order, which we confirm unambiguously using finite-size scaling.
On the other hand, wavefunctions with non-zero D-eccentricity
do not show such order.  
While we have not performed detailed studies, we expect that the case 
with closed Fermi surfaces is similar to the already described continuum 
DBL wavefunction, with a slight complication that one needs to do
proper Brillouin zone folding when considering singular surfaces in the 
momentum space.

\subsubsection{Open Fermi surfaces -- DLBL state}
Of particular interest is the case with open Fermi surfaces,
which can only be realized on the lattice.
Specifically, we studied a $24 \times 24$ system with $N=216$
bosons ($\rho = 0.375$ per site) and the fermion hopping parameters 
$t_\parallel = 1.35$, $t_\perp = 0.65$.  
In this case, the $d_1$ and $d_2$ Fermi surfaces have no parallel 
patches; we therefore expect that the boson Green's function decays 
exponentially.
When we measure $G_b(r)$, we find that it drops below our noise level 
already at 3 lattice spacings, so the corresponding plots are not 
particularly informative and are not shown here.  
Where we can measure reliably, the values after the projection are 
of the same order as the mean field values in the same system.
Since the latter decay exponentially at large distances,
we conjecture the same behavior in the DLBL wavefunction.

On the other hand, the boson density correlation along the $x$
and $y$ axes shows oscillations with a clear power law envelope, 
while the correlations are much smaller in the diagonal $x = \pm y$ 
directions; this behavior is again qualitatively consistent with the
mean field, since neither $d_1$ nor $d_2$ Fermi surfaces have normals 
in the diagonal directions, while one or the other has normals 
along the $x$ or $y$ axis.

\subsubsection{Flat Fermi surfaces -- extremal DLBL state}
\label{subsubsec:xtrmDLBL}
Finally, let us discuss the extremal case when the Fermi surfaces
are completely flat.  The $d_1$ fermions can move only along the
$x$-axis and $d_2$ fermions only along the $y$-axis.  
Using fermion orbitals that are localized on individual rows for $d_1$
(or columns for $d_2$), one can see that the boson wavefunction is 
nonzero only when the number of particles on each row (or on each column)
is the same.
Thus, the bosons can not propagate and their inter-site 
correlation is identically zero.
As we discuss in Sec.~\ref{sec:Energetics}, pure ring Hamiltonian
conserves boson number on each row/column, and the extremal DLBL
wavefunction may be useful in this context.

\begin{figure}
\centerline{\includegraphics[width=\columnwidth]{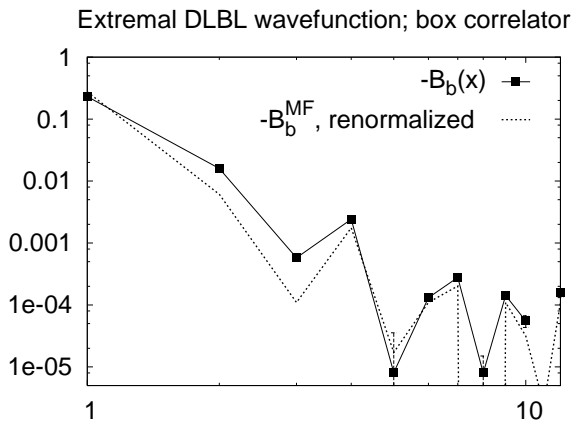}}
\vskip -2mm
\caption{Box correlator ${\cal B}_b(x)$ in the extremal DLBL wavefunction
with flat Fermi surfaces on a $24 \times 24$ lattice with $N=216$ bosons
(the lines are guide to the eye).
Note that the box correlator is negative in the mean field
and is found to be negative in the DLBL state; 
we plot $-{\cal B}_b(x)$ to be able to use the logarithmic scale.
The mean field Eq.~(\ref{Bb_MF}) is renormalized as explained in the
text using a numerical factor that removes double weighting of 
configurations.  
The measured values below $10^{-5}$ are at the noise threshold.
}
\label{fig:corrBOX}
\end{figure}

We can still use the box correlator, Eq.~(\ref{Bb_def}), to
characterize the state and see some ``gaplessness'' in the system.  
The measurement is shown in Fig.~\ref{fig:corrBOX}, where we also plot 
renormalized mean field result.  
The latter is obtained by dividing Eq.~(\ref{Bb_MF}) by 
$\rho^2 (1-\rho)^2$, and a crude justification for such 
procedure\cite{Zhang} is as follows:  
Each $d_1$ or $d_2$ mean field box calculation contains 
implicitly a weight of order $\rho^2 (1-\rho)^2$, since for the ring
operator to be nonzero, two specific sites need to be occupied
and two need to be empty.  However, after the projection it is enough 
to require that only the $d_1$ configuration is ``correct'' since the 
$d_2$ fermions are tied to $d_1$.
From Fig.~\ref{fig:corrBOX}, we see that the box correlator is negative 
and tracks closely the renormalized mean field values.

\begin{figure}
\centerline{\includegraphics[width=\columnwidth]{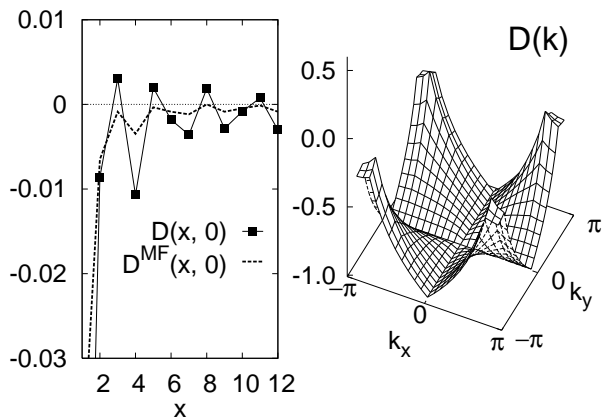}}
\vskip -2mm
\caption{Boson density correlation for the same extremal DLBL system
as in Fig.~\ref{fig:corrBOX}.  
Left panel:  Real space dependence along the $x$-axis, $D_b(x,0)$, 
together with the mean field expectation.
Right panel:  Density structure factor in the momentum space.  
}
\label{fig:DENScorr}
\end{figure}

We can also characterize the extremal DLBL state by measuring the 
density correlations and comparing with the mean field prediction, 
Eq.~(\ref{Dd_MF_XTRM}).  
The results are in Fig.~\ref{fig:DENScorr}.
The left panel shows $D_b(x,0)$ as a function of the distance
along the $x$-axis; it has stronger oscillatory component than the
mean field and swings back and forth across the zero line
while the mean field only touches it, but the overall magnitudes
are comparable and decay as $1/x^2$.
We also find that the density correlation in the $x = \pm y$ diagonal 
directions decays exponentially (not shown);
the mean field predicts zero correlation unless strictly along the axes, 
and we expect that after the projection this corresponds to exponential 
decay.

In the right panel of Fig.~\ref{fig:DENScorr}, we show the density
structure factor $D({\bf k})$.  Several features are clearly visible:
rising ``towers'' draw attention to the lines $(\pm 2 k_F, k_y)$ and 
$(k_x, \pm 2 k_F)$;
one can also see a ``cross'' formed by the lines $(0, k_y)$ and 
$(k_x, 0)$ that run along the axes.
As far as we can say, the character of the singularities across these
lines remains the same as in the mean field; there is an amplitude 
enhancement of the $2k_F$, but no qualitative difference otherwise. 
In particular, near the line $(0, k_y)$, we observe
$D_{b, {\rm sing}}(k_x \to 0, k_y) = A |k_x|$ with $A$ which is 
independent of $k_y$ as long as $|k_y| \gg |k_x|$.

We thus conjecture that the extremal DLBL wavefunction is adequately 
described by the free fermion mean field. 
Some of our findings, e.g., the cross singularity that has a long 
wavelength character, can be understood semi-analytically, 
since the absolute value of the wavefunction has a Jastrow form with a 
peculiar pseudo-potential,
$u(x, y) \sim -\ln|\sin(\pi x /L)| \delta_{y,0} 
              -\ln|\sin(\pi y /L)| \delta_{x,0}$,
it seems plausible that one can calculate other properties of this 
wavefunction analytically.
It is interesting to note that the above suggests a free fermion
description of this boson state, perhaps with some constraints that become
irrelevant in an infinite system.  
We also note that the extremal DLBL appears to be a relative of the 
so-called Excitonic Bose Liquid (EBL) phase predicted in a pure ring 
model on the square lattice\cite{Paramekanti} that we discuss in 
Sec.~\ref{sec:Energetics};
the above therefore suggests that there may be some such description
of the EBL in terms of fermions carrying fractional $1/2$ charge.

\section{Gauge Fluctuations}
\label{sec:Gauge_fluct}

We now study the gauge theory using analytic techniques,
focusing on the large $K$ limit where it is reasonable to
ignore the presence of magnetic monopoles in space-time and to 
expand the cosine term in Eq.~(\ref{Hgauge}) treating the 
gauge field as non-compact.  
Moreover, in the $K \to \infty$ limit, the fermions become free and 
one can put them into a Fermi sea state.
A Fermi surface of fermions minimally coupled to a non-compact $U(1)$ 
gauge field has a considerable history.\cite{Holstein, Reizer, PALee, IoffeKotliar, LeeNagaosa, Polchinski, Altshuler, Nayak, YBKim, LeeNagaosaWen, Senthil, Galitski_gauge}
The $2+1$D such system has been studied most notably as a theory for 
the spin sector in the uniform RVB phase in the slave boson approach
to the high-$T_c$ superconductors.
It has been argued that such fermion systems have a stable phase that 
in some crude aspects is similar to a Fermi liquid -- for example, 
it has a finite long-wavelength compressibility and spin susceptibility.
However, the system is strikingly different in other aspects and is 
described by a new non-Fermi liquid fixed point.
A scaling description of this fixed point was developed in 
Refs.~\onlinecite{Polchinski, Altshuler}.

While we largely follow the earlier work, we find that it is 
convenient to consider a slight reformulation in which the only 
uncontrolled approximation is to assume that the gauge field dynamics 
can be described correctly by the RPA approximation, retaining only 
terms quadratic in the gauge field with a singular quadratic kernel.
A virtue of this approach is that one thereby obtains a theory which has 
an $N$ flavor extension which is soluble at $N=\infty$.  Moreover, 
a controlled and systematic perturbation expansion in powers of $1/N$ 
can be implemented, which allows one to compute  physical properties in 
terms of non-universal ``bare" parameters such as the shapes of the 
Fermi surfaces.  We employ this approach to calculate the leading
$1/N$ behavior for both the boson correlator, $G_b({\bf r})$, 
and the density-density correlator, $D_b({\bf r})$, in the D-wave Bose 
Liquid.

In the next Sec.~\ref{sec:eff_FT}, we will consider an effective field 
theory approach, which allows us to check for the stability of the 
DBL phase that is present at large $N$.  Specifically, we study the 
effects of residual short range attractive interactions between the 
two fermion flavors.  
With vanishing D-eccentricity when the Fermi surfaces for both fermion
species become the same, the $s$-wave Cooper channel is ``nested", and a 
possible instability towards a paired BCS state seems likely. 
The resulting state is a bosonic condensate, since 
$\la b^\dagger \ra = \la d_1^\dagger d_2^\dagger \ra \ne 0$.
However, a non-vanishing D-eccentricity of the Fermi surfaces destroys 
the BCS nesting and the possible pairing instability.
Indeed, we will find that the DBL with a large D-eccentricity can exist 
as a stable phase.
(For smaller D-eccentricity, an instability towards a finite momentum 
Bose condensate or an incommensurate charge density wave co-existing 
with the DBL is a possibility.)
As is discussed in Sec.~\ref{sec:Wavefnc_props}, this is nicely in line 
with the properties of the associated Gutzwiller wavefunctions:
The $(\det)^2$ wavefunction that obtains in the limit of zero 
D-eccentricity appears to have off-diagonal long range order,
whereas the general $(\det)_x (\det)_y$ wavefunction exhibits power law 
correlations consistent with the DBL, as extracted from the gauge theory.

\subsection{Formulation}
\label{subsec:Fluct_formulation}

Being interested in the low energy properties, it is legitimate to 
focus on the fermions living near the Fermi surfaces, just as in 
Fermi liquid theory.  Moreover, the longitudinal Coulomb interactions
mediated by the gauge field are readily screened out within an RPA 
treatment, generating short-ranged screened density-density interactions.
But the transverse fluctuations of the gauge field -- the photon -- are 
incompletely screened by the fermion particle-hole excitations.  
It is thus adequate to just retain the transverse components of the 
vector potential.  Working in the Coulomb gauge with 
$\bm{\nabla} \cdot {\bf a} = 0$, the vector potential reduces to a 
scalar, e.g.,
${\bf a}({\bf k}) = a({\bf k}) \; {\rm sign}(k_y) 
(k_y \hat{\bf x} - k_x \hat{\bf y}) / |{\bf k}|$. 

\begin{figure}
\centerline{\includegraphics[width=\columnwidth]{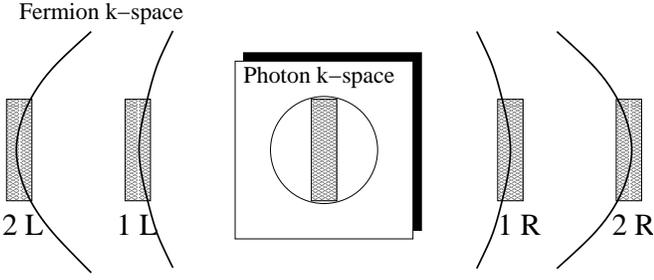}}
\vskip -2mm
\caption{Relevant momentum patches of the strongly coupled fermion -
transverse gauge field system.
For simplicity, here and throughout the text, we take the Fermi surface 
patches with normals in the $\hat{\bf x}$ direction; 
the required properties are the Fermi velocities and surface curvatures;
the latter are crucial since differently oriented patch systems
effectively decouple at low energies.
}
\label{fig:patches}
\end{figure}

Moreover, for each fermion species, it will be sufficient to focus on a 
pair of Fermi surface patches with normals parallel to some axis, 
say $\hat{\bf x}$, as shown in Fig.~\ref{fig:patches}.   
It is the $a_x$ component of the vector potential which is minimally 
coupled to these fermions.  The important wave vectors of the gauge field
$a_x({\bf k})$ that are strongly Landau damped by the fermions in these 
patches satisfy $k_x \ll k_y$ (see Fig.~\ref{fig:patches}), 
and in this region of momentum space $a_x({\bf k}) \approx a({\bf k})$.  
Conversely, the modes of the gauge field $a({\bf k})$ with $k_x \ll k_y$
feed back and scatter the patch fermions.
Thus, we can focus on fermion fields $d_1({\bf k}), d_2({\bf k})$ and 
the gauge field $a({\bf k})$ that are confined to their respective 
patches in momentum space, as shown schematically in 
Fig.~\ref{fig:patches}.

For concreteness, we assume that for each fermion species there are two 
relevant patches on the opposite sides of the Fermi surface that are 
labelled $s = R/L = +/-$ for the right/left patch as indicated in 
Fig.~\ref{fig:patches}.  We then decompose the Fermion fields into
right and left movers by writing,
\begin{equation}
d_{\alpha} ({\bf r}) \sim \sum_s 
e^{i s {\bf k}_{F_\alpha} \cdot {\bf r}} d_{\alpha s}({\bf r}) ~.
\end{equation}
Here, ${\bf k}_{F_\alpha}$ denote the locations on the two Fermi surfaces
which have normals aligned along the $x$-axis, and the fields 
$d_{\alpha s}({\bf r})$ are assumed to be slowly varying on the scale
of $k_F^{-1}$.

The full low energy action consists of three terms describing the 
dynamics of the gauge field, of the fermions, and of their interactions:
\begin{equation}
S^{(0)} = \int d^2{\bf r} \, d\tau \, 
[ {\cal L}_d + {\cal L}^{(0)}_a + {\cal L}_{int} ] ~.
\end{equation}
The ``patch" Lagrangian density for the fermions is simply,
\begin{equation}
{\cal L}_d =  \sum_{\alpha} \sum_{s=\pm} 
d^\dagger_{\alpha s} [ \partial_\tau + s v_\alpha (-i \partial_x) 
                       + \frac{v_\alpha c_\alpha }{2} (-i \partial_y)^2
                      ] d_{\alpha s} ~,
\end{equation}
which is characterized by the local Fermi velocities, $v_\alpha$, 
($\alpha=1,2$), and the local Fermi surface curvatures, $c_\alpha$.
The curvature is crucial in the fermion-gauge problem since there is an 
effective decoupling of the fermions outside the patch from the 
gauge field, $a = a_x$, which is parallel to the patch normals.
Here $s=\pm 1$ specifies the sign of the group velocity while $v$ will 
always denote the absolute value.  

The dynamics for the transverse gauge field describes the free photon,
\begin{equation}
{\cal L}_a^{(0)} = \frac{1}{2} [ (\partial_\tau a)^2 
                                 + \kappa (\bm{\nabla} a )^2 ] ~,
\end{equation}
whereas the gauge-fermion interactions are of the usual minimal coupling 
form:
\begin{equation}
\label{Lint}
{\cal L}_{int} = \sum_{\alpha, s} g_\alpha s v_\alpha 
d^\dagger_{\alpha s} \,a\, d_{\alpha s} ~.
\end{equation} 
Here, $g_1 = -g_2 \equiv g$ with $g=1$, but it will be convenient for 
bookkeeping purposes to retain $g$ as a dimensionless parameter in the 
theory.
We emphasize that both the fermion fields and the gauge field, 
which enter the above action in real space, are slowly varying fields
corresponding to momentum modes inside their respective patches;
in what follows, we always specify the wave vectors of the slow fields 
relative to their respective patch centers.

\subsection{RPA approximation for the gauge field dynamics}
\label{subsec:RPA}

The above theory is strongly coupled and cannot be treated perturbatively
in $g$.  As such, it is necessary to resort to approximations to make any
meaningful progress.  Here we make our central approximation, namely 
replacing the dynamics of the gauge field by a Landau damped form which 
results if one sums the RPA bubble of particle-hole excitations:
$S_a^{(0)} \to S_a$ with
\begin{equation}
\label{Sa}
S_a = \frac{1}{2} \int_{{\bf q}\omega} 
\left(\Gamma \frac{|\omega|}{|q_y|} + \chi |q_y|^2  \right) 
|a ({\bf q}, \omega)|^2 ~.
\end{equation}
The momentum/frequency integral is understood as 
$\int_{{\bf q} \omega} \equiv \int d^2{\bf q} d\omega / (2\pi)^3$,
with the momentum restricted to a region satisfying
$|q_x| < \Lambda_x, |q_y| < \Lambda_y$ with $\Lambda_x \ll  \Lambda_y$.

The magnitude of the Landau damping coefficient, $\Gamma$, will be 
determined by the low-energy fermions near the specified Fermi surface 
patches.  Within the RPA approximation one obtains,
\begin{equation}
\Gamma_{RPA} = \frac{g^2}{2\pi} \left( \frac{1}{c_1} + \frac{1}{c_2} \right) ~,
\label{Gamma_RPA}
\end{equation}
but we will retain $\Gamma$ as an independent parameter since beyond 
this point we will not be using RPA in any event.  
It will sometimes prove convenient to consider the limit in which the 
dimensionless gauge charge,
\begin{equation}
e^2 \approx \frac{g^2}{\Gamma c},
\end{equation}
is small.  Here $c$ is a characteristic Fermi surface curvature.
Indeed, within the $N$-flavor extension discussed below, while $1/N$ will
be the small parameter that leads to a controlled analysis, 
already at order $1/N$ there are some physical properties which we can 
only compute analytically when $e^2$ is simultaneously taken as a 
small parameter.

The parameter $\chi$ that enters in $S_a$ is a ``stiffness'' which
has to be positive for stability.  
Starting from the lattice gauge theory formulation, $\chi$ gives a 
measure of the energetic cost of setting up static internal fluxes.  
One can very crudely estimate $\chi$ using the expression for the 
diamagnetic response of free fermions,
\begin{eqnarray}
\chi &\approx & \frac{g^2}{12 \pi m_d} ~,
\label{chi}
\end{eqnarray}
where $m_d$ is some effective mass.
We reiterate that this stiffness comes from some energetics that is 
assumed to stabilize the Fermi surfaces of the $d$ fermions 
in the first place.  Therefore, from the low-energy perspective,
$\chi$ should be viewed as a phenomenological parameter that encodes 
some high-energy physics.  Furthermore, the stiffness $\chi$ is 
independent of the patch orientation in momentum space for the 
transverse gauge field $a({\bf k})$, since gauge invariance requires 
that the field energy be proportional to 
$\chi B^2 = \chi (\partial_x a_y - \partial_y a_x)^2$.  
In any event, the magnitude of $\chi$ will play a minor role below.
Indeed, for correlation functions with power law decay as we find below,
changing $\chi$ will only modify the amplitude but not the form of the 
correlators.

Before proceeding with an analysis of the full theory, 
$S_d + S_a + S_{int}$, it is important to note that within the above 
approximation the Landau damped dynamics of the gauge field described by 
the action $S_a$ is both singular and harmonic.  We have ignored possible
terms in the action which involve higher powers of the gauge field.
As such, at this stage we could integrate out the gauge field exactly 
generating a non-local and retarded four-fermion interaction.  
It is the absence of non-linearities for the gauge field dynamics which 
allows us to introduce a large-$N$ generalization which is soluble at 
$N=\infty$ and which facilitates a formal and systematic $1/N$ expansion,
as we now describe.

\subsection{$N$ flavor extension}
\label{subsec:N_flavor}

Purely for technical reasons, then, we generalize the theory to $N$ 
flavors of fermions at the two patches on each Fermi surface, 
$d_{\alpha s} \to d_{i \alpha s}$.
Here $i = 1, \dots, N$ labels the flavor index, while $s = \pm$ labels 
the patch location and $\alpha = 1, 2$ the two Fermion species.  
The action for the fermions is simply generalized as,
\begin{equation}
S_d^{(N)} = \sum_{i=1}^N \sum_{\alpha,s}  \int_{{\bf x} \tau} 
d^\dagger_{i\alpha s} ( \partial_\tau 
- i s v_\alpha \partial_x - \frac{v_\alpha c_\alpha}{2} \partial_y^2 ) 
d_{i\alpha s} ~.
\end{equation}
The real scalar gauge field $a({\bf r})$ is likewise generalized, 
but now as an $N \times N$ Hermitian matrix of complex fields, 
$a_{ij}({\bf r}) = a_{ji}^*({\bf r})$.  When expressed in momentum space,
each component of the matrix, $a_{ij}({\bf q})$, has dynamics
with a Landau damped form,
\begin{equation}
S_a^{(N)} = \frac{1}{2} \sum_{i,j=1}^N \int_{{\bf q} \omega} 
\left(\Gamma \frac{|\omega|}{|q_y|} + \chi |q_y|^2  \right) 
|a_{ij} ({\bf q}, \omega)|^2 ~.
\end{equation}
Since $a_{ij}({\bf q}, \omega) = a_{ji}^*(-{\bf q}, -\omega)$, 
this can also be re-written as a trace over the flavor indices of the 
matrix $\overleftrightarrow{a}({\bf q}, \omega)$:
\begin{equation}
S_a^{(N)} = \frac{1}{2} \int_{{\bf q}\omega} 
\left(\Gamma \frac{|\omega|}{|q_y|} + \chi |q_y|^2  \right) 
{\rm Tr} [ \overleftrightarrow{a}({\bf q}, \omega) 
           \overleftrightarrow{a}({\bf -q}, -\omega) ] ~.
\end{equation}
Finally, the fermion-gauge field interaction is taken as,
\begin{equation}
S^{(N)}_{int} = \sum_{\alpha,s} \frac{g_\alpha}{\sqrt{N}} 
s v_\alpha \sum_{i,j=1}^N \int_{{\bf x} \tau} 
d^\dagger_{i\alpha s} \, a_{ij} \, d_{j \alpha s} ~.
\end{equation}
Notice that the magnitude of the interaction strength (the charge) has
been taken to scale as $1/\sqrt{N}$.
At $N=1$ the theory reduces to that discussed in the previous subsection,
with the assumption of a harmonic Landau damped form for the gauge field 
dynamics as the only approximation.

It is worth pointing out that the full action has an enlarged global 
symmetry being invariant under, 
\begin{equation}
d_{i \alpha s} \to e^{i \theta_i} e^{i \phi_\alpha} d_{i \alpha s} ~;
\hskip0.5cm 
a_{ij} \to e^{i(\theta_i - \theta_j)} a_{ij} ~,
\end{equation}
for arbitrary phases $\theta_i$, $i=1, \dots, N$, and $\phi_{\alpha}$, 
$\alpha = 1,2$.  Thus, the densities 
$\sum_\alpha d_{i\alpha}^\dagger d_{i\alpha}$ are independently conserved
for any $i$, and also $\sum_i d_{i\alpha}^\dagger d_{i\alpha}$ are
independently conserved for any $\alpha$.
Note also that the dynamics of each of the $N$ flavor fermions $d_{i1}$ 
is identical to one another, and similarly for all the $d_{i2}$ fermions.

Before solving the full $N$-flavor theory, 
$S_d^{(N)} + S_a^{(N)} + S_{int}^{(N)}$, exactly at $N=\infty$,  
we define large $N$ generalizations of the various correlation functions.
As in the gauge mean field theory, we anticipate that the exact fermion 
Green's function for (any) one of the $N$ fermion flavors, which we 
denote $G^{(N)}_{d_\alpha}({\bf r})$, will be dominated at large 
distances by the patch fermions with normals parallel or antiparallel
to the observation direction $\hat{\bf r}$.
We can thus expand $G^{(N)}_{d_\alpha}({\bf r})$ for large $|{\bf r}|$ 
in terms of the ``patch" fermion Green's functions as,
\begin{equation}
G^{(N)}_{d_\alpha}({\bf r}, \tau) \approx \sum_s 
e^{i s {\bf k}_{F_\alpha} \cdot {\bf r}} G^{(N)}_{\alpha s}({\bf r},\tau)
~,
\end{equation}
with the definition,
\begin{equation}
G^{(N)}_{\alpha s} ({\bf r}-{\bf r}^\prime, \tau-\tau^\prime) = 
\la d^\dagger_{i\alpha s}({\bf r}, \tau) 
    d_{i\alpha s}({\bf r}^\prime, \tau^\prime) \ra ~.
\end{equation}
Note that there is no summation over $i$; throughout, any summation over
indices will be shown explicitly.

There are now $N$ flavors of bosons with creation operators,
\begin{equation}
b_i^\dagger({\bf r}) = d_{i1}^\dagger({\bf r}) d_{i2}^\dagger({\bf r}) ~,
\end{equation}
with all $N$ boson flavors having identical dynamics.
The large $N$ generalization of the boson correlator is then simply,
\begin{equation}
G_{b;ij}^{(N)} ({\bf r}-{\bf r}^\prime, \tau-\tau^\prime) = 
\la b_i^\dagger({\bf r}, \tau) b_j({\bf r}^\prime, \tau^\prime) \ra ~.
\end{equation}
Due to the global symmetries, the correlators which are off-diagonal in 
the flavor indices vanish, leaving a uniquely defined boson Green's 
function, $G_{b;ij}^{(N)} = \delta_{ij} G_{b}^{(N)}$.  
As desired, $G_b^{(N)}$ reduces at $N=1$ to the boson correlator 
defined in Sec.~\ref{subsec:MF_props}.

The boson density-density correlator, $D_b^{(N)}$, is readily defined in 
terms of (any) of the boson densities $\hat\rho_{ib}$ as in 
Eq.~(\ref{densitycorr:def}).  
Similarly, we can define the fermion density-density correlators for 
each species, $D^{(N)}_{d_\alpha}$, in terms of any one of the fermion 
densities, $\hat\rho_{i\alpha}$.
As before, we assume that the boson density-density correlator can be 
approximated as an average over $\alpha$ of the fermion density-density 
correlators, 
$D_b^{(N)} \approx \frac{1}{2} \sum_\alpha D^{(N)}_{d_\alpha}$.

In order to extract the above correlation functions, it is convenient to
define patch fermion particle-hole and particle-particle bubbles, 
which we denote as $\Pi^{(N)}_{ph}$ and $\Pi^{(N)}_{pp}$, respectively.
The exact bubble correlators can be formally expressed in terms of the 
exact patch fermion Green's functions and the exact particle-hole and 
particle-particle vertices,
\begin{widetext}
\begin{eqnarray}
\Pi^{(N)}_{ph;\, \alpha s; \alpha^\prime s^\prime}({\bf q}, \omega) &=& 
\int_{{\bf q}^\prime \omega^\prime} 
V^{(N)}_{ph;\, \alpha s; \alpha^\prime s^\prime}
({\bf q} + {\bf q}^\prime, \omega + \omega^\prime; 
 {\bf q}^\prime, \omega^\prime) \,
G^{(N)}_{\alpha s}({\bf q} + {\bf q}^\prime, \omega + \omega^\prime) \,
G^{(N)}_{\alpha^\prime s^\prime}({\bf q}^\prime, \omega^\prime) ~, \\
\Pi^{(N)}_{pp;\, \alpha s; \alpha^\prime s^\prime}({\bf q}, \omega) &=& 
\int_{{\bf q}^\prime \omega^\prime} 
V^{(N)}_{pp;\, \alpha s; \alpha^\prime s^\prime}
({\bf q} + {\bf q}^\prime, \omega + \omega^\prime; 
 {\bf q}^\prime, \omega^\prime) \,
G^{(N)}_{\alpha s}({\bf q} + {\bf q}^\prime, \omega + \omega^\prime) \,
G^{(N)}_{\alpha^\prime s^\prime}(-{\bf q}^\prime, -\omega^\prime) ~.
\label{bubble;vertex;Greens}
\end{eqnarray}
\end{widetext}
Here $V^{(N)}_{ph}$ and $V^{(N)}_{pp}$ denote the fully renormalized 
vertices, and $\alpha,s$ and $\alpha^\prime,s^\prime$ label the patch 
locations of the two fermion lines that come out of the vertex.  
The total momentum and frequency running through the bubble is 
${\bf q},\omega$, and is divided between the two fermions as shown above.
At $N=\infty$ there are no vertex corrections, 
$V^{(\infty)}_{ph} = 1$ and $V^{(\infty)}_{pp} = 1$, and each bubble
is just a convolution of the two Green's functions.

The boson correlator at large distances and long times can be readily 
expressed in terms of the bubbles,
\begin{equation}
G_b^{(N)}(s {\bf k}_{F_1} + s^\prime {\bf k}_{F_2} + {\bf q}, \omega) 
\approx \Pi^{(N)}_{pp;\, 1s; 2s^\prime}({\bf q}, \omega) ~,
\end{equation}
and similarly for the fermion density-density correlator,
\begin{equation}
D_{d_\alpha}^{(N)}(s {\bf k}_{F_\alpha} - s^\prime {\bf k}_{F_\alpha}
                   + {\bf q}, \omega) 
\approx \Pi^{(N)}_{ph;\, \alpha s; \alpha s^\prime}({\bf q}, \omega) ~.
\end{equation}

\subsection{Solution at $N=\infty$ }
\label{subsec:N_infinity}

At $N=\infty$ the fermion Green's function 
$G^{(N)}_{\alpha s}({\bf q}, \omega)$ can be readily obtained by 
summing the nested rainbow diagrams to give,
\begin{equation}
G^{(\infty)}_{\alpha s}({\bf q}, \omega)
= \frac{1} {i\omega \left(1 + |\Omega_\alpha /\omega|^{1/3} \right) 
            - s v_\alpha q_x - \frac{v_\alpha c_\alpha}{2} q_y^2 }
\label{N=infinity_Greens}
\end{equation}
with
\begin{equation}
\Omega_\alpha^{1/3} = \frac{v_\alpha}{2 \pi \sqrt{3}} 
\frac{g^2}{\chi^{2/3} \Gamma^{1/3}} ~.
\label{Omega}
\end{equation}
Because of the gauge interactions, each fermion species obtains an 
anomalous self-energy of the form $-i\omega |\Omega/\omega|^{1/3}$ 
which dominates over the bare term $-i\omega$ on energy scales below 
$\Omega$.  Physically, the fermions cease to propagate as free particles 
due to the strong scattering from the dynamical gauge field, 
becoming ``incoherent".

We first consider the long distance and time behavior of 
$G^{(\infty)}_{\alpha s}({\bf r},\tau)$.
As in the mean field Sec.~\ref{sec:MFT}, the Green's function is 
dominated at large distances by the patches of the Fermi surface with 
normals along $\pm \hat{\bf r}$.  
For the equal-time Green's function, we obtain,
\begin{equation}
G^{(\infty)}_{d_\alpha} ({\bf r}, 0) \approx \frac{3 \sqrt{\pi}}{4} 
\left( \frac{v_\alpha}{\Omega_\alpha} \right)^{1/2} 
\frac{1}{\sqrt{|{\bf r}|}} G_{d_\alpha}^{MF}({\bf r}) ~,
\end{equation}
where the velocity $v_\alpha$ and curvature $c_\alpha$ characterize the 
$d_\alpha$ Fermi surface patches with normals along $\pm \hat{\bf r}$, 
and $G_{d_\alpha}^{MF}({\bf r})$ is given in Eq.~(\ref{Gd_MF}).
On the other hand, the long time dependence of the local fermion Green's 
function is not affected by the anomalous self-energy and is determined 
solely by the density of states $\nu_0$ at the Fermi energy,
\begin{equation}
G^{(\infty)}_{d_\alpha} ({\bf 0}, \tau) = -\frac{\nu_0}{\tau} ~.
\end{equation}

Turning to the physical correlators,
at $N=\infty$ there are no vertex corrections to the fermion bubble, 
so we have simply,
\begin{equation}
D^{(\infty)}_{d_\alpha}({\bf r}) = - |G^{(\infty)}_{d_\alpha}({\bf r})|^2
~,
\end{equation}
which gives for the $N=\infty$ equal-time density correlator,
\begin{equation}
D^{(\infty)}_{b}({\bf r}) \approx \frac{1}{2} \sum_\alpha \frac{9\pi}{16}
\frac{v_\alpha}{\Omega_\alpha}
\frac{1}{|{\bf r}|} D^{MF}_{d_\alpha}({\bf r}) ~,
\end{equation}
where $D^{MF}_{d_\alpha}({\bf r})$ is given in Eq.~(\ref{DD_MF}).
Similarly, in the absence of vertex corrections at $N=\infty$ in the 
particle-particle bubble, one has for the equal-time boson Green's 
function, 
\begin{eqnarray}
G_b^{(\infty)} ({\bf r}) &=& G_{d_1}^{(\infty)} ({\bf r})  
G_{d_2}^{(\infty)} ({\bf r}) \\
&\approx& 
\frac{9\pi}{16}
\left( \frac{v_1 v_2}{\Omega_1 \Omega_2} \right)^{1/2} 
\frac{1}{|{\bf r}|} G_b^{MF}({\bf r}) ~,
\end{eqnarray}
where $G_b^{MF}({\bf r})$ is given in Eq.~(\ref{Gb_MF}).
However, the time dependence of the local boson Green's function 
$G_b({\bf 0}, \tau)$ is unchanged from the mean field Eq.~(\ref{Gtau}).
Finally, the box correlator is given by Eq.(\ref{Bb_MF}) with 
appropriate $N=\infty$ fermion Green's functions and decays
with a power law envelope of $-x^{-8}$.

Notice that both the boson and the boson density-density correlators
at $N=\infty$ fall off in space as $|{\bf r}|^{-4}$, which is faster 
than their mean field counterparts.  Injecting a boson corresponds to 
creating two fermions.  In the mean field, these fermions propagate 
independently and as free particles.  On the other hand, in the 
present $N=\infty$ theory that sums the nested rainbow diagrams, 
both fermions propagate ``incoherently" due to their scattering from the 
gauge field, which effectively reduces the ability of the created boson 
to propagate, thereby leading to a faster decay of the boson correlator.
At this level of approximation, the two injected fermions, while 
scattered by the gauge fluctuations, do not scatter off one another.  
The effects of interactions between the two injected fermions appear at 
order $1/N$, entering as vertex corrections.  Since the two fermion 
species have opposite gauge charge, one expects that they will attract 
one another.  As we shall see, though, they do not form a composite 
boson (a bound state, for example).
But their motion becomes strongly correlated allowing them to propagate 
more effectively when close by spatially.  The net result, as we find 
below, is that the boson can propagate more effectively when the 
motion of the fermion pair is correlated, leading to a slower decay 
of the correlator compared to the $N=\infty$ result.
Thus, the effect on the boson dynamics due to the decoherence 
experienced by the fermions from scattering off the gauge fluctuations
is compensated, perhaps only partially, by the attractive interaction 
between the pair of fermions.

\subsection{Fermion Green's function at order $1/N$}
\label{subsec:1_N_fermion_Greens}

The $1/N$ contributions to the fermion Green's function are obtained
by considering the nested rainbow diagrams and replacing one gauge field 
propagator 
$G_a({\bf q}, \omega) = (\Gamma |\omega|/|q_y| + \chi |q_y|^2)^{-1}$ 
with the bubble correction
$\delta G_a \sim \frac{1}{N} G_a^2 \Pi_{ph}^{(\infty)}$.
Formally, we can first calculate the rainbows using $G_a + \delta G_a$
everywhere and then extract the $1/N$ piece.
As we now argue, the functional form of the fermion Green's function 
in Eq.~(\ref{N=infinity_Greens}) remains unchanged except for finite 
$1/N$ shifts in the parameters.
Indeed, evaluating the $N=\infty$ particle/hole bubble at small 
wave vector and frequency gives 
$\Pi^{(\infty)}_{ph}({\bf q}, \omega) \sim |\omega|/|q_y|$, 
which only leads to a finite $1/N$ shift in $\Gamma$.
This is not surprising, since the bare singular gauge propagator
postulated at the outset in our theory is motivated by the free fermion 
particle-hole bubble, and one can verify that the bubble remains unchanged
also in the presence of an arbitrary fermion self-energy that depends 
on the frequency only.
Then, using this renormalized gauge propagator to evaluate the rainbow 
diagram gives a contribution to the fermion Green's function of the form,
$\delta G_{d_\alpha}^{(N)} \sim \frac{1}{N} \omega^{2/3} 
[G^{(\infty)}_{d_\alpha}]^2$.
This will at most give finite $1/N$ shifts to 
$\Omega_\alpha, v_\alpha, c_\alpha$. 
Next we consider the vertices at $1/N$, which undergo more dramatic 
modifications.

\subsection{Vertices at order $1/N$}
\label{subsec:1_N_vertex}

\subsubsection{Single photon contribution}
\label{subsubsec:V1a}

\begin{figure}
\centerline{\includegraphics[width=2.0in]{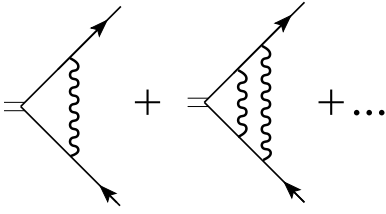}}
\vskip -2mm
\caption{
Ladder diagrams that contribute to $V_{ph}$ at $1/N$.
}
\label{fig:Vph}
\end{figure}

\begin{figure}
\centerline{\includegraphics[width=2.0in]{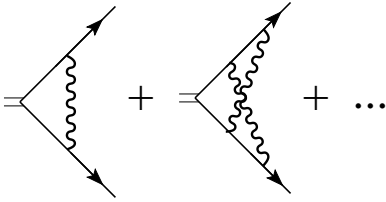}}
\vskip -2mm
\caption{
Crossed diagrams that contribute to $V_{pp}$ at $1/N$.
}
\label{fig:Vpp}
\end{figure}

Consider first dressing the bare vertices with a single gauge propagator.
We denote the value of the corresponding Feynman diagrams as
$V^{(1a)}_{ph}$ and $V^{(1a)}_{pp}$, where $(1a)$ stands for one 
(Landau damped) photon exchanged.  These diagrams are the
first ones in Figs.~\ref{fig:Vph}~and~\ref{fig:Vpp}.
As we will discuss below, the full structure at $1/N$ is more involved, 
but the calculation of $V^{(1a)}$ already captures the main qualitative
physics.

For simplicity, we first focus on the situation with zero external $y$ 
momentum and frequency, defining,
\begin{equation}
V^{(1a)}_{q_x^-} = 
V^{(1a)}({\bf q}, \omega; {\bf q}^\prime, \omega^\prime)|_{q_y=q_y^\prime=\omega=\omega^\prime=0} ~,
\end{equation}
where ${\bf q},{\bf q}^\prime$, $\omega,\omega^\prime$ denote 
the momentum/frequencies in and out of the two legs and 
$q_x^- = q_x - q_x^\prime$ is the momentum running through the vertex.  
For notational simplicity we have suppressed the corresponding flavor 
labels, $\alpha,s; \alpha^\prime,s^\prime$.
Evaluating the Feynman diagrams gives,
\begin{eqnarray}
V^{(1a)}_{ph;\, q_x^-} &=& 
\frac{1}{N} \,{\rm sign}(g_\alpha g_{\alpha^\prime})\, \delta_{s, -s^\prime}
\lambda_{ph} \ln(\Lambda_x/|q_x^-|) ~, 
\label{Vph1a}
\\
V^{(1a)}_{pp;\, q_x^-} &=& 
-\frac{1}{N} \,{\rm sign}(g_\alpha g_{\alpha^\prime})\, \delta_{s, -s^\prime}
\lambda_{pp} \ln(\Lambda_x/|q_x^-|) ~,
\label{Vpp1a}
\end{eqnarray}
revealing a logarithmic divergence cut off by the total $x$-momentum 
running through the vertex.
Explicit expressions for the dimensionless non-negative coefficients 
$\lambda_{ph}$ and $\lambda_{pp}$ are given below.

Notice that both the particle-hole and the particle-particle 
vertex corrections vanish unless the two patches have opposite group 
velocities.  Moreover, the sign of the correction is given by the 
sign of the product of the two charges and differs for the particle-hole 
and particle-particle channels.  The origin of these signs is essentially
Ampere's law familiar from electrodynamics. 
Two moving charges producing parallel charge currents experience an 
attractive interaction, bringing them closer together and enhancing their
interaction strength, which is encapsulated by the magnitude of the 
interaction vertex. 
For example, a particle and a hole of the same fermion species 
($\alpha = \alpha^\prime$) residing on the opposite Fermi surface patches
($s = -s^\prime$) have an enhanced interaction strength, which is
encoded in $V_{ph}^{(1a)} > 0$.
On the other hand, the interaction vertex is suppressed when the
charge currents are anti-parallel.
For example, a pair of particles with the same charge ($g = g^\prime$) 
experience a suppressed vertex interaction, $V_{pp}^{(1a)} < 0$, 
when their group velocities are antiparallel.

The absence of singular vertex corrections when the two patch fermions
have parallel group velocities, $s = s^\prime$, is, we believe, not just 
a peculiarity of this lowest order contribution, but will be valid 
generally for the exact vertices at arbitrary $N$.  This expectation is 
based on the specific form of the propagators: the fermion propagator 
has simple poles in the complex $q_x$ plane while the gauge field
propagator is independent of $q_x$, the latter a reflection of the 
transverse nature of the gauge field.
For any given Feynman diagram, it is the sign of the internal frequency
variables running around the loops together with $s, s^\prime$ 
which will determine whether the poles in the integrand will be in the 
upper or the lower complex $q_x$ planes.
For vertices with $s = s^\prime$ all of the fermion lines internal to 
any diagram will have the same group velocity, and with energy 
(frequency) conservation one expects that at least one of the 
$q_x$ variables will only have poles in the upper (or lower) half-plane, 
so that the contour can be appropriately deformed to show that the 
diagram vanishes.  
Thus, we henceforth restrict attention exclusively to vertices with
$s = -s^\prime$, and to ease the notation will drop the explicit
$s, s^\prime$ indices in the following.

The dimensionless magnitudes of the vertex enhancements in the 
particle-hole and the particle-particle channel, 
$\lambda_{ph}$ and $\lambda_{pp}$ respectively, are given by,
\begin{eqnarray}
\lambda_{ph} &=& 
\frac{g^2}{\sqrt{3}\pi \Gamma^{1/3} \chi^{2/3}
           \left( \frac{\Omega^{1/3}}{v} 
                 + \frac{\Omega^{\prime 1/3}}{v^\prime} \right)} 
{\cal E}[\zeta_{ph}] ~, 
\label{lambda_ph}
\\
\lambda_{pp} &=& 
\frac{g^2}{\sqrt{3}\pi \Gamma^{1/3} \chi^{2/3}
           \left( \frac{\Omega^{1/3}}{v} 
                 + \frac{\Omega^{\prime 1/3}}{v^\prime} \right)} 
{\cal E}[\zeta_{pp}] ~;
\label{lambda_pp}
\\
\lambda_{ph} &=& {\cal E}[\zeta_{ph}] ~,  \quad\quad
\lambda_{pp} = {\cal E}[\zeta_{pp}] ~.
\label{lambda_Ninfty}
\end{eqnarray}
In the last line, we specialized to the $N=\infty$ expressions for 
$\Omega$ and $\Omega^\prime$ (Eq.~\ref{Omega}),
which is valid for extracting the order $1/N$ vertex corrections.
Here, ${\cal E}[\zeta]$ is a dimensionless function of a dimensionless 
argument given by
\begin{widetext}
\begin{equation}
{\cal E}[\zeta] = \frac{3\sqrt{3}}{2\pi} \int_0^\infty \frac{t dt}{(1+t^3) (1 + \zeta^2 t^4)} 
 =  \frac{3\pi\zeta^5 + 8\pi(1-\zeta^4)/\sqrt{3}
- 3\sqrt{2}\pi(\zeta^{1/2} - \zeta^{7/2}) - 6\zeta^2 \ln\zeta}
{8\pi (1+\zeta^6) / \sqrt{3}} ~,
\label{Ffunc}
\end{equation}
\end{widetext}
and has been normalized so that ${\cal E}[0]=1$.  
This function is monotonically decreasing with increasing argument
varying as ${\cal E}[\zeta] \approx \frac{3 \sqrt{3}}{8 \zeta}$ for 
$\zeta \to \infty$.
The parameters $\zeta_{ph}$ and $\zeta_{pp}$ are given by,
\begin{eqnarray}
\zeta_{ph} &=& 
\frac{ \Gamma^{2/3} (c + c^\prime)}
     { 2 \chi^{2/3}
       \left( \frac{\Omega^{1/3}}{v} 
             + \frac{\Omega^{\prime 1/3}}{v^\prime} \right) } ~, 
\label{zeta_ph} \\
\zeta_{pp} &=& 
\frac{ \Gamma^{2/3} |c - c^\prime|}
     { 2 \chi^{2/3}
       \left( \frac{\Omega^{1/3}}{v} 
             + \frac{\Omega^{\prime 1/3}}{v^\prime} \right) } ~;
\label{zeta_pp} \\
\zeta_{ph} &=& \frac{\sqrt{3}\pi}{2 g^2} \Gamma (c + c^\prime) ~, 
\quad\;
\zeta_{pp} = \frac{\sqrt{3}\pi}{2 g^2} \Gamma |c - c^\prime| ~.
\label{zeta_Ninfty}
\end{eqnarray}
In the last line, we again used the $N=\infty$ values for 
$\Omega,\Omega^\prime$.
Notice that the vertex interaction strength between two fermions on 
opposing patches depends on their particular Fermi surface curvatures 
$c, c^\prime$.  Moreover, in the particle-particle channel the 
interaction strength is maximal when the curvatures are equal.  
This reflects a particle-particle nesting, which is responsible for the 
BCS pairing instability in the Fermi liquid context.

\vskip 2mm
\subsubsection{Photon ladders}

Having evaluated the one-photon corrections $V^{(1a)}$ to the bare 
vertices, we now discuss the full set of diagrams that appear at 
order $1/N$ in each case.
First consider the fermion particle-hole vertex which enters into the
density-density correlator: 
$V^{(N)}_{\rho_\alpha} = V_{ph;\, \alpha, \alpha}^{(N)}$.
The $1/N$ diagrams are shown in Fig.~\ref{fig:Vph} and contain an 
arbitrary number of non-crossing photon lines,\cite{footnote:N_flavor} 
beginning with the one-photon diagram $V_{ph}^{(1a)}$, Eq.~(\ref{Vph1a}).  
We shall denote the contribution from the diagram with $n$ photon lines 
as $V_{ph}^{(na)}$.  The magnitude of the singular contributions from 
each diagram in the ladder sum will depend on a dimensionless gauge 
charge,
\begin{equation}
e_\alpha^2 \equiv g^2/\Gamma c_\alpha ~.
\end{equation}
Above we found
$V_{ph}^{(1a)} = (1/N) {\cal E}[\sqrt{3}\pi/e_\alpha^2] \ln(1/q_x^-)
\equiv (1/N) A^{(1)}_1 \ln(1/q_x^-)$.
The singular contributions from the diagram with two photon lines 
can also be extracted from the Feynman diagram,
$V_{ph}^{(2a)} = (1/N) 
[ A^{(2)}_1 \ln(1/q_x^-) + A^{(2)}_2 \ln^2(1/q_x^-) ]$, 
with $A^{(2)}_1, A^{(2)}_2$ universal dimensionless functions
of $e_\alpha^2$.
From this we infer the general structure for the diagram with $n$ 
photon lines,
\begin{equation}
V_{ph}^{(na)} \sim \frac{1}{N} \sum_{m=1}^n A^{(n)}_m \ln^m(1/q_x^-) ~.
\end{equation}
This form suggests that the full set of $1/N$ ladder diagrams can
be exponentiated,
\begin{equation}
V_{\rho_\alpha}^{(N)} 
\sim \frac{1}{N} \exp[\gamma_\alpha \ln(1/q_x^-)] 
\sim \frac{1}{N} |q_x - q_x^\prime|^{-\gamma_\alpha} ~,
\label{ph_ladder_sum}
\end{equation}
with $\gamma_\alpha = \sum_{n=1}^\infty A^{(n)}_1$.
The validity of this exponentiation can be justified by formally 
summing the ladder series to obtain an integral equation for the full 
vertex.  The integral equation has a singular kernel and its solution is 
expected to have the power-law form with some exponent $\gamma_\alpha$ 
which will depend on the dimensionless charge $e_\alpha^2$.

Unfortunately, we have been unable to solve the integral equation
to obtain $\gamma_\alpha$.
However, we observe that for small charge $e_\alpha^2$,
$A^{(1)}_1 = 3 e_\alpha^2 / (8\pi) + O(e_\alpha^4)$.
This gives the leading behavior even though $A^{(1)}_1$ is not analytic 
in $e_\alpha^2$, as can be seen from Eq.~(\ref{Ffunc});
more generally, one finds $A^{(n)}_m = O(e_\alpha^{2n})$.
Thus, assuming small parameter $e_\alpha^2$, the leading contribution to 
$\gamma_\alpha$ comes just from the single photon diagram,
\begin{equation}
\gamma_\alpha = \frac{3}{8\pi} e_\alpha^2 + O(e_\alpha^4, 1/N) ~.
\label{anomalous;density;exponent}
\end{equation}

For $e_\alpha^2$ of order one, the exponent $\gamma_\alpha$, while 
obtained by summing diagrams at order $1/N$, is itself of order one.
It is the amplitude of the vertex in Eq.~(\ref{ph_ladder_sum})
which is of order $1/N$.
This reflects the particular structure of the large $N$ theory,
which, constructed to sum the nested rainbows for the fermion self energy
at leading order, also exponentiates the ladder sum for the 
particle-hole vertex at order $1/N$.
Notice that $\gamma_\alpha$ monotonically decreases as the curvature
of the Fermi surface is taken large, due to the increasingly restrictive 
phase space requirements on the particle-hole vertex in this limit.  
We expect that this trend is valid more generally, for larger 
$e_\alpha^2$ and for arbitrary $N$.

Summarizing the discussion of the particle-hole vertex, we can also 
write a scaling form for more general external momenta and frequencies,
\begin{equation}
V^{(N)}_{\rho_\alpha} \sim |q_x-q_x^\prime|^{-\gamma_\alpha}
\tilde{V}_{\rho_\alpha} 
\left(\frac{q_y^2}{q_x^-}, \frac{q_y^{\prime 2}}{q_x^-},
      \frac{\omega^{2/3}}{q_x^-}, \frac{\omega^{\prime 2/3}}{q_x^-} 
\right) ~.
\label{Vrho;scaling}
\end{equation}

\vskip 2mm

We next consider the vertex which enters in the boson-boson correlator, 
$V_b^{(N)} = V^{(N)}_{pp;\, 1,2}$.
The full set of diagrams that contribute to the particle-particle vertex 
are shown in Fig.~\ref{fig:Vpp} and contain an arbitrary number of 
maximally-crossed photon lines.\cite{footnote:N_flavor}
The first term is the one-photon diagram that we evaluated in 
Eq.~(\ref{Vpp1a}), 
$V_{pp;\, 1,2}^{(1a)} = (1/N) {\cal E}[\zeta_{pp}] \ln(1/q_x^-)
\equiv (1/N) B_1 \ln(1/q_x^-)$
(plus non-singular contributions).
An evaluation of the two-photon diagram gives the same form,
$V_{pp;\, 1,2}^{(2a)} = (1/N) B_2 \ln(q_x^-)$.
In contrast to the particle-hole case, here the two-photon diagram 
does not contain a log-squared term, which indicates that the crossed 
ladder series cannot be exponentiated.   While we cannot evaluate the 
higher order diagrams, we expect that each term will have a log 
contribution with some dimensionless amplitudes $B_n$.
If this is the case, the full structure of the particle-particle vertex 
up to order $1/N$ is,
\begin{equation}
V_b^{(N)} = 1 + \frac{\eta_1 \ln(\Lambda_x/q_x^-)}{N} + O(1/N^2) ~,
\label{bosonvertex_1/N}
\end{equation}
with $\eta_1 = \sum_{n=1}^\infty B_n$.
To check whether the latter series can be defined in principle,
we again take the limit of small dimensionless charges $e_\alpha^2$.
Consider the least favorable case with matched Fermi surface curvatures,
$c_1 = c_2$, so $e_1^2 = e_2^2 \equiv e^2 = g^2/(\Gamma c)$.
We find that the first term is independent of $e^2$,
$B_1 = 1$.  However, evaluating the two-photon diagram gives
$B_2 \sim (e^2)^2 \ln^3(1/e^2)$ for small $e^2$.
The diagrams are expected to be smaller when $c_1 \neq c_2$
(for example, even the first term $B_1 = {\cal E}[\zeta_{pp}]$
can be made small when $g^2 \ll \Gamma |c_1 - c_2|$).
We thus conjecture that the series converges for small enough
$e_\alpha^2$.

Provided it is legitimate to exchange the order of limits $N \to \infty$ 
and $q_x \to 0$ in Eq.~(\ref{bosonvertex_1/N}),
the  boson vertex can be written in a power law form,
\begin{equation}
V_b^{(N)}({\bf q}, \omega; {\bf q}^\prime, \omega^\prime)
|_{q_y=q_y^\prime=\omega=\omega^\prime=0} 
\sim |q_x - q_x^\prime|^{-\eta} ~,
\end{equation}
with
\begin{equation}
\eta = \eta_1/N + O(1/N^2) ~.
\label{eta_1/N}
\end{equation}  
We will discuss the legitimacy of this exponentiation procedure in 
Sec.~\ref{sec:eff_FT}.
For more general external momenta and frequencies the full boson 
vertex will satisfy a scaling form,
\begin{equation}
V_b^{(N)}({\bf q}, \omega; {\bf q}^\prime, \omega^\prime) \sim  
|q_x - q_x^\prime|^{-\eta} \,
\tilde{V}_b \left(\frac{q_y^2}{q_x^{-}}, \frac{q_y^{\prime 2}}{q_x^-},
                  \frac{\omega^{2/3}}{q_x^-}, 
                  \frac{\omega^{\prime 2/3}}{q_x^-} \right) ~,
\label{Vb;scaling}
\end{equation}
where we have used the fact that the vertex depends only on the
total $x$-momentum running through it, $q_x^- = q_x - q_x^\prime$.
The scaling function will vary as, $\tilde{V}_b(X,0,0,0) \sim X^{-\eta}$
when any of its arguments $X \to \infty$.  For example, one has,
$V_b \sim |q_y|^{-2\eta} \sim |\omega|^{-2\eta/3}$.

As will be detailed shortly, the singular vertices
Eq.~(\ref{Vrho;scaling}) and Eq.~(\ref{Vb;scaling}) will lead to 
anomalous decay exponents for the boson density-density correlator
and the boson Green's function.

\subsection{ Boson and density correlators at finite $N$}
\label{subsec:1_N_corr}

Since there are no singular $1/N$ corrections to the fermion 
Green's functions, to extract the boson and density correlators up to 
order $1/N$ one can use the $N=\infty$ fermion Green's functions when 
evaluating the bubble diagram.
Using Eq.~(\ref{N=infinity_Greens}) and the scaling form 
Eq.~(\ref{Vb;scaling}) for the $V_b$ vertex 
enables the particle-particle bubble to be expressed as,
\begin{eqnarray*}
\Pi^{(N)}_{pp} ({\bf q},\omega) \sim |q_x|^{-(2+\eta)}
\int_{{\bf q}^\prime \omega^\prime}
P \left( \frac{q_y^2}{q_x}, \frac{\omega^{2/3}}{q_x},
         \frac{q_x^\prime}{q_x}, 
         \frac{q_y^{\prime 2}}{q_x},\frac{\omega^{\prime 2/3}}{q_x} 
\right) ~,
\end{eqnarray*} 
where $P$ is an appropriate scaling function.
One can then scale out the $q_x$ dependence under the integrals 
to obtain,
\begin{equation}
\Pi^{(N)}_{pp}({\bf q},\omega) \sim |q_x|^{1-\eta} \,
\tilde{\Pi}_{pp}\left(\frac{q_y^2}{q_x}, \frac{\omega^{2/3}}{q_x} \right)
~.
\end{equation}
We thereby arrive at the desired expression for the boson correlator,
\begin{equation}
G_b^{(N)}({\bf k}_{F_1} - {\bf k}_{F_2} + {\bf q}, \omega) 
\sim |q_x|^{1-\eta} 
\tilde{G} \left( \frac{q_y^2}{q_x}, \frac{\omega^{2/3}}{q_x} \right) ~,
\label{Gb_scaling}
\end{equation}
with exponent $\eta$ given in Eq.~(\ref{eta_1/N}).
While this result was obtained systematically to order $1/N$, 
we expect it will remain valid more generally, and in particular will 
involve the same scaling combinations $q_y^2/q_x$, $\omega^{2/3}/q_x$.
The scaling function $\tilde{G}$ and the dependence of the exponent 
$\eta$ on the bare parameters such as the Fermi surface curvatures will 
presumably vary with $N$.
Near the other momentum ${\bf k}_{F_1} + {\bf k}_{F_2}$, 
the boson correlator will satisfy a similar scaling form except with 
$\eta=0$.
As argued before, we expect that the anomalous exponent at momentum
${\bf k}_{F_1} +{\bf k}_{F_2}$ will vanish in general,
not just to leading order in $1/N$.

The large distance behavior of the equal-time boson Green's function in 
the DBL follows directly from Eq.~(\ref{Gb_scaling}) and was already 
listed in the Introduction, Eq.~(\ref{Gb_final}).
We remark that simple ``power counting'' using Eq.~(\ref{Gb_scaling}) 
works here because the relevant frequencies and momenta are indeed in 
the ``scaling regime''; however, it should be kept in mind that we are 
dealing with line singularities and should exercise more caution with 
such arguments in general.
Once again, both the Fermi wave vectors ${\bf k}_{F_1}, {\bf k}_{F_2}$ 
and the scaling exponent $\eta$ will depend on the location and shape of 
the Fermi surface patches with normals along the observation direction 
$\hat{\bf r}$.   At wave vector ${\bf k}_{F_1} - {\bf k}_{F_2}$, the 
Green's function decay becomes more slow due to the Amperean attraction 
between the $d_1$ and $d_2$ fermions.
The fermions tend to move as pairs, but unlike a Cooper pair their 
motion is not phase coherent -- the pairs are uncondensed.
The boson correlator, while decaying more slowly with positive $\eta$,
does not exhibit ODLRO in the DBL phase.

As we found to leading order in $1/N$, Eq.~(\ref{eta_1/N}), $\eta$ is 
expected to be largest when the two Fermi surface curvatures are equal.
In the DBL phase which has closed Fermi surfaces, the characteristics
of the $d_1$ and $d_2$ patches with normals along either diagonal 
$\hat{\bf x} \pm \hat{\bf y}$ are always equal due to the symmetry of 
the square lattice.   Moreover, the decay of the boson Green's function 
$G_b({\bf r})$ will be non-oscillatory along the diagonals,
\begin{equation}
G_b({\bf r}) \sim \frac{1}{|{\bf r}|^{4-\eta}} ~, 
\hskip0.5cm \hat{\bf r} \parallel (\hat{\bf x} \pm \hat{\bf y}) ~,
\end{equation}
a prediction that we find consistent with the properties of the 
$(\det)_x (\det)_y$ wavefunctions studied in Sec.~\ref{sec:Wavefnc_props}
(see Fig.~\ref{fig:real-space-11}).
For $c_1=c_2$, a very rough estimate can be extracted from the leading 
behavior, $\eta = (1/N) + O(e^2/N)$ by setting $N=e^2=1$ giving 
$\eta \approx 1$.
This would shift the $N=\infty$ decay exponent back to its mean field 
value of $1/|{\bf r}|^3$.   
As noted earlier, the gauge fluctuations encapsulated already at 
$N=\infty$ strongly modify the motion of each fermion leading to a 
boson that moves less coherently decaying with a larger exponent than 
in the mean field theory.  The Amperean attraction between the pair of 
fermions which enters at order $1/N$ compensates this effect, 
leading to a slower decay of the boson correlator.  
For $\eta=1$ the two competing effects exactly compensate one another.

Next we consider the density-density correlator.
Just as for the boson correlator, the density correlator will
satisfy a scaling form:
\begin{equation}
D_{d_\alpha}^{(N)}(2{\bf k}_{F_\alpha} + {\bf q}, \omega) 
\sim |q_x|^{1-\gamma_\alpha} 
\tilde{D}_{2{\bf k}_{F_\alpha}} 
\left( \frac{q_y^2}{q_x}, \frac{\omega^{2/3}}{q_x} \right) ~,
\end{equation}
with an anomalous exponent that depends on the Fermi surface patches,
see Eq.~(\ref{anomalous;density;exponent}).
Back in real space, the dominant density correlator oscillates with 
wave vector $2{\bf k}_{F_\alpha}$, with an envelope decaying as a power 
law with the anomalous exponent,
\begin{equation}
D_{d_\alpha}^{(N)}({\bf r}) \sim 
- \frac{\cos[2{\bf k}_{F_\alpha} \cdot {\bf r} - 3\pi/2]}
       {|{\bf r}|^{4-\gamma_\alpha}}
- \frac{1}{|{\bf r}|^4} ~.
\end{equation}
We have also indicated that the zero momentum component is
not modified relative to the $N=\infty$ behavior.
Again, both ${\bf k}_{F_\alpha}$ and $\gamma_\alpha$ are particular to 
the Fermi surface patches with normals parallel to $\pm \hat{\bf r}$.
In contrast to the particle-particle channel, the particle-hole channel 
is not nested so that we suspect the exponents $\gamma_\alpha$ 
will be smaller than $\eta$ for the physically relevant case
with $N=1$ . 

In sum, at leading order in our systematic $1/N$ expansion, the boson 
correlator was found to exhibit power law decay with an exponent that 
depends on the bare Fermi surface curvatures, varying continuously 
around the Fermi surface.
Similar behavior was found for the density-density correlator.
Provided this qualitative behavior persists down to the physically 
relevant case of $N=1$, we conclude that the DBL phase is described 
by a manifold of scale invariant theories rather than an isolated 
fixed point.  Within the $1/N$ expansion, the power law form obtained 
for the boson correlator relied on the exponentiation of the logarithmic 
behavior on momentum and frequency of the leading $1/N$ correction.  
In the next section, we check the legitimacy of this procedure 
by revisiting the renormalization group approach developed earlier
to describe the low energy physics of a sea of fermions coupled
to a $U(1)$ gauge field.  

We conclude this section by pointing out that throughout we worked
on the assumption of having both $d_1$ and $d_2$ parallel patches
present as in Fig.~\ref{fig:patches}.  In the case with open Fermi 
surfaces shown in the right panel of Fig.~\ref{fig:introFS}, 
we encounter situations when for a given observation direction 
$\hat{\bf r}$ only one Fermi surface (or even none at all) has patches 
with normals in this direction.
In such cases, the boson correlator will decay exponentially; 
there is no inter-species gauge interactions (one species is simply
absent), but the fermions that are present are still strongly 
affected by the gauge field, and in particular the preceding analysis of 
their density correlations remains the same.
In the DLBL phase, there are no parallel patches on the entire two Fermi 
surfaces, so the $d_1$ and $d_2$ fermions effectively decouple from 
each other.  It is also interesting to remark that
in the limit of extreme D-eccentricity when the Fermi surfaces
are completely flat, we expect that the relevant gauge field
is very strongly damped.  This can be seen, e.g., from the naive
RPA approximation for the Landau damping coefficient $\Gamma$, 
Eq.~(\ref{Gamma_RPA}), in the limit of vanishing curvatures.
In this limit, the energy scale $\Omega$, Eq.~(\ref{Omega}), 
below which fermions become incoherent, goes to zero, 
which leads us to speculate that perhaps in this case the 
$d_1$ and $d_2$ fermions behave as essentially free.

\section{Effective field theory for the DBL: Criticality and Stability}
\label{sec:eff_FT}

Polchinski\cite{Polchinski} and others\cite{Altshuler, Nayak} 
argued that 2d fermions interacting with a $U(1)$ gauge field can be 
fruitfully studied within a renormalization group analysis of a 
particular effective field theory.  
In the present case, the fixed point theory has an action given by,
\begin{equation}
S_{\rm fixed\, point} = S_d + S_a + S_{int} ~,
\end{equation}
with
\begin{widetext}
\begin{equation}
S_d = \sum_{\alpha s} \int_{{\bf q}\omega} 
d^\dagger_{\alpha s}({\bf q},\omega) 
\left( -i \omega \left|\frac{\Omega_\alpha}{\omega}\right|^{1/3} 
       + s v_\alpha q_x + \frac{v_\alpha c_\alpha}{2} q_y^2 
\right) d_{\alpha s}({\bf q},\omega) ~,
\end{equation}
\end{widetext}
while the gauge field part $S_a$ is given in Eq.~(\ref{Sa}) and
the fermion-gauge coupling is specified in Eq.~(\ref{Lint}).
Within this fixed point ansatz, both the gauge field and the fermions 
have singular propagators, but their interaction is local.
The advantage of this effective field theory is that it is possible to 
define a simple renormalization group transformation which leaves the 
Gaussian part of the theory invariant, and under which the interaction 
strength $g$ is a marginal perturbation.  If $g$ is assumed to be small,
it can be treated via a conventional perturbative RG approach as we 
describe below.  One finds that the full effective field theory is 
invariant under the RG, i.e., it is at a fixed point.  This strongly 
suggests that exponentiating the logarithmic singularities present in the
large $N$ approach, as we did to order $1/N$ above, is a correct 
procedure at all orders in $1/N$.  If it is, then at order $1/N^2$ 
one must find $(\log)^2$ singularities with particular coefficients
such that the exponentiation procedure to obtain a power law behavior 
is consistent.

An unsettling drawback with this effective field theory approach is that 
it leaves unclear what constraints must be placed on additional 
interactions that can be added to the theory.
At the very least these interactions must be gauge invariant and 
consistent with momentum conservation, but is it legitimate to require 
that the interactions be local in the fermion fields?  
For example, can one require that four-fermion interactions be local 
with non-singular coefficients, or will singular interactions necessarily
be generated?  Indeed, if one were to formally integrate out the gauge 
field with its singular propagator, one would generate four-fermion 
interactions with a particular singular form.
In what follows we will ignore these subtleties, exploring the possible 
perturbative instabilities driven by non-singular fermion interactions.  
Specifically, we will consider all non-singular quartic interactions 
involving four fermions living near the Fermi surfaces that are 
consistent with the relevant symmetries, most importantly momentum 
conservation.

The fixed point action, as it stands, has 9 parameters,
\begin{eqnarray}
\Gamma, \chi, v_\alpha, c_\alpha, \Omega_\alpha, g ~.
\label{all_params}
\end{eqnarray}
But with an appropriate rescaling of the fermion fields and the gauge 
field together with the momenta and frequency, it is possible to set 
5 of these parameters to unity, 
$\tilde{\Gamma} = \tilde{\chi} = \tilde{g} = \tilde{v}_\alpha = 1$.  
The effective field theory is then specified by 4 dimensionless 
parameters, $\tilde{c}_\alpha = c_\alpha \Gamma / g^2$ and 
$\tilde{\Omega}_\alpha = \Omega_\alpha \chi^2 \Gamma / (g^2 v_\alpha^3)$.
Evidently, this theory is not describing a fixed ``point" per se,
but constitutes a four-dimensional manifold of theories which are 
invariant under the RG.  Establishing definitively that a particular 
bare (lattice) gauge theory Hamiltonian is attracted to this manifold is 
exceedingly difficult.  Arguably, it is even harder to deduce where on 
this manifold the theory flows.  
The values of the dimensionless parameters which are obtained from the 
leading large $N$ analysis can, nevertheless, be used as a rough guide
in addressing both questions.

The RG analysis of the fermion-gauge action, $S_{\rm fixed\, point}$, 
proceeds as follows.  At each stage, the fermion fields reside in the 
appropriate momentum space patches 
$|q_x| < \Lambda$, $|q_y| < (\Lambda/c)^{1/2}$,
where $\Lambda$ is the shell width in the direction normal to the Fermi 
surface.  The corresponding restriction on the frequency is
$|\omega| < (v \Lambda)^{3/2} / \Omega^{1/2}$.
The gauge fields reside in similar momentum-frequency regions but 
centered around zero momentum, and the overall setup is illustrated in 
Fig.~\ref{fig:patches}.
If the dimensionless parameters exhibited earlier are of order one
(which is the case if the parameters are taken from the large-$N$ 
analysis), the corresponding regions are roughly similar for all fermion
and gauge fields.  
Also, in practical RG calculations beyond the tree level it is 
convenient to keep the cutoff only on the frequencies and perform 
unrestricted integrations over the momenta.

We integrate out the high-energy fields from the shell between 
$\Lambda$ and $\Lambda/b$ and then rescale the momenta and
frequencies in order to recover the initial cutoff:
$q_x = q_x'/b$, $q_y = q_y'/b^{1/2}$, and $\omega = \omega'/b^{3/2}$.
We also perform appropriate rescaling of the fields:
$d = b^2 d'$, $a = b^2 a'$.
Upon such tree-level scaling, the fixed-point action remains 
exactly as before, i.e., all couplings remain unchanged.
It is also useful to bear in mind that even though we nominally
restore all cutoffs as we proceed, in terms of the original momenta 
the patches become more and more elongated in the $\hat{\bf y}$ 
direction in Fig.~\ref{fig:patches}.  
In particular, any overlap between two patches that correspond to nearby 
but non-parallel tangents $\hat{\bf y}$ and $\hat{\bf y}'$ 
goes to zero in the low-energy limit.

To proceed with the analysis beyond the tree level, we will work
perturbatively in the (dimensionless) fermion-gauge coupling, 
assuming that it is small.
Quite generally, since the momentum shell RG cannot produce terms that 
are singular at small frequencies and momenta, the couplings 
$\Omega_\alpha$ and $\Gamma$ will not renormalize at any order in the 
perturbation expansion.  On the other hand, the parameters 
$v_\alpha$, $c_\alpha$, $\chi$, and $g$, can potentially flow.
However, a direct examination shows that they do not renormalize at the 
lowest (one-loop) order, and we strongly suspect that this remains true 
at all orders.  The implication is that the fermion-gauge interaction is 
exactly marginal and that $S_{\rm fixed\, point}$ describes a 
manifold of RG fixed points parameterized by 4 dimensionless couplings.
Although the fermion-gauge coupling need not be small, we expect that 
this RG-invariant manifold will extend outside the perturbatively 
accessible regime.
This manifold describes the putative DBL phase.

In order to establish the stability of the DBL, however, we need to 
consider the effects of all symmetry allowed perturbations that can be 
added to $S_{\rm fixed\, point}$.
Stability requires that all such perturbations are irrelevant 
under the RG.  Restricting ourselves to local terms, we focus now on the 
four-fermion interactions which are most likely to destabilize the 
fixed point manifold.

\subsection{Short-range fermion interactions}
\label{subsec:4_fermion}
Let us consider a general quartic term,
\begin{equation}
W\; d^\dagger_{\alpha_1 s_1} d^\dagger_{\alpha_2 s_2}
    d_{\alpha_3 s_3} d_{\alpha_4 s_4} ~,
\end{equation}
which can contain different fermion species labelled by $\alpha$ and 
involve different patches labelled by $s$.
The fermion/patch labels are assumed to be such as to satisfy the 
momentum conservation.  Also, we do not show explicitly the (conserved) 
frequency and momentum of the fermion fields.
The amplitude for this four-fermion interaction, $W$, measures the 
strength of a direct scattering between two fermions. 

At the tree level in the RG, all such short-range interactions
are irrelevant,
\begin{equation}
W \to W' = W/b ~,
\label{Wtree}
\end{equation}
scaling towards zero at low energies.  Physically, the incoherent motion 
of the fermions, manifest in their $\omega^{2/3}$ self energy, leads to 
a reduced phase space for direct two-body interactions relative to the 
case for free fermions.

\begin{figure}
\centerline{\includegraphics[width=2.5in]{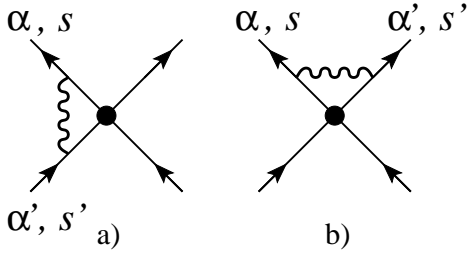}}
\vskip -2mm
\caption{One-loop diagrams needed to calculate renormalization
of four-fermion interactions.
}
\label{fig:4ferm}
\end{figure}

At one loop, a gauge propagator can connect any two legs of the quartic 
vertex which involve fermions located on patches with parallel or 
antiparallel Fermi surface normals.
Two such physically distinct diagrams are shown in Fig.~\ref{fig:4ferm}.
Both processes correspond to an interaction between the two fermions
mediated by the Landau damped gauge field.  The one-loop integrals
needed to evaluate these diagrams are identical to those for the 
vertex corrections in subsection~\ref{subsubsec:V1a}, and give the 
following contributions to the four-fermion interaction before any 
rescaling:
\begin{eqnarray}
\text{a):} \quad
\delta W &=& \text{sign}(g_\alpha g_{\alpha^\prime})
\delta_{s, -s^\prime} \lambda_{ph} \, W \, \ln(b)  ~,
\label{Wph}
\\
\text{b):} \quad
\delta W &=& - \text{sign}(g_\alpha g_{\alpha^\prime})
\delta_{s, -s^\prime} \lambda_{pp} \, W \, \ln(b)  ~.
\label{Wpp}
\end{eqnarray}
Here, the primed and unprimed parameters refer to the possibly distinct 
species of the two fermions connected by the gauge propagator.
The anomalous exponents $\lambda_{ph}$ and $\lambda_{pp}$ are given in 
Eqs.~(\ref{lambda_ph})~and~(\ref{lambda_pp})
-- these formulas are general and do not assume any relations among the 
parameters, as appropriate in the present RG setup; 
sometimes, which will be indicated explicitly, we will use the 
$N=\infty$ expressions Eq.~(\ref{lambda_Ninfty}) to get a crude 
estimate of the numerical values.

Similarly to the analysis in Sec.~\ref{subsubsec:V1a},
either diagram gives zero if the fermions involved have 
parallel group velocities, whereas a nontrivial contribution is found 
when the two fermion fields reside on the opposite patches with 
antiparallel group velocities.
The sign is again determined by Ampere's law:  Two fermions emerging 
from a scattering process with parallel charge currents attract one 
another, spend more time close together, and consequently enhance the 
amplitude for the scattering strength $W$.  Conversely, the interaction 
strength is suppressed when the pair of scattered fermions have 
anti-parallel charge currents.  
The difference in the absolute values of the contributions from the 
diagrams a) and b) is similar to the distinction between the 
particle-hole (CDW) and particle-particle (BCS) processes in a Fermi 
liquid.  Thus, in the situation when $c$ and $c^\prime$ have the same 
sign, the particle-hole process is not ``nested'' because of the 
curvature of the Fermi surface.  On the other hand, the 
particle-particle process is somewhat better nested and becomes 
perfectly nested when $c=c^\prime$, which is the familiar phase space 
reasoning in the theory of the BCS instability.
This distinction between the processes can be clearly seen from the 
arguments $\zeta_{ph}$ and $\zeta_{pp}$, 
Eqs.~(\ref{zeta_ph}) and (\ref{zeta_pp}),
of the function ${\cal E}[\zeta]$ in the two cases.

With the above general expressions in hand, we now consider specific 
four-fermion interactions.  Of interest is the one-loop contribution
$\Upsilon \propto g^2$ to the eigenvalue of such perturbations at the 
fixed point,
\begin{equation}
\frac{dW}{dl} = (-1 + \Upsilon) W + O(W^2) ~.
\end{equation}
The $O(W^2)$ terms, which in general mix different four-fermion terms,
can be deduced similarly to the RG treatment of quartic interactions 
around the Fermi liquid fixed point\cite{Shankar, Polchinski_FL}
and in fact have very similar structure.  In the Fermi liquid case, 
the interactions are marginal and there is no term linear in $W$, 
so the $O(W^2)$ terms determine the physics.
In the fermion-gauge case, on the other hand, the interactions can
be either relevant when $\Upsilon > 1$ or irrelevant when $\Upsilon < 1$.
For $\Upsilon < 1$, the DBL phase is stable to weak perturbations,
but sufficiently strong bare interactions might still drive the 
DBL through a quantum phase transition into another phase.  
In this case the $O(W^2)$ terms could be helpful in discerning the 
nature of the new phase.

As in Fermi liquid theory, the strongest constraint on the allowed
quartic interactions comes from momentum conservation, since all four 
fermions have to live near the Fermi surfaces.
In Fermi liquid theory there are only two allowed types of vertices,
the forward scattering interactions which contribute to the Landau 
parameters and the Cooper vertices.  
We first consider interactions in which all four fermions are of the 
same species.  A general forward scattering interaction then takes the 
form,
\begin{equation}
W^{f}_{{\bf k}, {\bf k}^\prime} 
d^\dagger_{{\bf k}\alpha} d_{{\bf k}\alpha} 
d^\dagger_{{\bf k}^\prime \alpha} d_{{\bf k}^\prime \alpha} ~,
\end{equation}
where ${\bf k}, {\bf k}^\prime$ are two wave vectors on the species 
$\alpha$ Fermi surface.
Consider first the generic case with ${\bf k} \ne \pm {\bf k}^\prime$.
Since this interaction does not involve opposite patches on the 
Fermi surface, the one-loop contribution will vanish, $\Upsilon^f = 0$.
This is also the case when ${\bf k} = {\bf k}^\prime$.
The forward scattering vertex with ${\bf k} = -{\bf k}^\prime$
is part of the Cooper channel to which we now turn.

For a single species of fermion, there exists only the odd angular 
momentum Cooper pairing channel (``triplet pairing''),
\begin{equation}
W^t_{{\bf k}, {\bf k}^\prime} 
d^\dagger_{{\bf k} \alpha} d^\dagger_{-{\bf k} \alpha}
d_{-{\bf k}^\prime \alpha} d_{{\bf k}^\prime \alpha} ,
\end{equation}
with 
$W^{t}_{{\bf k}, {\bf k}^\prime} = -W^{t}_{-{\bf k}, {\bf k}^\prime} 
= -W^{t}_{{\bf k}, -{\bf k}^\prime}$.  
The one loop contribution is negative,
$\Upsilon^t_{{\bf k}, {\bf k}^\prime} = 
- (\lambda_{pp}^{\bf k} + \lambda_{pp}^{{\bf k}^\prime})$
for ${\bf k} \ne \pm {\bf k}^\prime$, due to the Amperean repulsion
between the two particles (and between the two holes) with antiparallel 
group velocities.  
For the special case ${\bf k} = \pm {\bf k}^\prime$ one has 
$\Upsilon^t_{{\bf k}, \pm {\bf k}} = 
- 2 (\lambda_{pp}^{\bf k} - \lambda_{ph}^{\bf k})$ 
since there is now also an Amperean attraction between particles 
and holes on opposite patches of the Fermi surface.  
But since we expect $\lambda_{pp}^{\bf k} \ge \lambda_{ph}^{\bf k}$ 
due to the nesting in the particle-particle channel, one has 
$\Upsilon^t \le 0$ in this case as well.
Thus, the one-loop contribution makes this triplet pairing vertex 
more irrelevant.  Physically the repulsive gauge interaction between 
two $d_\alpha$ fermions is unfavorable for their pairing.

We now turn to vertices which involve two $d_1$ fermions and two $d_2$ 
fermions.  For simplicity, we first focus on the situation with 
vanishing D-eccentricity, so that the two Fermi surfaces coincide.
Consider first the forward scattering interactions coupling the 
densities of the two fermion species,
\begin{equation}
W^{f;1,2}_{{\bf k}, {\bf k}^\prime} 
d^\dagger_{{\bf k} 1} d_{{\bf k} 1} 
d^\dagger_{{\bf k}^\prime 2} d_{{\bf k}^\prime 2} ~.
\label{forward_scattering}
\end{equation}
The only non-vanishing one loop correction is when
${\bf k} = -{\bf k}^\prime$, and gives 
$\Upsilon^{f;1,2}_{{\bf k}, -{\bf k}} = 
2 (\lambda^{\bf k}_{pp} - \lambda^{\bf k}_{ph}) \geq 0$.
If this is greater than one, it would signal a possible instability, 
but in any event we will see that this is smaller than the eigenvalue 
in the conventional BCS ``singlet" pairing channel that we consider next.

The singlet BCS interaction is of the form,
\begin{equation}
W^{s}_{{\bf k}, {\bf k}^\prime} 
d^\dagger_{{\bf k} 1} d^\dagger_{-{\bf k} 2}
d_{-{\bf k}^\prime 2} d_{{\bf k}^\prime 1} ~.
\end{equation}
With zero D-eccentricity as assumed here, the BCS interaction is nested.
Moreover, since the two particles (or the two holes) here carry opposite 
gauge charge and therefore have parallel gauge currents, there will be 
an Amperean attraction giving a positive contribution to the Cooper 
vertex: 
$\Upsilon^{s}_{{\bf k}, {\bf k}^\prime} = 
\lambda_{pp}^{\bf k} + \lambda_{pp}^{{\bf k}^\prime}$,
for ${\bf k} \ne \pm {\bf k}^\prime$.
For a circular Fermi surface, symmetry dictates that 
$\lambda_{pp}^{\bf k}$ will be ${\bf k}$-independent,
but this is actually the case generally for such one-loop contribution
in the case with matched $d_1$ and $d_2$ Fermi surfaces since the 
curvatures on the opposing patches are equal.  
Thus, as claimed earlier, the eigenvalue of the singlet BCS pairing 
interaction, $\Upsilon^s = 2\lambda_{pp}$, is larger than the 
corresponding eigenvalue of the forward scattering interaction coupling 
the densities of the two species on opposite sides of the Fermi surface, 
$\Upsilon^s > \Upsilon^{f;1,2}_{{\bf k}, -{\bf k}}$.
For the special case ${\bf k} = - {\bf k}^\prime$, there will be an 
additional positive contribution to the eigenvalue of the singlet BCS 
pairing interaction due to the Amperean attraction between particles and 
holes,
$\Upsilon^s_{{\bf k}, -{\bf k}} =  
2 (\lambda_{pp}^{\bf k} + \lambda_{ph}^{\bf k})$.  
But being a set of measure zero, this will not contribute to a BCS-type 
pairing instability.
At the present it is not clear what state would be preferred by a 
large interaction of this form and what significance this might have.

It is interesting to use the $N=\infty$ parameters to deduce an 
approximate value for the one-loop contribution to $\Upsilon^s$.  
With zero D-eccentricity, the two Fermi surfaces coincide, 
so the curvatures are equal, $c_1=c_2$, and 
$\lambda_{pp}^{\bf k} = {\cal E}[0] = 1$ for all ${\bf k}$.
This gives  $\Upsilon^s = 2$ and implies that the BCS interaction is 
strongly relevant,
\begin{equation}
\frac{d W^s} {d l} = (-1 + \Upsilon^s) W^s = W^s ~,
\end{equation}
within this approximation.  Since we expect some bare short-range 
attraction between the $d_1$ and $d_2$ particles, the runaway flows will 
lead to BCS pairing.  The composite boson 
$b^\dagger = d_1^\dagger d_2^\dagger$ will condense, resulting in a 
superfluid phase.  Thus, based on this analysis, we suspect that the 
putative S-type Bose liquid phase accessed from the gauge theory with 
identical Fermi surfaces for $d_1$ and the $d_2$ will be generically 
unstable towards superfluidity.

We now consider the situation with non-zero D-eccentricity.
In this case, $d_1$ and $d_2$ have different Fermi surfaces and the 
Cooper vertex will no longer be nested.  This precludes a conventional 
weak coupling BCS instability for the D-wave Bose liquid phase.
The remaining channels that could potentially drive a weak coupling 
instability are the forward scattering interactions given in 
Eq.~(\ref{forward_scattering}) with the locations ${\bf k}$ 
and ${\bf k}^\prime$ on the two Fermi surfaces tuned to have 
anti-parallel patch normals.

For bosons on the square lattice, with increasing D-eccentricity the two 
closed Fermi surfaces will eventually open up, and above some critical 
eccentricity the requirement of anti-parallel $d_1$ and $d_2$ Fermi 
surface normals will be no longer possible.
In this large D-eccentricity regime, when the two Fermi surfaces are so
dissimilar, there are no such forward scattering interactions coupling
the two species which can be enhanced by the gauge fluctuations.
Sometimes one can find interactions that can be enhanced via the 
particle-hole attraction within the same species, e.g., 
$d_{{\bf k} 1}^\dagger d_{{\bf -k} 1} 
 d_{{\bf k}^\prime 2}^\dagger d_{{\bf k}^{\prime\prime} 2}$,
which requires tuning both ${\bf k}^\prime$ and ${\bf k}^{\prime\prime}$
to satisfy momentum conservation but this is not always possible.
In any event, the corresponding one-loop $\Upsilon = \lambda_{ph}^{\bf k}$ 
is not likely to change the irrelevance of this term from the tree level;
e.g. the $N=\infty$ estimate of $\lambda_{ph}^{\bf k}$, 
Eq.~(\ref{lambda_Ninfty}), is always smaller than 1.
Regarding other interactions, none of the intra-species four fermion 
terms have positive eigenvalues.  
We conclude that in this large D-eccentricity regime the D-wave Local 
Bose liquid (as specified in Sec.~\ref{subsec:openFS}) will exist 
as a stable phase.

Returning to the case with smaller D-eccentricity, the stability of the 
DBL phase will depend on the eigenvalues $\Upsilon^f(\hat{\bf n})$ of 
the forward scattering interaction that couples a patch on one Fermi 
surface with normal $\hat{\bf n}$ to a patch on the other Fermi surface 
with anti-parallel normal.  The particles on opposing patches 
experience a strong Amperean attraction, whereas a particle and hole 
experience a weaker repulsion, so that the eigenvalues 
$\Upsilon^f(\hat{\bf n}) = 2(\lambda_{pp} -\lambda_{ph})$ 
will be positive for all orientations of $\hat{\bf n}$.  
However, the magnitude of $\Upsilon^f(\hat{\bf n})$ depends on the 
Fermi surface curvatures and will vary with orientation.
This can be made explicit by evaluating the one-loop contribution using 
the $N=\infty$ values of the various parameters, thereby we obtain an 
approximate expression for the eigenvalue,
\begin{eqnarray*}
\Upsilon^f(\hat{\bf n}) \approx 
2{\cal E} \left[ \frac{\sqrt{3}\pi}{2 g^2} \Gamma |c_1 - c_2| \right]
- 2{\cal E} \left[ \frac{\sqrt{3}\pi}{2 g^2} \Gamma (c_1 + c_2) \right]
~.
\end{eqnarray*}
The first contribution is maximal and equal to $2$ when the curvatures 
on the two Fermi surfaces coincide, which they will when $\hat{\bf n}$ 
is along a diagonal of the square lattice.
We have no such estimate of the second contribution since 
$g, \Gamma, c_\alpha$ are independent in our $N=\infty$ theory; 
if we use in addition a crude RPA approximation Eq.~(\ref{Gamma_RPA}), 
we get the total
$\Upsilon^f = 2 - 2{\cal E}[\sqrt{3}] = 1.53 > 1$.
In this case, the D-wave Bose liquid in this regime of D-eccentricity 
will presumably be unstable, driven by the forward scattering 
interactions with $\hat{\bf n} = \hat{\bf x} \pm \hat{\bf y}$ 
along the diagonals of the square lattice.

The above numerical estimate is very uncontrolled and is given only 
to see possible trends.  Still, let us speculate what the resulting
phase might if there is indeed such instability.
Most naively, there are two guesses, obtained by a mean field 
decoupling of the four fermion interaction
$d^\dagger_{{\bf k} 1} d_{{\bf k} 1} 
 d^\dagger_{{\bf k}^\prime 2} d_{{\bf k}^\prime 2}$,
where ${\bf k}$ and ${\bf k}^\prime$ are the locations of the two
patches with normals $\hat{\bf n}$ and $-\hat{\bf n}$ that
give the largest $\Upsilon^f$.
We can assume a non-vanishing particle-particle or particle-hole 
condensate, 
$\la d^\dagger_{{\bf k} 1} d^\dagger_{{\bf k}^\prime 2} \ra $ or
$\la d^\dagger_{{\bf k} 1} d_{{\bf k}^\prime 2} \ra$, respectively.
Both condensates carry non-zero momentum,
${\bf k} + {\bf k}^\prime$ for the particle-particle condensate, and 
${\bf k} - {\bf k}^\prime$ for the particle-hole.  
But since the interaction is repulsive in the particle-hole channel,
such a condensate seems rather unlikely.
Also, the combination $d_1^\dagger d_2$ carries a non-zero gauge charge, 
so is unphysical and cannot serve as a legitimate order parameter
[but the product 
$\la d^\dagger_{{\bf k} 1} d_{{\bf k}^\prime 2} 
     d^\dagger_{-{\bf k}^\prime 2} d_{-{\bf k} 1} \ra$ 
is gauge invariant, and if condensed would correspond to
an energy density wave at momentum $2 ({\bf k} - {\bf k}^\prime)$].

If there is an instability, we think it is more likely to occur in the 
particle-particle channel.  The order parameter 
$\la d^\dagger_{{\bf k} 1} d^\dagger_{{\bf k}^\prime 2} \ra $ 
is gauge invariant and correspond physically to a finite momentum 
Bose condensate.  
The situation is analogous to the FFLO problem, and as there, provided 
the condensate is not too strong, it will only gap out parts of the 
Fermi surfaces.  As such, this would still be a very unusual type of 
superfluid, which in a mean field description would have residual 
gapless fermionic excitations.  Beyond the mean field, the gauge 
fluctuations would still scatter the gapless fermions, rendering them 
incoherent.  This state would thus correspond to finite momentum 
Bose condensation co-existing with the DBL.

In sum, whether or not the DBL for small D-eccentricity is unstable 
to such finite momentum pairing is a quantitative issue which will 
depend on the values of the anomalous dimensions.  
An analysis of the actual wavefunctions in Sec.~\ref{sec:Wavefnc_props}
reveals no such tendencies, so might be taken as an indication against 
the instability.  
Indeed, as we saw in Sec.~\ref{sec:Wavefnc_props}, the S-wave Bose 
liquid wavefunction does appear generically unstable towards a 
conventional superfluid, consistent with the expectation of a 
zero momentum BCS instability in the gauge theory.   
In the same spirit, we will see in the next section that for hard-core 
bosons at half-filling the DBL wavefunction in the extreme 
D-eccentricity limit (corresponding to fermions which can only hop in 
one of the two directions on the square lattice) 
reveals a co-existence of a commensurate $(\pi, \pi)$ CDW with a gapless 
DBL Fermi surface.
This indicates that the Gutzwiller wavefunctions for the DBL, 
at least in some special instances, can also reveal translational 
symmetry breaking instabilities,
while we repeat that no such instabilities are observed for generic DBL
wavefunctions.

This concludes our discussion of the effective field theory description
and possible instabilities of the DBL phase.
As we have seen, the main potential instability involves pairing 
$d_1$ and $d_2$ fermions moving with opposite group velocities,
but such BCS channel is suppressed for mismatched Fermi surfaces
and is completely eliminated in the DLBL regime, which appears to
be particularly stable.

\section{Ring Hamiltonian Energetics for the DBL phase}
\label{sec:Energetics}

In this section we ask what energetics may stabilize the DBL or DLBL
phases.  Specifically, we motivate and study the $J - K_4$ Hamiltonian 
Eq.~(\ref{Hring}) with competing boson hopping and ring exchange terms.
When $J>0$ and $K_4>0$, this Hamiltonian does not satisfy the Marshall 
sign conditions, so one expects the ground state wavefunction to take 
both positive and negative values.

The Hamiltonian~(\ref{Hring}) is motivated by considering the 
gauge theory description Eq.~(\ref{HU1}) of the DBL phase in the 
strong coupling limit of the gauge theory, 
$h \gg K, t_\parallel, t_\perp$.  
In this limit, one can perturbatively eliminate the gauge field;
the resulting boson Hamiltonian contains, among other terms, both $H_J$ 
and $H_4$ with 
$J = A\, t_\parallel t_\perp / h$ and
$K_4 = A'\, K t_\parallel^4 / h^4 
- A''\, t_\parallel^2 t_\perp^2 / h^3$ 
with positive numerical coefficients $A, A', A''$.
Only the signs of the contributions to $J$ and $K_4$ are of 
interest here, since in what follows both $J$ and $K_4$ are taken as 
free parameters of the Hamiltonian Eq.~(\ref{Hring}).
Coming from the gauge theory, we have $J>0$ and $K_4>0$ in the regime of 
primary interest when $K \sim h$ and $t_\parallel \gg t_\perp$.
To define the system, we also need to specify the boson density per site 
$\rho$.  The hard-core boson model has particle-hole symmetry,
so it is enough to consider $\rho \leq 1/2$.

Before proceeding, we note that the above Hamiltonian with $J>0$ and 
$K_4<0$ was introduced and analyzed in Ref.~\onlinecite{Paramekanti}.
Since there is no sign problem in this case, extensive quantum Monte 
Carlo studies of Refs.~\onlinecite{Sandvik, Melko, Rousseau} were
able to map out the phase diagram of the model.
In addition to a bosonic superfluid phase, the numerical studies 
found two ordered phases at half-filling, a bond-ordered stripe phase 
and a staggered charge density wave (CDW).  When the Monte Carlo was 
performed at fixed chemical potential,\cite{Melko} varying the chemical 
potential drove first-order transitions out of the commensurate solid 
phases into the superfluid.
On the other hand, when the Monte Carlo was performed at fixed 
density,\cite{Rousseau} the superfluid phase was observed as soon as the 
density dropped below half-filling, while a phase separation occurred
when the density was decreased further.

The case $J>0$ and $K_4>0$ was not studied in Monte Carlo since there is 
a sign problem.  Below, we present a rudimentary variational energetics 
study of this regime.  
The results are summarized in Fig.~\ref{fig:tKphased} and lend some 
support that the DLBL phase of Sec.~\ref{subsec:openFS} may be 
stabilized by such competing hopping and ring exchange terms.

\subsection{Model with ring exchanges only}
Consider first the Hamiltonian with the ring term only.
In this case, we can change the sign of $K_4$ by dividing the 
square lattice into four sublattices and performing a transformation 
$b \to -b$ on one sublattice.  
The cited quantum Monte Carlo studies did not consider the pure ring 
model, but extrapolating these to $J=0$, it seems likely that at 
half-filling the ground state is the $(\pi, \pi)$~CDW.
However, we also expect that there is a regime away from half-filling
that realizes a novel ``Excitonic Bose Liquid'' (EBL) phase
discovered and studied by Paramekanti\etal,\cite{Paramekanti}
which we now discuss.

The EBL was accessed by
considering a rotor version of the ring Hamiltonian,
\begin{eqnarray*}
H_{\rm rotor} &=& \frac{\cal U}{2} \sum_{\bf r} (n_{\bf r} - \bar{n})^2
\\
&-& |{\cal K}|
\sum_{\bf r} \cos(\phi_{\bf r}
                  - \phi_{{\bf r} + \hat{\bf x}}
                  + \phi_{{\bf r} + \hat{\bf x} + \hat{\bf y}} 
                  - \phi_{{\bf r} + \hat{\bf y}}) ~,
\end{eqnarray*}
where the phase $\phi_{\bf r}$ and the boson density $n_{\bf r}$ are 
canonically conjugate.
When the cosine in the ring term is expanded, i.e., in the phase with 
no topological defects, one finds a quasi--one-dimensional dispersion 
relation for the excitons (density waves),
$\omega_{\bf k}^2 \sim \sin^2(k_x/2) \sin^2(k_y/2)$.
This gapless normal Bose state is expected to be stable 
against becoming an insulator when $|{\cal K}|$ dominates over 
${\cal U}$.

The ring only model on the square lattice has special conservation laws: 
particle numbers on each column and each row are individually conserved.
It is this property that is responsible for the vanishing of 
$\omega_{\bf k}$ along the lines $k_x=0$ and $k_y=0$ in the EBL phase.  
Paramekanti\etal\cite{Paramekanti} showed that due to the gapless lines,
EBL is a critical (power-law) 2d quantum phase with continuously
varying exponents.

For our energetics study, it is useful to have a good trial wavefunction 
for the EBL phase of hard-core bosons.  Motivated by the described 
``spin wave'' theory, we write the following wavefunction
\begin{equation}
\Psi_{\rm EBL}^{(K_4 < 0)}({\bf r}_1, {\bf r}_2, \dots)
\;\;\propto\;\; \exp[-\sum_{i<j} u_{\rm EBL}({\bf r}_i - {\bf r}_j)] ~,
\end{equation}
with
\begin{equation}
u_{\rm EBL}({\bf r}-{\bf r}') = 
\frac{1}{V} \sum_{\bf k} \frac{\cal W}{4 |\sin(k_x/2)| |\sin(k_y/2)|} 
e^{i {\bf k} \cdot ({\bf r} - {\bf r}')} ~. 
\end{equation}
In the spin wave theory, we have ${\cal W} = \sqrt{\cal U/|K|}$, 
but more generally ${\cal W}$ is treated as a variational parameter.
The summation over ${\bf k}$ excludes the lines where either
$k_x=0$ or $k_y = 0$.
Such wavefunction can be defined in any sector with fixed number
of particles on each row and column as is appropriate in the study
of the ring-only model.
We find that for moderate ${\cal W}$ and away from half-filling,
this wavefunction has a density structure factor of the form,
$\la \hat\rho(-{\bf k}) \hat\rho({\bf k}) \ra \sim |k_x| |k_y|$
at long wavelengths as expected in the EBL.
Furthermore, the EBL theory predicts that the box correlator defined in 
Eq.~(\ref{Bb_def}) decays as a power law with a continuously varying
exponent, and this is what we observe for the wavefunction.
In the energetics study, the parameter ${\cal W}$ is varied to minimize
the ring energy $H_4$.

Returning to the model with $K_4 > 0$, the trial wavefunction is
\begin{equation}
\Psi_{\rm EBL}^{(K_4 > 0)}({\bf r}_1, {\bf r}_2, \dots)
\propto (-1)^{N_{\rm I}} 
\exp[-\sum_{i<j} u_{\rm EBL}({\bf r}_i - {\bf r}_j)] ~,
\label{Psi_EBL}
\end{equation}
where $N_{\rm I}$ is the number of bosons on the sublattice ${\rm I}$ of 
the four sublattices.  One can readily translate the preceding results
to this wavefunction, which has the sign structure appropriate 
for the pure ring exchange model with $K_4 > 0$.

Interestingly, the DLBL wavefunction in the extreme D-eccentricity
limit, which was introduced towards the end of Sec.~\ref{subsec:openFS}
and discussed further in Sec.~\ref{subsubsec:xtrmDLBL},
also has the correct sign structure for the ring Hamiltonian.
This follows from an observation that this wavefunction takes 
opposite signs for any two boson configurations that are connected by 
$H_4$.  It is important here that $H_4$ contains only unit square 
plackets, and one subtlety is that such unit ring moves are not ergodic 
in the sectors mentioned earlier with fixed boson numbers on each row 
and column.  The signs of the EBL and extremal DLBL wavefunctions 
as written will not agree everywhere.
However, in the study of the ring model, it is natural to consider 
smaller sectors comprising states that are connected by repeated 
applications of $H_4$, and in such sectors the signs of the two 
wavefunctions agree.  
Thus, one needs to carefully specify what sector restriction is being 
made when working with these wavefunctions.  As far as the $K_4$ 
energetics is concerned, we find essentially no difference between the 
wavefunctions obtained by restricting to the smaller (ergodic) sectors 
or the larger sectors which fix only the boson numbers on each row and 
column.
The presented results are all for the latter choice of sectors;
the Monte Carlo sampling is performed using arbitrary rectangular 
ring moves, which are ergodic in this sector.

\begin{figure}
\centerline{\includegraphics[width=\columnwidth]{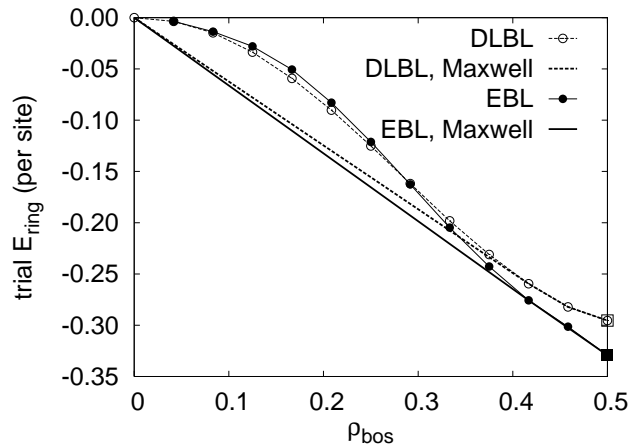}}
\vskip -2mm
\caption{
Trial ring energy in the optimized EBL wavefunction (filled circles)
and extremal DLBL wavefunction (open circles).
Measurements are performed on a $24 \times 24$ system;
the boson number is always a multiple of 24 since the extremal DLBL 
requires that the number of particles is exactly the same in each 
row/column.
Away from half-filling the optimized EBL and DLBL wavefunctions
produce uniform liquids, but exactly at half-filling we find a long-range
$(\pi,\pi)$ CDW order in both states, which is emphasized with 
large box symbols.
The thin lines are guides to eye, while the thick lines are the result 
of the Maxwell construction for the two trial energies that addresses
the possibility of the phase separation at low densities.  
From this construction, the uniform liquids appear to be stable in the 
the density windows  $0.42 < \rho < 0.50$.
}
\label{fig:etrial_ring}
\end{figure}

Figure~\ref{fig:etrial_ring} shows the ring exchange energy per site 
for the optimized EBL and extremal DLBL wavefunctions plotted vs boson 
density.  The two trial energies are fairly close; the EBL state has 
somewhat lower energy for $\rho \gtrsim 1/4$, and the difference is 
largest near $\rho = 1/2$.
Note that the extremal DLBL wavefunction has no variational 
parameters, while the EBL state has one parameter.
The trial DLBL energy can be somewhat improved by maintaining the 
same sign structure but taking a variable power of the determinants; 
such trial energies (which we do not show) approach closer to the 
optimal EBL energies.
Figure~\ref{fig:etrial_ring} suggests that the DLBL wavefunction 
not only has the right sign structure, but also has similar short range 
correlations to the EBL state.
Of course, additional Jastrow-type factors can be used to further
improve the DLBL energy, but here and below we emphasize the good sign 
structure and correlations that are already present in the 
determinantal construction itself.

The similarity between the EBL and extremal DLBL wavefunctions is also 
revealed in their density structure factors, 
see also Sec.~\ref{subsubsec:xtrmDLBL}.
In both states away from half-filling, for small $|k_x| \ll |k_y|$, 
we can write $D_b(k_x, k_y) = A(k_y) |k_x|$.  
In the DLBL state, $A(k_y)$ appears to be independent of $k_y$.  
On the other hand, in the EBL state, $A(k_y) \sim |k_y|$.  
Thus, the character of the $k_x \to 0$ singularity is the same in both 
states for any non-zero $k_y$, and it is only in the limit 
$k_x, k_y \to 0$ that the two differ.
Interestingly, both wavefunctions also exhibit ``$2 k_F$'' singularities
in the density structure factor.
Such singularities are not apparent in the naive continuum ``spin wave'' 
EBL theory, but manifest themselves in the lattice hard-core boson 
EBL wavefunction.

The study of density correlations also reveals that these optimized 
states at half-filling in fact have spontaneous $(\pi, \pi)$ 
charge order, while away from half-filling no order is observed.
Thus, even though in the present crude energetics work we do not consider
directly conventional ordered states of bosons such as charge or bond 
density waves, the EBL and DLBL wavefunctions provide some access to 
the energetics of such ordered states.  
(Note, however, that these trial states likely have some degree of 
EBL-ness, e.g., finite compressibility; this is analogous to the 
well-known observation that when a conventional Jastrow wavefunction 
is driven into a regime of charge order, it in fact represents a 
supersolid that also supports off-diagonal long range order.)
As we have already mentioned, Monte Carlo studies\cite{Melko, Rousseau}
suggest staggered CDW at $\rho = 1/2$, and our finding of
the same tendency in the considered wavefunctions lends some support
to the goodness of the variational approach.
Conventional boson orders are potentially important near other 
commensurate boson densities but are less important at generic 
$\rho$, and we will not consider charge ordered states further.

Leaving the charge orders aside, an important feature is prominent 
in the energy plots in Fig.~\ref{fig:etrial_ring} for the considered 
uniform liquids.  
We see regions where $\epsilon(\rho)$ is concave, which signals an 
instability towards phase separation.  Using the Maxwell construction 
for the available data, we conclude that the system phase separates
for boson densities $\rho \lesssim 0.42$.  
At low densities, the system will spontaneously separate into an empty 
phase and the EBL phase at density $\rho = 0.42$.
In the region near $\rho = 1/2$, the considered uniform liquids
appear to be stable, except perhaps very close to half-filling.

The tendency in the ring model towards phase separation at low densities
was noted by Rousseau\etal\cite{Rousseau} 
Indeed, if we consider two bosons on an otherwise empty lattice, 
the lowest ring energy is achieved when the two reside on the same 
square placket, since only in this case the ring term is nonzero.
Thus, two bosons can be bound together by the ring energy.
By performing exact diagonalization of small clusters up to $4 \times 4$ 
with open boundary conditions and for different boson numbers, 
we find that the energy per particle is the lowest when the density is 
near half-filling; also, larger open clusters achieve lower energies per 
boson.  
One can construct eigenstates of the ring Hamiltonian at low densities 
composed of such isolated clusters, and the discussed tendencies suggest 
that not only two but many bosons tend to clump together in the 
presence of the ring exchanges.

One more notable feature in the energy plots occurs near half-filling.  
We show only the $\rho \leq 1/2$ part, while the $\rho \geq 1/2$ part 
would be the mirror image due to particle-hole symmetry.
Therefore, $\epsilon(\rho)$ shows a cusp precisely at half-filling --
at least in the EBL case and for the system sizes studied.
Since $\mu = d\epsilon/d\rho$, the cusp may be interpreted as an 
evidence for a charge gap at this density, which appears to be in line 
with the proposal of the CDW order by the Monte Carlo studies.

We do not know whether the above observations persist in the limit
of large systems.  Thus, it may also happen that there will be
no regime of the EBL phase in the specific model if the system chooses to
phase separate into empty space and a half-filled charge insulator.
If there is a stable EBL regime over some density window near 
half-filling, then the phase separation at lower densities is likely to 
be into empty space and an EBL phase of an appropriate density 
determined from the Maxwell construction.
The existence of an EBL phase near half-filling could presumably be 
checked by careful Monte Carlo simulations for the pure ring model.
It is also plausible that the regime of the stable uniform EBL phase 
could be extended by adding interactions that disfavor phase separation, 
but we have not explored this in the present work.

\subsection{Full $J$-$K_4$ Hamiltonian}
We now turn to the case with non-zero boson hopping $J > 0$.
Consider first small $J$, and suppose we start in the uniform
EBL or DLBL regime assuming such exists.  As we have discussed, 
the hopping and ring terms frustrate one another.  
At present, we do not know how to include the effects of
hopping if we begin with the EBL.
Thus, the EBL wavefunction is defined in sectors with fixed 
boson numbers on each row/column, but what relative signs should 
we take for the different sectors that are now mixed by the boson
hopping?  If, for example, we naively extend Eq.~(\ref{Psi_EBL}) to all 
boson configurations, with perhaps modified Jastrow factors, the
expectation value of $H_J$ remains precisely zero.
It seems that one needs to change the sign structure of the wavefunction
in order to treat the frustration accurately.
 
On the other hand, the DLBL wavefunction provides a natural 
continuation from the extremal case of the ring only model, 
and we find that it is indeed capable of utilizing both the ring energy 
and the hopping energy.  More precisely, for non-zero $J$ the optimized 
D-eccentricity is finite and varies as $J$ is increased,
moving towards zero D-eccentricity at large $J$.
Within our restricted variational energetics study this is the strongest 
evidence we can offer that the ground state of the ring Hamiltonian in 
this regime is in the DLBL phase.

\begin{figure}
\centerline{\includegraphics[width=\columnwidth]{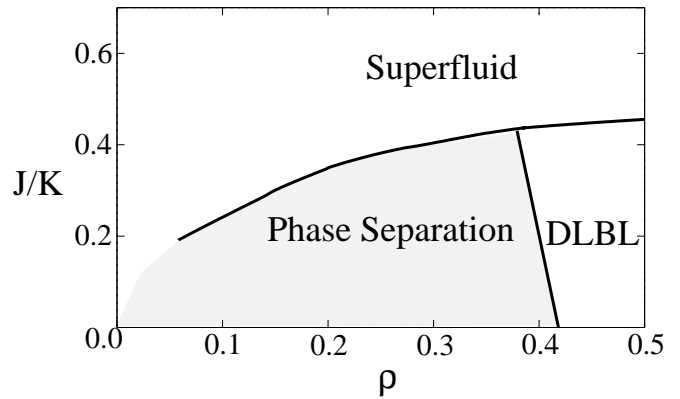}}
\vskip -2mm
\caption{
Schematic ``variational phase diagram'' of the $J - K_4$ model.
Numerics is done for the same sizes as in Fig.~\ref{fig:etrial_ring}.
The phase boundaries are obtained by examining the DBL state, allowing 
for the possibility of phase separation, and the superfluid state
(the latter appears to be always uniform in our study).
The optimal values of $t_\perp / t_\parallel$ throughout the whole 
DBL phase are such that the fermions have open Fermi surfaces,
i.e., this is the DLBL phase of Sec.~\ref{subsec:openFS}.
}
\label{fig:tKphased}
\end{figure}

Some details of the variational study are summarized in the 
``phase diagram'' in Fig.~\ref{fig:tKphased}.
At a given density, the trial energy of the DBL state is optimized
using the ratio $t_\perp / t_\parallel$ as a variational parameter.  
Similarly to the pure ring model, we also analyze the possibility of 
phase separation, which is done in the spirit of 
Fig.~\ref{fig:etrial_ring}.
We find a stable uniform DBL phase in the density range roughly 
similar to the ring only model.  
Physically, we expect the phase separation to be suppressed when the
hopping becomes non-zero, but our study is too limited to explore this 
systematically.

The phase diagram Fig.~\ref{fig:tKphased} also includes a conventional 
superfluid state of bosons, which is expected in the regime of large $J$.
We take the Jastrow-type form for the trial wavefunction,
Eq.~(\ref{Psi_Jastrow}), and use the pseudopotential 
$u({\bf r} - {\bf r}') = W/|{\bf r} - {\bf r}'|^p$ 
with two variational parameters $W$ and $p$.
Allowing more parameters such as an independent nearest-neighbor 
pseudopotential does not visibly change the optimized energies.
Interestingly, as deduced from the energetics, the superfluid-DBL phase 
boundary roughly coincides with the limit of vanishing D-eccentricity
$t_\perp = t_\parallel$.
The resulting $(\det)^2$ wavefunction is positive, and, as we discussed
at length in Sec.~\ref{sec:Wavefnc_props}, this state has off-diagonal 
long range order.  
Of course, it does not have the same long-wavelength correlations 
as the superfluid (e.g., it has unusual $2k_F$ density correlations),
but it appears to have reasonable short-range correlations.

Finally, we also remark that in the present $J - K_4$ study, the 
optimal DBL phase when it wins over the superfluid is in fact the DLBL 
phase with open Fermi surfaces discussed in Sec.~\ref{subsec:openFS} 
and is in the regime where the boson Green's function decays 
exponentially in all directions.
As we discussed in Secs.~\ref{sec:MFT} and \ref{sec:Wavefnc_props},
despite such ``locality'' in the boson correlator, this unusual phase 
is still critical with many low energy excitations.

Summarizing, the DBL wavefunction is able to interpolate 
between the EBL and the superfluid regimes in the frustrated 
$J - K_4$ model.  This is the main argument supporting our proposal 
that the ground state of the ring Hamiltonian in the intermediate regime 
is in the DLBL phase.

\section{Summary and Discussion}
\label{sec:concl}

Our main results were summarized in the Introduction,
so here we only highlight the most interesting points and discuss
possible future directions.
We are searching for examples of uncondensed quantum boson liquids that respect 
time reversal and occur at generic continuously varying densities.
In this paper, we proposed several wavefunctions that produce such
liquids and studied in detail the specific states dubbed DBL and DLBL.
While the initial wavefunction construction appears somewhat ad-hoc,
we argued that it is no more so than the slave particle construction
of spin liquid states in frustrated antiferromagnets, which has been
developed to maturity over the last two decades.  
We introduced a particular slave fermion treatment of hard core bosons
on a square lattice that corresponds to the proposed DBL wavefunctions.
This approach allows to go beyond the trial ground state wavefunction,
suggesting natural excitations, and eventually leads to a gauge theory 
description of the low-energy properties of the putative boson liquid 
phase.
Using techniques previously developed for the so-called uniform RVB
spin liquid with a Fermi sea of spinons,
we argued for the possible stability of the new DBL phases to gauge 
fluctuations and found the DLBL regime to be particularly stable.
Within the gauge theory formulation, we also analyzed boson and boson
density correlators at long distances, which reveal special singular 
surfaces in the momentum space.

The singular surfaces were confirmed by measuring the properties of
the proposed DBL/DLBL wavefunctions directly; in the same numerics,
no potential orders were observed, indicating stability of the states.
We remark here that the Gutzwiller wavefunctions and the gauge theory 
should not be viewed as identical descriptions, and, in particular,
the detailed long-distance properties may be different.
The Gutzwiller wavefunctions can be used for crude energetics
and qualitative characterization, but in our opinion the gauge theory 
is more complete as far as the actual quantum phase is concerned.
Indeed, the wavefunctions as constructed do not incorporate any
dynamics of the gauge fluxes, which are clearly additional low-energy 
``variational'' degrees of freedoms that the quantum system is going to 
explore and utilize.

From the singular momentum surfaces in the boson properties, 
one can in principle recover the underlying Fermi surfaces of the 
slave particle construction; the latter have correct volumes for 
the boson density, offering a tantalizing possibility of some 
``Luttinger theorem'' for uncondensed liquids of interacting
bosons.

Some striking thermodynamic properties of the proposed boson liquids
follow from their fermion - gauge character.\cite{Holstein, Reizer, PALee, IoffeKotliar, LeeNagaosa, Polchinski, Altshuler, Nayak, YBKim, LeeNagaosaWen, Senthil} 
In both the DBL and DLBL phases, because of the gapless Fermi surfaces
and the gauge interactions, the specific heat at low temperatures is 
expected to behave as $C \sim T^{2/3}$, while the resistivity is 
expected to vanish as $R_b \sim T^{4/3}$.
The last result in the DBL phase with closed Fermi surfaces is simply a 
quote from earlier studies of such Fermi sea - gauge 
systems,\cite{IoffeKotliar, LeeNagaosa}
while the case with open Fermi surfaces requires slightly more care.  
Our argument is based on the application of the Ioffe-Larkin 
rule\cite{IoffeLarkin} which gives the boson resistivity as 
$R_b = R_{d_1} + R_{d_2}$.
For the Fermi surfaces as in the right panel of Fig.~\ref{fig:introFS},
and the conductivity measured, say, in the $\hat{\bf x}$ direction,
the small-angle scattering by the gauge field can effectively degrade 
the $d_2$ fermion current, therefore the standard result 
$R_{d_2} \sim T^{4/3}$.
On the other hand, such scattering cannot completely degrade the $d_1$ 
fermion current, so $R_{d_1} \ll R_{d_2}$, and $R_b \sim T^{4/3}$ 
follows. 
(Parenthetically, the $d_1$ current can be degraded by quartic or more 
particle interactions, but such contributions vanish as $T^2$ or faster.)

\begin{figure}
\centerline{\includegraphics[width=3in]{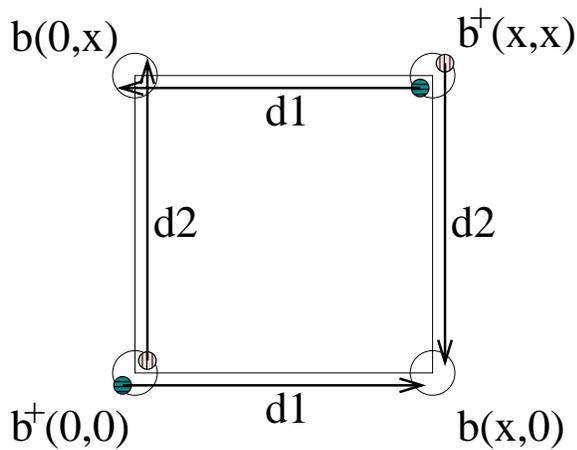}}
\vskip -2mm
\caption{Boson box correlator 
$\la b^\dagger(0,0) b^\dagger(x,x) b(x,0) b(0,x) \ra$
in the DLBL phase reveals the fermionic character of the partons,
decaying as power law and having negative sign from the fermion
exchange.
}
\label{fig:partons}
\end{figure}

Intriguingly, in the DLBL phase, the boson correlators are expected 
to be short-ranged despite such manifest thermodynamic signatures of 
gaplessness and criticality.
On the other hand, the boson box correlator in the DLBL decays as a 
power law and is negative, $-x^{-8}$, offering perhaps a more physical 
glimpse of the fermionic partons in this phase.
Indeed, consider inserting bosons at $(0,0)$ and $(x,x)$
and removing at $(x,0)$ and $(0,x)$.  The dominant contribution
to the box correlator comes from the $d_1$ fermions propagating
$(0,0) \to (x,0)$ and $(x,x) \to (0,x)$, while the $d_2$ paths
are $(0,0) \to (0,x)$ and $(x,x) \to (x,0)$,
which is illustrated in Fig.~\ref{fig:partons}.
The minus sign of the box correlator is then due to one fermionic
exchange needed so that the two pairs of injected fermions are removed 
with the same pairing.

The proposed DBL/DLBL wavefunctions do not satisfy Marshall signs,
and their interesting sign structure is brought out by the nodal 
pictures in the continuum such as Fig.~\ref{fig:nodes},
which give some caricature of the wavefunction signs also on the 
lattice even though the nodes are not sharply defined in this case.
Clearly, these wavefunctions cannot be ground states of boson models 
without frustration.
To address the question of what Hamiltonians may stabilize such phases,
we considered a particular frustrated hard core boson model with
competing hopping and four-site ring exchange terms, 
and our energetics study suggests that the DLBL state is a good 
candidate in this model near half-filling.  
Unfortunately, the frustrated nature of the boson motion
makes this model not suitable for large system quantum Monte Carlo 
studies.  We still suggest it as an interesting model for numerical
studies such as exact diagonalization of small systems or DMRG; 
the gapless nature of the boson liquid may make results hard to 
interpret, but the short-range character of the correlations and 
particularly the quasi-local nature of the DLBL phase can perhaps 
facilitate DMRG to access larger systems.

Looking ahead, we now describe one of the main drives behind the
study of uncondensed boson liquids, despite its own intrinsic interest.
We are searching for electronic conducting non-Fermi liquids,
and the ideas of the present work suggest some avenues in this
direction.
As a specific example, consider the slave boson approach that is
popular in the context of the $t-J$ model of high-$T_c$ superconductors.
The electron operator is written as 
$c^\dagger_\sigma = b^\dagger f^\dagger_\sigma$,
leading to a theory of spinons and slave bosons strongly coupled via
an emergent fluctuating gauge field.
Phenomenologically, a Fermi sea of spinons is very
appealing in the strange metal, but what about the slave bosons?
If they condense, we would recover the conventional Fermi liquid.
It has been argued that the fluctuating gauge field frustrates the 
motion of the bosons and suppresses the tendency to 
condense.\cite{IoffeLarkin, IoffeKotliar, LeeNagaosa, Feigelman, LeeNagaosaWen}
If this suppression could persist to zero temperature, we would
obtain a conducting non-Fermi liquid state.
Such a possibility is precisely what we are trying to establish in the 
present work, and our approach would be to write the slave boson in 
terms of new (second generation) slave fermions, 
$b^\dagger = d_1^\dagger d_2^\dagger$.

On the level of wavefunctions, we would then write an electronic state 
of the form:
\begin{eqnarray*}
\Psi_{\rm electron}(\up, \dn)
= [(\det)_x \times (\det)_y] (\up, \dn) \,\, \det(\up) \, \det(\dn) ~,
\end{eqnarray*}
where schematically $(\up)$ or $(\dn)$ denotes locations
of all electrons with spin-up or spin-down respectively.
Each determinant in the slave boson wavefunction 
$(\det)_x \times (\det)_y$ evaluates appropriate $d$-particle
orbitals at the locations of all electrons irrespective of their
spin, which assures the no-double occupancy constraint.  
There is significant freedom in specifying these orbitals,
and whether such wavefunctions are useful for any electronic system
requires detailed energetics studies of specific Hamiltonians.
One can nevertheless develop a low-energy theory of such a $T=0$ phase 
in the spirit of the present paper, thereby obtaining an itinerant 
non-Fermi liquid conducting phase.
If the DBL or DLBL bosonic states studied in this work appear as
useful caricatures of the behavior of the electronic charge, 
we may even call the resulting phase a ``D-wave metal''!
Such all-fermion description of electrons on the square lattice with 
no double occupancy has some phenomenological appeal, but detailed 
explorations are left for future work.

\acknowledgments
We would like to acknowledge discussions with T.~Senthil and
A.~Vishwanath, and thank Jason Alicea for help with some calculations.
The work at KITP was supported by the National Science Foundation 
through grants PHY-9907949 and DMR-0529399.




\begin{thebibliography}{10}
\bibitem{footnote:N_flavor}
The structure of the $1/N$ diagrams for the particle-hole and 
particle-particle vertices, figures \ref{fig:Vph} and \ref{fig:Vpp}, 
is specific to our $N$-flavor extension defined in Sec.~\ref{subsec:N_flavor}, 
while different extensions can lead to different organization of 
diagrams.
\end{thebibliography}
\end{document}